\documentclass[12pt]{article}

\usepackage{epsf,epsfig}
\usepackage{graphics}

\usepackage{cite}
\usepackage{amsmath,amssymb,booktabs}
\renewcommand{\arraystretch}{1.2}

\def\slash#1{#1 \hskip-0.45em /}

\def\be{\begin{equation}}
\def\ee{\end{equation}}
\def\bea{\begin{eqnarray}}
\def\eea{\end{eqnarray}}

\def\nnb{\nonumber}
\def\eps{\epsilon}
\def\as{\alpha_s}

\def\np{n_+}
\def\nm{n_-}
\def\DB0{\partial B_0}

\def\Cl2{\mbox{Cl}_2}

\def\slash#1{#1 \hskip-0.45em /}

\newcommand{\PL}[2]{\, {\rm Li}_{#1}\!\left({#2}\right)}

\newcommand{\braket}[1]{\langle #1 \rangle}

\newcommand{\loopint}[1]{\int \!\!\! \frac{d^D #1}{\left(2\pi\right)^D}\!}
\newcommand{\ESGamma}{S_{\Gamma}}
\newcommand{\hs}[1]{\hspace*{#1 pt}}
\newcommand{\vs}[1]{\vspace*{#1 pt}}
\newcommand{\MeijerG}[7]{G^{#1}_{#2} \bigg( #3 \left|
\begin{array}{c}
\left\{ #4 \right\} \, , \, \left\{ #5 \right\} \\
\left\{ #6 \right\} \, , \, \left\{ #7 \right\}
\end{array}
\right)}
\newcommand{\MB}[2]{\hs{-12} \int\limits_{\hs{15}_{ #1 -i \,
\infty}}^{\hs{15}^{ #1 +i\, \infty}} \hs{-15} \frac{d #2}{2\pi i}}
\newcommand{\pFq}[5]{\, \! _{#1} F_{#2}( #3 \, ; \, #4 \, ; \, #5 )}

\def\eps{\epsilon}

\begin{document}
\thispagestyle{empty}

\begin{flushright}
{\small
PITHA~09/28\\
SFB/CPP-09-106\\
SI-HEP-2009-17\\
November 18, 2009}
\end{flushright}

\vspace{0.8cm}
\begin{center}
\Large\bf\boldmath
NNLO vertex corrections to non-leptonic\\ 
$B$ decays: Tree amplitudes
\unboldmath
\end{center}

\vspace{0.8cm}
\begin{center}
{\sc M.~Beneke,
T.~Huber{\begingroup\def\thefootnote{\fnsymbol{footnote}}\footnote[1]{
Address after September 30, 2009:~Fachbereich~7, Universit\"at~Siegen, 
Walter-Flex-Str.~3, 57068~Siegen, Germany}\endgroup}} 
and {\sc Xin-Qiang~Li}{\begingroup\def\thefootnote{\fnsymbol{footnote}}\footnote[2]{Alexander-von-Humboldt Fellow}\endgroup}\\

\vspace{0.7cm}
{\sl Institut f\"ur Theoretische Physik E, RWTH Aachen University\\
D--52056 Aachen, Germany}
\end{center}

\vspace{1.3cm}
\begin{abstract}
\vspace{0.2cm}\noindent
The colour-suppressed tree amplitude in non-leptonic $B$ decays is 
particularly sensitive to perturbative and non-perturbative 
corrections. We calculate the two-loop (NNLO) vertex corrections to the
colour-suppressed and colour-allowed tree amplitudes in QCD factorization. 
Our results are given completely analytically, including the full 
dependence on the charm quark mass. We then update theoretical 
predictions for a range of interesting observables derived from 
$\pi\pi$, $\pi\rho$ and $\rho\rho$ final states that do not 
depend (significantly) on penguin contributions, and hence are 
now available with NNLO accuracy. We observe good agreement with 
experimental data within experimental and theoretical errors,  
except for observables involving the $\pi^0\pi^0$ branching fraction.
\end{abstract}

\newpage
\setcounter{page}{1}

\newpage
\allowdisplaybreaks[2]

\section{Introduction}
\label{sec:intro}

Non-leptonic $B$ decays are among the primary observables at the $B$ 
factories. Their sheer number of more than a hundred final states
offers a large number of observables: branching ratios, CP asymmetries, 
polarizations, as well as certain well-motivated combinations
thereof. The large dataset accumulated by the $B$ factories and the 
Fermilab Tevatron therefore allows for a rich phenomenology and for
precise determinations of CKM and unitarity triangle parameters. 
In the future LHCb and a possible SuperB factory will further
increase the amount of data; and they have the potential to discover yet 
unobserved channels and allow for an even more precise
determination of quark flavour parameters.

Despite the fact that non-leptonic $B$ decays are theoretically not as 
clean as for instance rare or radiative $B$ decays, the amount of 
available experimental data justifies the need for precise theoretical 
predictions. A successful attempt to gain control over the
complicated QCD effects in non-leptonic $B$ decays is QCD factorization
(QCDF)~\cite{Beneke:1999br,Beneke:2000ry,Beneke:2001ev,Beneke:2004bn}. 
Within this approach hadronic matrix elements are represented as
convolutions of perturbative objects (hard-scattering kernels) with 
non-perturbative quantities (light-cone distribution amplitudes,
LCDA). It therefore constitutes a framework that systematically 
disentangles perturbative from non-perturbative physics. The factorization 
formula,
valid at leading order in $\Lambda_{\rm QCD}/m_b$ and to all orders 
in $\alpha_s$, reads
\bea
\label{factformula}
\langle M_1 M_2 | Q_i | \bar{B} \rangle & = &  
i m_B^2 \,\bigg\{f_{+}^{BM_1}(0)
\int_0^1 \! du \; T_{i}^I(u) \, f_{M_2}\phi_{M_2}(u) \nnb \\
&& \hspace*{-1.5cm}
+ \,\int_0^\infty \! d\omega \int _0^1 du dv \; 
T_{i}^{II}(\omega,v,u) \, \hat f_B \phi_B(\omega)  \; f_{M_1}\phi_{M_1}(v) \;
 f_{M_2} \phi_{M_2}(u) \bigg\}.
\eea
The hard-scattering kernel $T^I$ starts
from ${\cal O}(1)$ and contains all vertex interactions, whereas $T^{II}$ 
starts from ${\cal O}(\alpha_s)$ and comprises hard 
spectator-interactions. On the basis of the flavour structure of the 
operators one distinguishes several ``topological amplitudes'', which 
are known completely to next-to-leading order (NLO), 
i.e.\ through ${\cal
O}(\alpha_s)$. The importance of accounting for radiative corrections to 
the hard-scattering kernels can be clearly seen from the 
expression for the colour-suppressed tree amplitude, which at 
NLO reads~\cite{Beneke:1999br}
\bea
 \alpha_2(\pi\pi) &=&   0.220 - \left[ 0.179+ \, 0.077 \, i\right]_{\rm NLO}
+ \left[\frac{r_{\rm sp}}{0.445}\right] 
\left\{\left[0.114\right]_{\rm LOsp}+\left[0.067\right]_{\rm tw3}\right\}.
\label{a2NLO}
\eea
One recognizes the large cancellation between the tree-level and the 
one-loop vertex correction (denoted by ``NLO''), which is due to the 
fact that the Born term is small because of colour suppression. 
This suppression is lifted 
at one-loop due to the appearance of additional
colour structures.\footnote{This suppression can be lifted only once, and 
all following terms should be natural in size as expected from a 
perturbative series. Hence, the size of the one-loop correction 
does not indicate a breakdown of
perturbation theory.} The last term proportional to $r_{\rm sp}$ 
is another colour-unsuppressed NLO correction that originates from 
hard spectator-scattering, which turns out to be significant as well, 
and opposite in sign compared to the one-loop vertex correction.

From this discussion it is clear that the colour-suppressed tree 
amplitude is particularly sensitive to the next-to-next-to-leading order 
(NNLO), which is one of the
reasons for going beyond NLO. This is also motivated by phenomenology 
since for certain quantities one observes deviations from
QCDF predictions with experiment, especially in decay channels which are 
dominated by $\alpha_2$, such as the branching ratio and direct
CP asymmetry of $\bar B\to\pi^0\pi^0$~\cite{Beneke:2003zv}. It is therefore 
interesting and desirable to explore whether the NNLO QCDF
corrections turn the amplitudes into the right direction to cure 
discrepancies between theory and experiment. Moreover, since direct CP
asymmetries only start from ${\cal O}(\alpha_s)$, the 
${\cal O}(\alpha_s^2)$ terms constitute only the first correction and 
are therefore needed in order to decrease the scale uncertainty 
stemming from the leading-order terms.

Part of these ${\cal O}(\alpha_s^2)$ contributions is already complete, 
namely the one-loop hard spectator corrections to $T^{II}$, for
which both the 
tree~\cite{Beneke:2005vv,Pilipp:2007mg,Kivel:2006ki} and 
penguin~\cite{Beneke:2006mk,Jain:2007dy}
topologies have been worked out. Later, the imaginary part of the 
vertex kernel $T^I$~\cite{Bell:2006tz,Bell:2007tv} was computed. 
There is clearly a 
demand for the complete ${\cal O}(\alpha_s^2)$ terms since 
leaving out one of the two terms in the formula (\ref{factformula}) 
in a given order may lead to large unphysical effects, as is evident 
at NLO from (\ref{a2NLO}).
In the present paper we compute the two-loop NNLO corrections to 
the  topological tree amplitudes
$\alpha_1$ and $\alpha_2$, including the real and imaginary part, 
in a fully analytic form.
Recently, the so far missing real part also appeared in~\cite{Bell:2009nk}. 
What we add, besides confirming a technically highly non-trivial 
calculation, is an analytic expression for the charm and bottom mass 
dependence arising from massive
quark-loop insertions into the gluon propagator. This should be  
useful for the NNLO calculation of the remaining penguin amplitudes 
that we plan to address in a later work~\cite{BBHLpenguin}. We 
also discuss the numerical impact of the two-loop correction  
on $B\to \pi\pi, \pi\rho, \rho\rho$ branching fractions and branching 
fraction ratios, updating some results of~\cite{Beneke:2003zv}, which 
are now available with NNLO accuracy. A complementary analysis of these 
modes at NNLO can be found in~\cite{Bell:2009fm}.

Our article is organized as follows. In Section~\ref{sec:matching} we 
introduce the theoretical framework and explain
the matching of QCD onto SCET. In Section~\ref{sec:methods} we describe 
the techniques applied during the two-loop calculation. In
Section~\ref{sec:masterformulae} we give the master formulas for the 
hard-scattering kernels, and Section~\ref{sec:gegenbauer} contains 
the results of their convolution in Gegenbauer moments. 
The non-leptonic $B\to \pi\pi, \pi\rho, \rho\rho$ branching fraction 
and ratio analysis is found in Section~\ref{sec:pheno}.
We will conclude in Section~\ref{sec:conclude}. The lengthy analytic 
expressions for the unintegrated hard-scattering kernels and a discussion 
of the two-loop master integrals can be found in two appendices.

\section{Theoretical framework, matching QCD to SCET}
\label{sec:matching}
The starting point of our calculation is the effective weak 
Hamiltonian where the top quark and the heavy gauge bosons are integrated 
out~\cite{Buchalla:1995vs}. In order to compute the ${\cal O}(\alpha_s^2)$ 
corrections to $\alpha_1$ and $\alpha_2$ we need the current-current 
operators of the Hamiltonian. Throughout this paper
we work in the Chetyrkin-Misiak-M\"unz (CMM) operator 
basis~\cite{Chetyrkin:1997gb} since it allows to consistently use the 
naive dimensional regularization (NDR) scheme with  
anticommuting $\gamma_5$. We then have
\be
\mathcal{H}_{eff} = \frac{G_F}{\sqrt{2}} \, V_{ud}^\ast \, V_{ub} 
\left(C_1 Q_1 + C_2 Q_2 \right) + {\rm h.c.} \, ,
\ee
where
\bea
Q_1 &=&  \bar u \gamma^\mu T^a (1-\gamma_5) b \;\, 
\bar d \gamma_\mu T^a (1-\gamma_5) u  \, ,
\nonumber \\
Q_2 &=&  \bar u \gamma^\mu (1-\gamma_5) b \;\, 
\bar d \gamma_\mu (1-\gamma_5) u \, .
\label{Q12}
\eea
Since we will use dimensional regularization the operator basis has 
to be extended by so-called evanescent operators.
They are only non-vanishing for $D \neq 4$ space-time dimensions,
and are needed in order to close the operator basis under
renormalization. In our case they read
\bea
E_1^{(1)} &=&\bar u \gamma^\mu\gamma^\nu\gamma^\rho T^a (1-\gamma_5) b \;\, 
\bar d \gamma_\mu\gamma_\nu\gamma_\rho T^a (1-\gamma_5) u - 16 Q_1 \, ,
\nonumber\\
E_2^{(1)} &=&\bar u \gamma^\mu\gamma^\nu\gamma^\rho (1-\gamma_5) b \;\, 
\bar d \gamma_\mu\gamma_\nu\gamma_\rho (1-\gamma_5) u- 16 Q_2  \, , 
\nonumber\\
E_1^{(2)} &=&\bar u \gamma^\mu\gamma^\nu\gamma^\rho\gamma^\sigma
\gamma^\lambda T^a (1-\gamma_5) b \;\, \bar d \gamma_\mu\gamma_\nu
\gamma_\rho\gamma_\sigma\gamma_\lambda T^a (1-\gamma_5) u-
20 E_1^{(1)}-256 Q_1 \, ,
\nonumber\\
E_2^{(2)} &=&\bar u \gamma^\mu\gamma^\nu\gamma^\rho\gamma^\sigma
\gamma^\lambda (1-\gamma_5) b \;\, \bar d \gamma_\mu\gamma_\nu
\gamma_\rho\gamma_\sigma\gamma_\lambda (1-\gamma_5) u - 
20 E_2^{(1)} - 256 Q_2\, .
\eea

Our computation amounts to performing a matching calculation from QCD onto
SCET~\cite{Bauer:2000yr,Bauer:2001yt,Beneke:2002ph,Beneke:2002ni}, which in 
the context of spectator scattering has been worked out 
in~\cite{Beneke:2005vv}. Adopting the notation of that paper for 
the calculation of the vertex correction means that we determine the 
matching coefficients of SCET four-quark operators~\cite{Chay:2003ju}. 
There are two cases to consider depending on how the quark 
flavours from the operators (\ref{Q12}) flow into the final state. 
The first possibility is that the $d$ and $\bar u$ quarks from 
the $(\bar d u)_{\rm V-A}$ current move in nearly the same direction, 
and are therefore described by the same type of collinear fields. 
This gives rise to the colour-allowed topological tree amplitude 
$\alpha_1(M_1 M_2)$, in which the second meson $M_2$ carries the flavour 
quantum numbers of  $[\bar u d]$. We refer to this as the 
``right insertion'' of the operators (\ref{Q12}) \cite{Beneke:2005vv}, 
and express the matrix elements of $Q_i$ as
\begin{equation}
\braket{Q_i} = \sum\limits_a \, H_{ia}  \, \braket{O_a}. 
\label{eq:RI}
\end{equation}
The basis of non-local SCET four-quark operators on the 
right-hand side is given by
\bea
O_1 &=& \bar \chi \, \frac{\slash{n}_{-}}{2} (1-\gamma_5) \chi \;\,
 \bar \xi \, \slash{n}_{_+} (1-\gamma_5) h_v \, , 
\nonumber\\
O_2 &=& \bar \chi \, \frac{\slash{n}_{-}}{2} (1-\gamma_5)
\gamma_{\perp}^{\alpha}\gamma_{\perp}^{\beta}\chi  \;\,
 \bar \xi \, \slash{n}_{_+} (1-\gamma_5)
\gamma_{\perp\beta}\gamma_{\perp\alpha} h_v \, , 
\nonumber \\
O_3 &=&
 \bar \chi \, \frac{\slash{n}_{-}}{2} (1-\gamma_5)
\gamma_{\perp}^{\alpha}\gamma_{\perp}^{\beta} \gamma_{\perp}^{\gamma}
\gamma_{\perp}^{\delta}\chi \;\,
\bar \xi \, \slash{n}_{_+} (1-\gamma_5)
\gamma_{\perp\delta}\gamma_{\perp\gamma}\gamma_{\perp\beta}
\gamma_{\perp\alpha} h_v\, .
\label{scetbasis1}
\eea
Our conventions are as follows. Meson $M_1$, which picks up the 
spectator antiquark from the $\bar B$ meson, moves into the direction 
of the light-like vector $n_-$. The collinear quark field for 
this direction is $\xi$, with $\slash{n}_{-} \xi = 0$. The 
second meson $M_2$ moves into the opposite light-like direction 
$n_+$. The collinear quark field for this direction is 
$\chi$, satisfying  $\slash{n}_{+} \chi =0$. The effective 
heavy-quark field is labelled by the time-like vector 
$v=(n_++n_-)/2$ with $v^2=1$ as usual, and satisfies 
$\slash{v} h_v = h_v$. The perpendicular component of any four-vector, 
including Dirac matrices, is defined by
\be
p^{\mu}  = \np \cdot p \, \frac{\nm^\mu}{2}+ 
\nm \cdot p \,  \frac{\np^\mu}{2} + p_{\perp}^\mu .
\ee
The SCET operators $O_i$ are actually non-local on the light-cone, 
the $n_+$-collinear part being of the form 
$\bar \chi(t n_-)[\ldots]\chi(0)$. The matching coefficients $H_{ia}$ 
in (\ref{eq:RI}) are therefore functions of a single variable, 
usually chosen to be the Fourier conjugate of $t$, and the 
product $H_{ia}  \, \braket{O_a}$ must be interpreted as a convolution 
product. The non-locality, 
as well as the collinear Wilson lines that render the non-local 
product $\bar\chi(t n_-)[\ldots]\chi(0)$ gauge-invariant, and 
also the quark flavour, are not indicated explicitly 
in (\ref{scetbasis1}). We refer to \cite{Beneke:2005vv} for further 
technical details concerning these issues.

The second possibility, called ``wrong insertion'', arises when the 
$u$ quark from the  $(\bar u b)_{\rm V-A}$ current 
and $\bar u$ quark from the $(\bar d u)_{\rm V-A}$ current move in 
nearly the same direction. The corresponding amplitude is 
the colour-suppressed tree amplitude $\alpha_2(M_1 M_2)$ and 
meson $M_2$ is now made up of  $[\bar u u]$. To obtain this 
amplitude, the QCD operators are matched to SCET via 
\begin{equation}
\braket{Q_i} = \sum\limits_a \, \tilde H_{ia}  \, \braket{\tilde O_a}\, ,
\label{eq:WI}
\end{equation}
where now the basis consists of 
\bea
\tilde O_1 &=& \bar \xi \, \gamma_{\perp}^{\alpha} (1-\gamma_5) \chi \;\,
 \bar \chi (1+\gamma_5) \gamma_{\perp\alpha} h_v \, , 
\nonumber \\
\tilde O_2 &=& \bar \xi \, \gamma_{\perp}^{\alpha}\gamma_{\perp}^{\beta}
\gamma_{\perp}^{\gamma} (1-\gamma_5)\chi  \;\,
 \bar \chi (1+\gamma_5) \gamma_{\perp\alpha}\gamma_{\perp\gamma}
\gamma_{\perp\beta} h_v \, , \nonumber \\
\tilde O_3 &=&
 \bar \xi \,
\gamma_{\perp}^{\alpha}\gamma_{\perp}^{\beta}\gamma_{\perp}^{\gamma}
\gamma_{\perp}^{\delta}\gamma_{\perp}^{\epsilon} (1-\gamma_5)\chi \;\,
 \bar \chi (1+\gamma_5)
\gamma_{\perp\alpha}\gamma_{\perp\epsilon}\gamma_{\perp\delta}
\gamma_{\perp\gamma}\gamma_{\perp\beta} h_v \, .
\label{scetbasis2}
\eea
Note that the $\chi$ fields, which carry the flavour of $M_2$, 
now stand in two different fermion bilinears, hence the name ``wrong 
insertion''. The remarks above on non-locality apply to these 
operators as well.

The SCET operators in~(\ref{scetbasis1}) and~(\ref{scetbasis2}) are constructed such that all operators 
with indices $2$ and 
$3$ are evanescent, i.e.\ they vanish in $D=4$ dimensions. Moreover, 
$\tilde O_1$ is Fierz equivalent to $O_1$ in $D=4$ dimensions. Hence, we
treat the difference $\tilde O_1 - O_1$ as another evanescent operator, 
and regard $O_1$ as the only physical SCET operator. Due to the absence 
of soft-gluon interactions between collinear fields in different directions 
in the leading-power SCET Lagrangian after a 
field redefinition~\cite{Bauer:2001yt}, $O_1$ factorizes into a 
$(\bar\chi\chi)$ part, whose SCET matrix element is a light-cone 
distribution amplitude, and a $(\bar \xi h_v)$ part, which defines 
the soft part of a heavy-to-light form 
factor~\cite{Charles:1998dr,Beneke:2000wa}.

\begin{figure}[t]
\centerline{\includegraphics[width=0.6\textwidth]{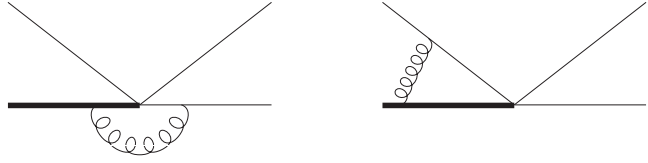}}
\vskip0.2cm
\caption{Examples of factorizable (left) and non-factorzable (right) 
diagrams. By definition, in the former gluon lines do not connect 
the horizontal with the sloped lines (quarks that go into $M_2$), 
in the latter they do connect them. The bold line denotes the bottom quark.}
\label{fig:facnonfac}
\end{figure}

On the QCD side of the matching relations (\ref{eq:RI}), (\ref{eq:WI}) 
we can write the perturbative expansion of the renormalized 
matrix elements as
\bea
\braket{Q_i} &=& \bigg\{ A_{ia}^{(0)} + \frac{\as}{4\pi} 
\left[ A_{ia}^{(1)} + Z_{ext}^{(1)} \, A_{ia}^{(0)} + 
Z_{ij}^{(1)} A_{ja}^{(0)}\right]  \nnb \\
&& \hspace*{0.4cm}+ \,\left(\frac{\as}{4\pi}\right)^2 \left[ A_{ia}^{(2)} + 
Z_{ij}^{(1)} A_{ja}^{(1)} + Z_{ij}^{(2)} A_{ja}^{(0)} + 
Z_{ext}^{(1)} \, A_{ia}^{(1)} + Z_{ext}^{(2)} \, A_{ia}^{(0)}\right. \nnb 
\\
&&\hspace*{0.4cm} 
\left. + \,Z_{ext}^{(1)} \, Z_{ij}^{(1)} A_{ja}^{(0)} + 
Z_{\alpha}^{(1)} A_{ia}^{(1)} + 
\, (-i) \, \delta m^{(1)} \, A^{\prime (1)}_{ia}\right] + 
{\cal O}(\as^3) \bigg\} \, \braket{O_a}^{(0)}
\label{eq:QCDside}
\eea
for the right insertion, and for the wrong insertion we simply replace 
$A \to \tilde A$ and $O \to \tilde O$. In~(\ref{eq:QCDside}), and
throughout the paper, $\alpha_s$ denotes the strong coupling 
in the $\overline{\rm MS}$ scheme with five active quark flavours at 
the scale $\mu$, and a superscript on any quantity $S$ will label the 
order in $\alpha_s$ according to
\bea
S &=& S^{(0)} + \frac{\alpha_s}{4\pi} \, S^{(1)} + 
\left(\frac{\alpha_s}{4\pi}\right)^2 \, S^{(2)} + {\cal O}(\alpha_s^3) \, .
\label{exps}
\eea
The renormalization factors $Z_{ij}$, $Z_{\alpha}$, $\delta m$, 
and $Z_{ext}$ account for operator, coupling, mass, and wave-function
renormalization, respectively. We will give more details on them in 
Section~\ref{sec:masterformulae}.
The $A_{ia}^{(\ell)}$ in~(\ref{eq:QCDside}) denote bare $\ell$-loop 
on-shell matrix elements of the operators $Q_i$ in the effective weak
Hamiltonian. For $\ell=2$ we only need $i=1,2 \,$, but for $\ell < 2$ 
we need in addition the tree-level and one-loop matrix elements of the 
evanescent operators due to terms such as $Z_{ij}^{(1)} A_{ja}^{(1)}$ 
in (\ref{eq:QCDside}). The amplitudes $A_{ia}^{(\ell)}$ can be 
further split up 
into factorizable and non-factorizable diagrams, see
Figure~\ref{fig:facnonfac}, according to
\bea
A_{ia}^{(\ell)}&=& A_{ia}^{(\ell){\rm f}} + A_{ia}^{(\ell){\rm nf}} \, ,
\eea
which turns out to be convenient since in the matching procedure the 
factorizable diagrams will cancel to a large extent.

On the SCET side of the matching relation the matrix elements of
 $O_a$ assume a similar form to~(\ref{eq:QCDside}), but there is no mass counterterm for a
HQET quark, and the coupling is renormalized in the four-flavour theory 
(denoted by a hat on the coupling and renormalization constant),
\bea
\braket{O_a} &=& \left\{ \delta_{ab} + \frac{\hat{\alpha}_s}{4\pi} 
\left[ M_{ab}^{(1)} + Y_{ext}^{(1)} \, \delta_{ab} + 
Y_{ab}^{(1)} \right] \right.
+ \left(\frac{\hat{\alpha}_s}{4\pi}\right)^2 \left[ M_{ab}^{(2)} + 
Y_{ac}^{(1)} M_{cb}^{(1)} + Y_{ab}^{(2)} \right. \nnb \\
&& \left. \left. + Y_{ext}^{(1)} \, M_{ab}^{(1)} + 
Y_{ext}^{(2)} \, \delta_{ab} + Y_{ext}^{(1)} \, Y_{ab}^{(1)} + 
\hat Z_{\alpha}^{(1)}
M_{ab}^{(1)} \right] + {\cal O}(\as^3) \right\} \, \braket{O_b}^{(0)} \, .
\label{eq:SCETside}
\eea
The amplitudes are now labelled $M_{ab}^{(\ell)}$, and $Y$ are the 
SCET renormalization constants. Again, the corresponding formula
for the wrong insertion is obtained by obvious replacements. When 
dimensional regularization is used as IR regulator the on-shell
renormalization constants are equal to unity and the bare matrix 
elements are zero except for the two-loop diagrams with a massive charm
quark-loop insertion,
\bea
\braket{O_a} &=& \left\{ \delta_{ab} + 
\frac{\hat{\alpha}_s}{4\pi} \, Y_{ab}^{(1)} + 
\left(\frac{\hat{\alpha}_s}{4\pi}\right)^2 \left[ M_{ab}^{(2)} +
Y_{ab}^{(2)}\right] + {\cal O}(\as^3) \right\} \, \braket{O_b}^{(0)} \, .
\label{eq:SCETsidedimreg}
\eea
This expression assumes that we determined the operator 
ultraviolet renormalization factors $Y_{ab}$ from the general 
expression~(\ref{eq:SCETside}), using another regulator than 
dimensional for the infrared divergences, 
but then use~(\ref{eq:SCETsidedimreg}) in the matching relation since 
the QCD side (\ref{eq:QCDside}) is calculated with dimensional 
regularization as infrared regulator.  
In this relation we eliminate the four-flavour coupling by the 
$D$-dimensional relation $\hat\alpha_s = \xi_{45}^{-1} 
\alpha_s$, where $\xi_{45}=1+\mathcal{O}(\alpha_s)$ is 
expanded as in (\ref{exps}). Since the SCET operators depend on 
a variable, the renormalization factors $Y_{ab}=Y_{ab}(u,u')$ are functions 
of two variables and the product in (\ref{eq:SCETsidedimreg}) is a 
convolution.

Just as for four-quark operators, we can write down a matching equation 
for QCD {\textit{currents}} to SCET {\textit{currents}}. This is
convenient since in this way we can organize the calculation such that 
the factorizable diagrams are cancelled, at least for the right
insertion. We consider the relations\footnote{The first relation is 
somewhat schematic, since we suppressed the 
non-locality of the operator and the Wilson line connecting the 
quark fields on both sides of the equation.} 
\bea
\bar q \, \frac{\slash{n}_{-}}{2} (1-\gamma_5) q &=& 
C_{\bar q q} \; \bar \chi \, \frac{\slash{n}_{-}}{2} (1-\gamma_5) \chi \, , 
\nonumber \\
\bar q \, \slash{n}_{+} (1-\gamma_5) b &=& 
C_{FF} \; \bar \xi \, \slash{n}_{_+} (1-\gamma_5) h_v \, . 
\label{eq:current2}
\eea
Since a single collinear sector in SCET is 
equivalent to full QCD~\cite{Beneke:2002ph} we 
have $C_{\bar q q}=1+{\cal O}(\as^2)$ because diagrams with massive
bottom quark loops (not contained in SCET) arise only 
at two loops and beyond. The matching coefficient 
$C_{FF}$ can be determined from matching calculations for 
the $b\to u$ transition~\cite{Bauer:2000yr,Beneke:2004rc,Bonciani:2008wf,Asatrian:2008uk,Beneke:2008ei,Bell:2008ws}. 
In the notation of~\cite{Beneke:2008ei} we have
\bea
C_{FF} &=& C_1 + \frac{u}{2} \, C_2 + C_3 {}_{\big|_{u=1}}
= 1 - \frac{\as}{4\pi} \, \frac{C_F}{12} \left(6 L^2+30 L+\pi^2+72\right) + 
{\cal O}(\as^2) \, ,
\eea
with
\be
L \equiv \ln\left(\frac{\mu^2}{m_b^2}\right).
\ee
From~(\ref{eq:current2}) one can construct the 
following factorized QCD operator
\bea
O_{{\rm QCD}} &\equiv& \big[\bar q \, 
\frac{\slash{n}_{-}}{2} (1-\gamma_5) q\big] \, 
\big[\bar q \, \slash{n}_{+} (1-\gamma_5) b\big]
= C_{FF} \, C_{\bar q q} \,O_1 \, ,
\eea
which is {\em defined} by the product of the two currents given above. 
By construction the matrix element of the factorized QCD operator
is the product of a light-cone distribution amplitude and the 
{\em full QCD} form factor. In terms of this operator the matrix 
elements of $Q_i$ read for the right and wrong insertion, respectively,
\bea
\braket{Q_i} &=& T_i \, \braket{O_{{\rm QCD}}} + \sum\limits_{a>1} H_{ia} 
\braket{O_a} \, , 
\label{eq:meOQCD1} \\
\braket{Q_i} &=& \widetilde T_i \, \braket{O_{{\rm QCD}}} + \tilde H_{i1} 
\braket{\tilde O_1 - O_1} + \sum\limits_{a>1} \tilde H_{ia} 
\braket{\tilde O_a}\, . \label{eq:meOQCD2}
\eea
Comparing~(\ref{eq:meOQCD1}) and~(\ref{eq:meOQCD2}) with~(\ref{eq:RI}) 
and~(\ref{eq:WI}) one finally obtains for the hard-scattering
kernels
\bea
T_i &=& \frac{H_{i1}}{C_{FF} \, C_{\bar q q}} \, , \qquad 
\widetilde T_i = \frac{\tilde H_{i1}}{C_{FF} \, C_{\bar q q}} \, . 
\eea


\section{Computational methods}
\label{sec:methods}

The calculation of the two-loop QCD amplitude involves the evaluation of 
the 62 non-factorizable diagrams shown in Figures~15 and~16 
of~\cite{Beneke:2000ry}. In addition, there are four diagrams which 
involve a gluon self-energy,
and in each of these the self-energy loop comprises massless quarks, 
gluons, ghosts, and the massive charm and bottom quarks.

We work in dimensional regularization 
with $D=4-2\eps$, where UV and IR (soft and collinear) divergences
appear as poles of up to the fourth order in $\eps$. Since we work in 
the CMM basis we can
adopt for $\gamma_5$ consistently the NDR scheme with fully
anticommuting $\gamma_5$.

The amplitude of the diagrams is reduced by techniques that have become 
standard in multi-loop calculations. We apply a 
Passarino~--~Veltman~\cite{Passarino:1978jh} reduction to the  vector and 
tensor integrals. The Dirac and color algebra is then performed by
means of an in-house Mathematica routine. The dimensionally 
regularized scalar integrals are further reduced to a small set of 
master integrals using the Laporta 
algorithm~\cite{Laporta:1996mq,Laporta:2001dd} based on 
integration-by-parts~(IBP) identities~\cite{Tkachov:1981wb,Chetyrkin:1981qh}. 
To this end we use 
the package AIR~\cite{Anastasiou:2004vj}.
 
The techniques we apply during the evaluation of the master 
integrals are manifold. The easier integrals can be written in closed 
form in terms of $\Gamma$-functions and hypergeometric functions and 
can subsequently be expanded in $\eps$ with the package {\tt
HypExp}~\cite{Huber:2005yg,Huber:2007dx}. In more complicated cases we 
derive Mellin-Barnes representations by means of the package
AMBRE~\cite{Gluza:2007rt}. We perform the analytic
continuation to $\eps=0$ with the package MB~\cite{Czakon:2005rk}, 
which is also used for numerical cross-checks. We then apply Barnes'
lemmas and the theorem of residues to the multiple Mellin-Barnes integrals, 
and insert integral representations of hypergeometric
functions as well as $\psi$-functions and Euler's $B$-function where 
appropriate. As a third technique we apply the method of
differential equations~\cite{Kotikov:1990kg,Kotikov:1991hm,Kotikov:1991pm} 
and evaluate the boundary condition with the Mellin-Barnes
technique. Eventually, the master integrals are evaluated as 
Laurent series in $\eps$.

The master integrals of all diagrams except the ones in which the gluon 
self-energy contains a charm-quark loop are obtained in a
fully analytic form. The coefficient functions of the Laurent series in 
$\eps$ are logarithms, polylogarithms and harmonic
polylogarithms~\cite{Remiddi:1999ew} of maximum weight four. Two master 
integrals of the diagrams with a charm loop in the gluon self-energy
cannot be displayed in a purely analytic form, see~(\ref{eq:I44}) 
and~(\ref{eq:I45}). Hence the charm-dependent part of the 
hard-scattering kernels still contains two-fold auxiliary integrals, 
see~(\ref{eq:Tcre}),~(\ref{eq:f44}), and~(\ref{eq:f45}). However, after
convolution of the hard-scattering functions with the Gegenbauer 
expansion of the light-cone distribution amplitudes, 
we obtain fully analytic results, including the charm-mass dependence,
for the topological tree amplitudes $\alpha_1$ and $\alpha_2$.

The master integrals have been calculated already in~\cite{Bell:2006tz} 
and they have been used in~\cite{Bell:2007tv}. 
We find perfect agreement on the
expressions given in~\cite{Bell:2006tz}. Nevertheless, we give in 
Appendix~\ref{ap:masters} explicitly the results of a selected subset 
of master integrals.
The reason for this is that we found closed forms valid to all orders 
in~$\eps$, or that we need the structure of the result for the hard-scattering 
kernels and the topological tree amplitudes derived in the subsequent 
sections.

\section{Master formulas}
\label{sec:masterformulae}
From the equations given in Section~\ref{sec:matching} we can deduce the 
master formulas for the hard-scattering kernels. 
Expanding them according to
\bea
T_i &=& T_i^{(0)} + \frac{\alpha_s}{4\pi} \, T_i^{(1)} + 
\left(\frac{\alpha_s}{4\pi}\right)^2 \, T_i^{(2)} + {\cal O}(\alpha_s^3) \, ,
\eea
where $\alpha_s$ denotes the five-flavour coupling in the 
$\overline{{\rm MS}}$ scheme, the master formula for the hard-scattering
kernels for the right insertion reads
\bea
T_i^{(0)} &=& A^{(0)}_{i1} \, , 
\nonumber \\
T_i^{(1)} &=& A^{(1){\rm nf}}_{i1}+ Z_{ij}^{(1)} \, A^{(0)}_{j1} \, , 
\nonumber \\
T_i^{(2)} &=& A^{(2){\rm nf}}_{i1} + Z_{ij}^{(1)} \, A^{(1)}_{j1} + 
Z_{ij}^{(2)} \, A^{(0)}_{j1} + Z_{\alpha}^{(1)} \, A^{(1){\rm nf}}_{i1}+ \, 
(-i) \, \delta m^{(1)} \, A^{\prime (1){\rm nf}}_{i1}\nnb \\
&& - \,T_i^{(1)} \big[C_{FF}^{(1)} +Y_{11}^{(1)}-Z_{ext}^{(1)}\big] - 
\sum_{b>1} H_{ib}^{(1)} \, Y_{b1}^{(1)} \, . \label{eq:hsk2loopRI}
\eea
The actual calculation amounts to evaluating the two-loop 
on-shell matrix element of the transition $b(p)\to 
q_1(p') q_2(u q)\bar q_3(\bar u q)$ with kinematics $p=p'+q$,  
$p^2=m_b^2$, $p^{\prime \,2}=q^2=0$, and $u$ ($\bar u=1-u$) the momentum 
fraction of the quark (antiquark) in $M_2$. Thus, $T_i^{(\ell)}$, 
$A_{k1}^{(\ell)}$ and $H_{ib}^{(\ell)}$ depend on $u$, and a term 
involving $Y_{ab}^{(\ell)}$ such as $H_{ib}^{(1)} \, Y_{b1}^{(1)}$ 
must be interpreted as the convolution product
$\int_0^1 du' \,H_{ib}^{(1)}(u') \, Y_{b1}^{(1)}(u',u)$.

While the individual terms on the right-hand side of 
(\ref{eq:hsk2loopRI}) may be divergent, the 
hard-scattering kernels are free of poles in $\eps$. 
In general, we need tree-level and one-loop quantities
such as  $A_{j1}^{(0)}$, $T_i^{(1)}$, $A_{j1}^{(1)}$ etc. to order
${\cal O}(\epsilon)$ (and even ${\cal O}(\epsilon^2)$ 
for $A_{j1}^{(0)}$ and $T_i^{(1)}$),
since they multiply divergent renormalization 
constants in the expression for the two-loop hard-scattering 
function $T_i^{(2)}$. The $Z_{ij}$ are the operator
renormalization constants from the effective weak Hamiltonian. 
Their matrix expressions read~\cite{Gambino:2003zm,Gorbahn:2004my}
\bea
Z^{(1)} &=& \frac{1}{\eps} \left(\begin{array}{cccccc}
\; -2 \; & \; \frac{4}{3}  \; &  \; \frac{5}{12}  \; & 
\;  \frac{2}{9}  \; &  \; 0  \; & \;  0  \; \\
\; 6  \;  &     \;    0   \;    &    \;      1    \;   &   
\;    0   \;      &  \; 0 \;  &  \; 0 \; 
\end{array}\right) \; , 
\nonumber \\[0.2cm]
Z^{(2)} &=&\frac{1}{\eps^2}
\left(\begin{array}{cccccc}
\; -\frac{4T_f n_f}{3} +17 \; &  
\; \frac{8T_f n_f}{9}-\frac{26}{3}  \; &  
\; \frac{5T_f n_f}{18}-\frac{25}{6}  \; & 
\; \; \frac{4T_f n_f}{27}-\frac{31}{18} \;  &  
\; \frac{19}{96} \;  &  \; \frac{5}{108}  \; \\
\; 4 T_f n_f - 39 \;  &   
\; 4 \; &  
\; \frac{2T_f n_f}{3} -\frac{31}{4} \;  &  
\; 0 \;  &  
\; \frac{5}{24} \;  &  
\; \frac{1}{9} \; 
\end{array}\right) \\
&& +\frac{1}{\eps}
\left(\begin{array}{cccccc}
\; \frac{8T_f n_f}{9} +\frac{79}{12} \; &  
\; \frac{20T_f n_f}{27}-\frac{205}{18}  \; &  
\; -\frac{5T_f n_f}{108}+\frac{1531}{288}  \; & 
\; \; -\frac{2T_f n_f}{81}-\frac{1}{72} \;  &  
\; \frac{1}{384} \;  &  
\; -\frac{35}{864}  \; \\
\; \frac{10 T_f n_f}{3} +\frac{83}{4} \;  &   
\; 3 \; &  
\; -\frac{T_f n_f}{9} +\frac{119}{16} \;  &  
\; \frac{8}{9} \;  &  
\; -\frac{35}{192} \;  &  
\; -\frac{7}{72} \; 
\end{array}\right) , 
\nnb 
\eea
where $n_f = n_l+1=5$ is the total number of quark flavours, and $T_f=1/2$. 
The row index 
of the matrices $Z^{(1)}$ and $Z^{(2)}$ labels 
$(Q_1,Q_2,E_1^{(1)},E_2^{(1)},E_1^{(2)},E_2^{(2)})$,
whereas the column index stands for $(Q_1,Q_2)$. Hence in all terms of 
the form $Z_{ij} A_{j1}$ we also have to consider the matrix elements 
of the evanescent operators, including 
the factorizable diagrams. 
The terms proportional to
$Z_{\alpha}^{(1)}$, $\delta m^{(1)}$, and $Z_{ext}^{(1)}$ account for 
coupling, mass, and wave function renormalization, respectively.
The coupling is renormalized in the $\overline{{\rm MS}}$ scheme with 
five active quark flavours, whereas the mass and the
wave-functions are renormalized in the on-shell scheme.
The term $C_{FF}$ was already discussed in the previous section. 
The SCET renormalization kernel $Y_{11}^{(1)}$ factorizes according
to
\bea
Y_{11}^{(1)}(u,u') &=&  Z_{J}^{(1)}\,\delta(u-u') + 
Z_{BL}^{(1)}(u,u')
\eea
into the universal renormalization factor for SCET currents,
\bea
Z_{J}^{(1)}&=& C_F \left\{-\frac{1}{\epsilon^2} + 
\frac{1}{\epsilon}\left[-L-\frac{5}{2}\right]\right\}\, ,
\eea
and the contribution from the ERBL~\cite{Lepage:1980fj,Efremov:1979qk} 
kernel, for which we use the convention of Eq.~(51) in~\cite{Beneke:2005gs}.
The last term in~(\ref{eq:hsk2loopRI}) emerges from the fact that the 
SCET evanescent operators $O_{2,3}$ mix into $O_1$. Renormalizing 
the evanescent operator matrix elements to zero introduces the 
finite off-diagonal renormalization constants $Y_{b1}^{(1)}$. 
As a consequence, we also have to calculate the one-loop hard-scattering 
kernels of the evanescent operators, $H_{ib}^{(1)}$, $b>1$, for 
which only $b=2$ arises at this order. We find that only the 
diagram with gluon exchange between the $n_+$-collinear quark and 
antiquark lines contributes to the off-diagonal renormalization constant 
$Y_{21}^{(1)}$, as was the case in 
a related calculation performed for the spectator-scattering 
kernels in~\cite{Beneke:2005vv}. The required convolution results in
\bea
- \sum_{b=2,3} H_{1b}^{(1)} \, Y_{b1}^{(1)} &=& - \frac{64\,L}{9} 
- \frac{800}{27}   + 
  \frac{16\,{\pi }^2}{27\,(1-u)} + \frac{16\,{\pi }^2}{27\,u} 
+ \frac{32\,\ln(1 - u)}{9} + 
  \frac{32\,\ln(u)}{9} \nnb \\
  && - \frac{32\,\PL{2}{1 - u}}{9\,u} -
\frac{32\,\PL{2}{u}}{9\,(1-u) }- \frac{32\,i \pi}{9}
\eea
for $i=1$; it vanishes for $i=2$.

The master formula for the expansion of the wrong insertion 
hard-scattering kernels reads
\bea
\widetilde T_i^{(0)} &=& \widetilde A^{(0)}_{i1} \, ,
\nonumber \\
\widetilde T_i^{(1)} &=& \widetilde A^{(1){\rm nf}}_{i1}+ Z_{ij}^{(1)} \, 
\widetilde A^{(0)}_{j1}
  + \underbrace{\widetilde A^{(1){\rm f}}_{i1} - A^{(1){\rm f}}_{21} \, 
\widetilde A^{(0)}_{i1}}_{{\cal{O}}(\eps)}
    - \underbrace{[\widetilde Y_{11}^{(1)}-Y_{11}^{(1)}]\, 
\widetilde A^{(0)}_{i1}}_{{\cal{O}}(\eps)} \, , 
\nonumber \\
\widetilde T_i^{(2)} &=& \widetilde A^{(2){\rm nf}}_{i1} + 
Z_{ij}^{(1)} \, \widetilde A^{(1)}_{j1} + Z_{ij}^{(2)} \, 
\widetilde A^{(0)}_{j1}
+ Z_{\alpha}^{(1)} \, \widetilde A^{(1){\rm nf}}_{i1}\nnb \\
&& + \, (-i) \, \delta m^{(1)} \, \widetilde A^{\prime (1){\rm nf}}_{i1}
+ Z_{ext}^{(1)} \, \big[\widetilde A^{(1){\rm nf}}_{i1}+ Z_{ij}^{(1)} \, 
\widetilde A^{(0)}_{j1}\big]\nnb \\
&& - \,\widetilde T_i^{(1)} \big[  C_{FF}^{(1)} + \widetilde Y_{11}^{(1)}\big] 
- \sum_{b>1} \widetilde H_{ib}^{(1)} \, \widetilde Y_{b1}^{(1)} \nnb\\
&& + \,[\widetilde A^{(2){\rm f}}_{i1} - A^{(2){\rm f}}_{21} \, 
\widetilde A^{(0)}_{i1}] + \, (-i) \, \delta m^{(1)} \, 
[\widetilde A^{\prime (1){\rm f}}_{i1}
- A^{\prime (1){\rm f}}_{21} \, \widetilde A^{(0)}_{i1}]\nnb\\
&& + \,(Z_{\alpha}^{(1)}+Z_{ext}^{(1)})\,
[\widetilde A^{(1){\rm f}}_{i1} - A^{(1){\rm f}}_{21} \, 
\widetilde A^{(0)}_{i1}]\nnb \\
&& - \,[\widetilde M^{(2)}_{11} - M^{(2)}_{11} ] \, 
\widetilde A^{(0)}_{i1} \nnb \\
&& - \,(C_{FF}^{(1)}-\xi_{45}^{(1)})\, 
[\widetilde Y_{11}^{(1)}-Y_{11}^{(1)}] \, \widetilde A^{(0)}_{i1} - 
[\widetilde Y_{11}^{(2)}-Y_{11}^{(2)}]\,
\widetilde A^{(0)}_{i1} \, . \label{eq:master2loopWI}
\eea
All terms in the first three lines of $\widetilde T_i^{(2)}$ have a 
corresponding term in the master formula (\ref{eq:hsk2loopRI}) 
for the right insertion.
The fourth and fifth lines describe Fierz differences of factorizable 
QCD diagrams not contained in the QCD form factor or LCDA. The
sixth line stems from two-loop SCET matrix elements with a massive 
charm-loop insertion. The terms in the last line
of~(\ref{eq:master2loopWI}) emerge from the fact that the difference 
$\tilde O_1 - O_1$ is evanescent and hence we must renormalize its
matrix element to zero. Similar terms already appear in the 
one-loop expression $\widetilde T_i^{(1)}$, in which case, however, they 
are $\mathcal{O}(\eps)$, as indicated, and can be dropped. 

\begin{figure}[t]
\centerline{\includegraphics[width=0.8\textwidth]{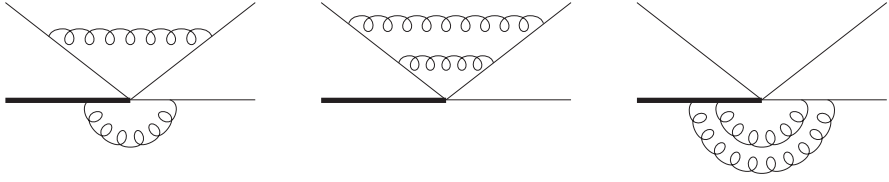}}
\vskip0.2cm
\caption{Examples of factorizable diagrams of class~A (left),~B (center), and~C (right).}
\label{fig:classABC}
\end{figure}

In the two-loop expression, we here find $- \sum_{b>1} 
\widetilde H_{ib}^{(1)} \, \widetilde Y_{b1}^{(1)} = 0$, since 
the mixing of $\tilde O_2$ into $\tilde O_1$ at one-loop is 
of higher order in $\epsilon$. The terms in the fourth to the last line 
of the expression for $\widetilde T_i^{(2)}$ in (\ref{eq:master2loopWI}), 
which appear only for the wrong insertion,  
are given explicitly for $i=1$\footnote{The corresponding terms for 
$i=2$ are in these
cases obtained by removing a factor of $C_F=4/3$.} as follows:
\bea
\widetilde A^{(2){\rm f}}_{11} - A^{(2){\rm f}}_{21} \, 
\widetilde A^{(0)}_{11} &=& \frac{64}{81\, \eps} + \frac{128\,L}{81}  
+ \frac{32}{81}\,n_l \,T_f - \frac{32}{81} \, T_f + \frac{56}{81} \, , 
\qquad  \label{eq:2loopFierzWI} \\
(Z_{\alpha}^{(1)}+Z_{ext}^{(1)})[\widetilde A^{(1){\rm f}}_{11} - 
A^{(1){\rm f}}_{21} \, \widetilde A^{(0)}_{11}] &=& -  
\frac{64}{81}\, n_l \,T_f - \frac{64}{81} \, T_f + \frac{208}{27} \, ,  
\label{eq:ZaZextWI} \\
(-i) \, \delta m^{(1)} \, [\widetilde A^{\prime (1){\rm f}}_{11} - 
A^{\prime (1){\rm f}}_{21} \, \widetilde A^{(0)}_{11}] &=& {\cal O}(\eps)\, ,  \label{eq:massCTWI}\\
- [\widetilde M^{(2)}_{11} - M^{(2)}_{11} ] \, \widetilde A^{(0)}_{11} 
&=& \frac{32}{81} \, T_f \, , \label{eq:2loopMWI} \\
- (C_{FF}^{(1)}-\xi_{45}^{(1)}) [\widetilde Y_{11}^{(1)}-Y_{11}^{(1)}] \,
\widetilde A^{(0)}_{11} &=&{\cal O}(\eps)\, , \label{eq:CFFxi45WI} \\
\big[(\widetilde Y_{11}^{(1)}-Y_{11}^{(1)}) \, \widetilde Y_{11}^{(1)} - 
(\widetilde Y_{11}^{(2)}-Y_{11}^{(2)})\big]\,
\widetilde A^{(0)}_{11} &=& - \frac{32}{81} \, n_l \, T_f + 
\frac{136}{81} \, . \label{eq:2loopY11}
\eea
Let us elaborate more on~(\ref{eq:2loopFierzWI}) --~(\ref{eq:2loopY11}).
In~(\ref{eq:2loopFierzWI}) we have to calculate the factorizable
two-loop diagrams both, for the right and the wrong insertion. 
Since the two are related by a Fierz transformation, many terms drop out, 
resulting in the simple expression (\ref{eq:2loopFierzWI}). 
There are three classes of two-loop diagrams that contribute 
to~(\ref{eq:2loopFierzWI}). They are labelled
class A, B, and C, respectively, and examples are shown in 
Figure~\ref{fig:classABC}.
Diagrams of class~A do not contribute to the on-shell matrix elements since
scaleless integrals vanish in dimensional regularization. For the same reason,
only those diagrams of class~B contribute that contain a massive quark loop
insertion in the gluon propagator. Diagrams of class~C have to be 
evaluated explicitly, and their right-insertion result can be checked 
against the two-loop heavy-to-light current matching calculation 
at the kinematic
endpoint~\cite{Bonciani:2008wf,Asatrian:2008uk,Beneke:2008ei,Bell:2008ws}.
In~(\ref{eq:ZaZextWI}) the renormalization constants multiply the 
Fierz difference of the one-loop factorizable diagram, and 
in~(\ref{eq:massCTWI}) we evaluate the Fierz difference of the 
one-loop factorizable diagram with a mass counterterm
insertion on the heavy line; the latter turns out to be of ${\cal O}(\eps)$.
Eq.~(\ref{eq:2loopMWI}) comprises the Fierz difference of two-loop on-shell
SCET diagrams with a charm-loop insertion into a gluon propagator. 
In order to determine the SCET renormalization constants $Y_{11}^{(\ell)}$
and $\widetilde Y_{11}^{(\ell)}$ in~(\ref{eq:CFFxi45WI})
and~(\ref{eq:2loopY11}), we must use an IR 
regulator different from the dimensional one. The difference 
$\widetilde Y_{11}^{(1)}-Y_{11}^{(1)}$ must then be a finite, 
regulator-independent constant. However, it turns out that this 
difference is ${\cal O}(\epsilon)$, hence there is no contribution 
to $\widetilde T_i^{(2)}$ from~(\ref{eq:CFFxi45WI}).

The last term~(\ref{eq:2loopY11}), which does not appear in this 
form in~(\ref{eq:master2loopWI}), requires further explanation. While 
$Y_{11}^{(\ell)}$ is a $\overline{\rm MS}$ renormalization constant for 
the physical SCET operator $O_1$, $\widetilde Y_{11}^{(\ell)}$ is a 
non-minimal renormalization factor determined by the requirement 
that the matrix element of the evanescent operator $\tilde O_1-O_1$ 
vanishes. As a consequence, $\widetilde Y_{11}^{(1)} -Y_{11}^{(1)}$ 
depends on the  (non-dimensional) infrared regulator at ${\cal O}(\eps^2)$, 
and so does the finite term of the two-loop renormalization factor 
$\widetilde Y_{11}^{(2)} -Y_{11}^{(2)}$, for example from 
a diagram such as the first one in Figure~\ref{fig:classABC}. 
The infrared regulator 
dependence of  $\widetilde Y_{11}^{(1)} -Y_{11}^{(1)}$ enters 
the ${\cal O}(\epsilon^2)$ terms of $\widetilde T_i^{(1)}$ as can be 
seen from~(\ref{eq:master2loopWI}), and subsequently the 
finite term of $\widetilde T_i^{(2)}$ through 
$\widetilde T_i^{(1)} \widetilde Y_{11}^{(1)}$. Of course, the 
final result for  $\widetilde T_i^{(2)}$ must be independent of 
this regulator. To see this we extract the infrared-sensitive 
${\cal O}(\epsilon^2)$ term from  $\widetilde T_i^{(1)}$  by 
decomposing it in the form $\widetilde T_i^{(1)} = 
\widetilde T_i^{(1)F} - [\widetilde Y_{11}^{(1)} -Y_{11}^{(1)}] 
\,\widetilde T_i^{(0)}$ and combine the subtracted term 
with $ - [\widetilde Y_{11}^{(2)}-Y_{11}^{(2)}]\,
\widetilde A^{(0)}_{i1}$ from~(\ref{eq:master2loopWI}). This results 
in the left-hand side of~(\ref{eq:2loopY11}). We then calculate 
the combination of renormalization factors that appears in this 
equation and obtain the infrared-finite result given above. For this 
calculation we used two different methods. One applies the method of 
IR-rearrangement similar to~\cite{Misiak:1994zw,Chetyrkin:1997fm}. 
The other combines terms algebraically such that no infrared regulator 
has to be introduced to calculate~(\ref{eq:2loopY11}). In this way 
we also verify that the finite term is regulator independent. 
Summing all terms we find that  $\widetilde T_i^{(2)}$ is infrared 
and ultraviolet finite as required by the factorization theorem. 

The final result for the hard-scattering kernels is lengthy and 
complicated. We therefore relegate the explicit expressions for the
right-insertion kernels $T_i^{(1)}$ and $T_i^{(2)}$ ($i=1,2$) to 
Appendix~\ref{ap:hsk}.\footnote{Note that the appendix contains the 
one-loop kernels explicitly only through order ${\cal O}(\eps^0)$, 
although their higher order terms enter the two-loop master formulas.} 
It turns out that the result of the calculation of 
the wrong-insertion kernels can be written as linear combination of 
the right-insertion ones and a residual constant as follows:
\bea
\widetilde T_1^{(1)} &=&  -\frac{1}{3} \, T_1^{(1)} -\frac{4}{3}\, , 
\nonumber \\
\widetilde T_2^{(1)} &=&  2 \, T_1^{(1)}\, , 
\nonumber \\
\widetilde T_1^{(2)} &=& -\frac{1}{3} \, T_1^{(2)} 
+\frac{4}{9} \, T_2^{(2)} - \frac{20}{27} \, n_l \, T_f - 
\frac{20}{27} \, T_f-\frac{49}{9} \, ,
\nonumber \\
\widetilde T_2^{(2)} &=& 2 \, T_1^{(2)} +\frac{1}{3} \, T_2^{(2)} + 
6 \,  T_1^{(1)}  \, .
\label{eq:wronglinearright}
\eea


\section{Convolution in Gegenbauer moments}
\label{sec:gegenbauer}

For the topological tree amplitudes to NNLO we adopt the following notation
\bea
\label{alphaiexpansion}
\alpha_i(M_1M_2)&=& \sum\limits_j C_j \, V^{(0)}_{ij} + 
\sum\limits_{l\ge 1} \left(\frac{\alpha_s}{4\pi}\right)^l \, 
\left[\frac{C_F}{2N_c}\sum\limits_j C_j \, V^{(l)}_{ij}+P_i^{(l)}\right] + 
\ldots \, ,
\eea
where the ellipsis stands for contributions from hard 
spectator-scattering which are not subject of the present paper. As before 
$\alpha_s$ denotes
the five-flavour coupling in the $\overline{{\rm MS}}$ scheme. 
The quantities $V_{ij}$ are convolutions of the hard-scattering kernels
with the light-cone distribution amplitude of a light meson,
\bea
V_{1j}^{(0)} &=& \displaystyle\int_0^1 \!\!du \; T_{j}^{(0)} \; 
\phi_{M}(u) \, , 
\nonumber \\
V_{2j}^{(0)} &=& \displaystyle\int_0^1 \!\!du \; \widetilde T_{j}^{(0)} \; 
\phi_{M}(u) \, , 
\nonumber \\
\frac{C_F}{2N_c}V_{1j}^{(l)} &=& \displaystyle\int_0^1 \!\!du \; 
T_{j}^{(l)}(u) \; \phi_{M}(u) \, , 
\nonumber \\
\frac{C_F}{2N_c}V_{2j}^{(l)} &=& \displaystyle\int_0^1 \!\!du \; 
\widetilde T_{j}^{(l)}(u) \; \phi_{M}(u) \, .
\label{ticonv}
\eea
The light-cone distribution amplitude of a light meson, $\phi_{M}$, is 
expanded into the eigenfunctions of
the one-loop renormalization kernel,
\bea
\phi_{M}(u) &=& 6u(1-u) \left[1+\sum_{n=1}^{\infty}a_n^{M} \, 
C^{(3/2)}_n(2u-1)\right] \, ,
\label{gegenbauerexp}
\eea
where $a_n^{M}$ and $C^{(3/2)}_n(x)$ are the Gegenbauer moments and 
polynomials, respectively, and the integrals (\ref{ticonv}) are then 
performed term by term. Below we provide integrated results truncating 
the Gegenbauer expansion (\ref{gegenbauerexp}) at $n=2$, which is 
sufficient in practice. The light-cone distribution amplitude and 
Gegenbauer moments are scale-dependent, and our notation is such 
that $a_n^{M}$ refers to the scale $\mu$ that is also the scale of 
$\alpha_s$. Since $\phi_{M}(u)$ is 
normalized to unity and
the tree-level scattering kernels are constant, we simply have
\bea
V_{11}^{(0)} &=& 0 \, , \qquad   
V_{21}^{(0)} =\displaystyle \frac{4}{9} \, , 
\nonumber \\
V_{12}^{(0)} &=& 1 \, , \qquad   
V_{22}^{(0)} =\displaystyle \frac{1}{3} \, . 
\eea
At the one-loop level, one obtains
\bea
V_{11}^{(1)} &=& -\frac{45}{2} -6 L-3 i \pi  + 
\left( \frac{11}{2}-3 i \pi \right) a_1^{M} -\frac{21}{20} \, a_2^{M}
\, , 
\nonumber \\
V_{12}^{(1)} &=& 0 \, , 
\nonumber \\
V_{21}^{(1)} &=& -\frac{1}{3} \, V_{11}^{(1)} - 6 \, , 
\nonumber \\
V_{22}^{(1)} &=& 2 \, V_{11}^{(1)} \, .
\eea
Note that from one-loop onwards, the constants in the $V_{ij}$ are 
scheme dependent and hence look different in the CMM basis compared
to those in the traditional basis~\cite{Buchalla:1995vs} employed in the 
earlier QCD factorization calculations. This explains the different constant 
in $V_{11}^{(1)}$ compared to the corresponding expression $V_M$ 
in \cite{Beneke:2001ev}. 

At two loops the expressions get more complicated. 
In order not to spoil the simple structure of
the result we split the $V^{(2)}_{ij}$ according to
\bea
V^{(2)}_{ij} &\equiv& V^{(2)}_{ij} {}_{\big|_{m_c=0}} + 
\, T_f \, \Delta V^{(2)}_{ij} {}_{\big|_{m_c}} \, .
\eea
The first term on the right-hand side describes the result for a massless
charm quark. The second term is the difference between the contribution of a
quark of mass $m_c$ and a massless quark, see also~(\ref{eq:hskT22}). 
This construction ensures that we set $n_l=4$ irrespective of
whether we treat the charm quark as massive or massless. Explicitly, we have
\bea
V_{11}^{(2)} {}_{\big|_{m_c=0}} \!\!&=& \left\{ -39 L^2-234 L 
+8194 \zeta (5)-2028 \pi ^2 \zeta (3)-\frac{100862 \zeta (3)}{15}+
44 \pi ^2\ln(2)  \right. \nnb \\
             && \hs{29}-\frac{55 \pi ^4}{216}+\frac{43909 \pi ^2}{18}-
\frac{172597}{180} +n_l \, T_f \left(4 L^2+\frac{70 L}{3}-\frac{4 \pi
	     ^2}{3}+\frac{493}{9}\right) \nnb \\
             && \hs{29}+ T_f\left(4 L^2+\frac{70 L}{3}+
\frac{56 \pi ^4}{75}-\frac{164 \pi ^2}{3}+\frac{1321}{9}+
\frac{4128}{5} \zeta(3)\right.\nnb \\
	     && \hs{60}\left.-\frac{92}{5} \pi^2\sqrt{5} -
\frac{288}{5} \pi ^2 \ln(\textstyle\frac{1}{2}+
\frac{\sqrt{5}}{2}\displaystyle)\right) \nnb \\
	     && +\,i\pi \left[-39 L-\frac{20\zeta (3)}{3}+
\frac{1301 \pi ^2}{270}-\frac{1591}{9}+
n_l \, T_f \left(4 L+\frac{44}{3}\right) \right. \nnb \\
	     && \hs{27.5}\left.\left. + 
T_f \left(4L+\frac{932}{3}-184\sqrt{5}
\ln(\textstyle\frac{1}{2}+\frac{\sqrt{5}}{2}\displaystyle)
	        -288 \ln^2(\textstyle\frac{1}{2}+\frac{\sqrt{5}}{2}\displaystyle)-\frac{192}{5} \zeta (3)\right)\right]\right\} \nnb \\
             &&  + \,a_1^{M}\left\{ \frac{1639 L}{18}-50964 \zeta (5)+12618 \pi ^2\zeta (3)+\frac{208846 \zeta (3)}{5}-324 \pi ^2 \ln (2) \right. \nnb \\
	     && \hs{38}+\frac{1163 \pi^4}{120} -\frac{541129 \pi ^2}{36}+\frac{3604333}{1080} +n_l \, T_f \left( -\frac{22 L}{3}-4 \pi ^2-\frac{80}{3}\right) \nnb \\
             && \hs{38}+ T_f\left(-\frac{22 L}{3}+\frac{19532}{3}+ \! 316 \pi ^2-\frac{544 \pi ^4}{25}
	        -\frac{29856}{5} \zeta (3) \right. \nnb \\
             && \hs{69} \left. -\frac{716}{5}\pi^2\sqrt{5}+\frac{2976}{5}\pi ^2 \ln(\textstyle\frac{1}{2}+\frac{\sqrt{5}}{2}\displaystyle)\right)\nnb \\
	     && +\,i\pi \left[-\frac{149 L}{3} +\frac{103 \pi^2}{10}-\frac{5923}{36} +n_l \, T_f \left(4 L+14\right) + T_f \left(4 L+990-\frac{576}{5}\zeta(3)\right. \right. \nnb \\
	     && \hs{29}-1432 \sqrt{5} \ln(\textstyle\frac{1}{2}+\frac{\sqrt{5}}{2}\displaystyle)
	     +2976\ln^2(\textstyle\frac{1}{2}+\frac{\sqrt{5}}{2}\displaystyle)\bigg)\bigg]\bigg\} \nnb \\
	     &&  + \,a_2^{M}\left\{ -\frac{1169 L}{60}+170274 \zeta (5)-42138 \pi ^2 \zeta (3)-\frac{4896383 \zeta (3)}{35}+\frac{2392}{3} \pi ^2 \ln (2)  \right. \nnb \\
	     && \hs{38}-\frac{199 \pi^4}{12}+\frac{13604203 \pi ^2}{270}-\frac{733820723}{75600}+n_l \, T_f \left( \frac{7 L}{5}-2 \pi ^2+\frac{8059}{300} \right) \nnb \\
	     && \hs{38}+ T_f \left(\frac{7 L}{5} - \frac{1585223}{900}-\frac{3322}{3} \pi ^2
	     +\frac{1112}{25} \pi ^4 +\frac{36288}{5} \zeta (3)\right.\nnb \\
	     && \hs{70} \left. +\frac{2768}{5} \pi ^2\sqrt{5}-\frac{13248}{5} \pi ^2\ln(\textstyle\frac{1}{2}+\frac{\sqrt{5}}{2}\displaystyle) \right) \nnb \\
	     && +\,i\pi \left[ \frac{953 \pi^2}{210}-\frac{47929}{720}+ \frac{6}{5} \, n_l \, T_f  + T_f \left(-\frac{39182}{15}-\frac{1152}{5} \zeta (3) \right.\right. \nnb \\
	     &&\hs{29} +5536 \sqrt{5}\ln(\textstyle\frac{1}{2}+\frac{\sqrt{5}}{2}\displaystyle)
               -13248\ln^2(\textstyle\frac{1}{2}+\frac{\sqrt{5}}{2}\displaystyle)\bigg)\bigg]\bigg\} \, .
\label{eq:T211mc0}
\eea
The corresponding charm-mass dependent part reads
\bea\label{eq:Deltacharm}
\Delta V^{(2)}_{11} {}_{\big|_{m_c}} \!\!&=&\left\{ -\frac{4}{3}z^2 q_3(z) +16  z^2 \ln (z) q_4(z)+24 z^2 q_5(z)-\frac{2}{3} (22z+1)q_6(z)\right. \nnb \\
	     && \hs{26}+18 z (4 z-1) q_7(z)-4 \sqrt{z} \, (28 z-1) q_8(z) +16 z^2 q_{10}(z) +2 z^2 q_{11}(z) \nnb \\
	     && \hs{26}-16 z^{3/2} q_{12}(z)+92 z -2\pi^2 \left(\!23 z^2+3 z+\sqrt{z}-\frac{1}{3}\!\right)\!  -16 \sqrt{z}\, \text{Li}_2\left(1-\sqrt{z} \right) \nnb \\
	     && \hs{26}+2 \left(3 z^2-3 z-1\right) \ln^2(z)+ 76 z \ln (z)+4 \left(3 z^2-3 z+\sqrt{z}+1\right) \text{Li}_2(1-z)\nnb \\
	     &&+\,i\pi\bigg[24 z^2 (2\ln (z)-3) \ln ^2(\eta ) -32 z^2 q_1(z) +296 z \nnb \\
	     && \hs{26}-4 \ln (z)+4 (22 z+1) q_2(z) \bigg]\bigg\}\nnb \\
	     &&  + \,a_1^{M}\bigg\{\!-288 z^2 \ln ^2(\eta ) \text{Li}_2(\eta )-4 z^2 q_3(z) + 8 z^2 \left(12 z^2+16 z+3+6 \ln (z)\right)q_4(z) \nnb \\
	     && \hs{35}-\frac{2}{3} \left(72 z^3+84 z^2+22 z+1\right) q_6(z)-\frac{4}{15 z}\left(45 z^3-5 z^2+3\right) \text{Li}_2(1-z) \nnb \\
	     && \hs{35}+\frac{2}{z} \left(36 z^4-312 z^3+\frac{467}{3} z^2-z+\frac{1}{5}\right) q_7(z)+ 576 z^3-336 z^2  \nnb \\
	     && \hs{35}-\frac{4}{3\sqrt{z}} \left(216 z^4-504 z^3-44 z^2-32 z+\frac{3}{5}\right) q_8(z)+\frac{18892}{3} z \nnb \\
	     && \hs{35}+8 z \left(3 z^3-8 z^2-6 z-3\right) q_{10}(z)-12 z^2 q_{11}(z) +112 z^{3/2} q_{12}(z)\nnb \\
	     && \hs{35}+\frac{2}{15 \sqrt{z}} \left(20 z^2+20 z+3\right) \left(8\text{Li}_2\left(1-\sqrt{z}\right)-2\text{Li}_2(1-z)+\pi^2\right) \nnb \\
	     && \hs{35}+\frac{2 \pi ^2}{15z}\left(2325 z^3+5 z^2+30 z-3\right) + 4 z \left(144 z^2-72 z+613\right) \ln (z)\nnb \\
	     && \hs{35}-\frac{2}{15 z} (45 z^3-5z^2+3) \ln ^2(z)\nnb \\
	     &&	 +\,i\pi\bigg[ 288 z^3+24 z^2 \left(12 z^2+16 z+6 \ln (z)+3\right) \ln ^2(\eta )-96 z^2 q_1(z)+360 z^2 \nnb \\
	     && \hs{35} +328 z-4 \ln (z)+4 \left(72 z^3+84 z^2+22 z+1\right) q_2(z)\bigg]\bigg\} \nnb \\ 
	     &&  + \,a_2^{M}\bigg\{\! -8 z^2 \left(-3 \ln ^2(\eta )-12 \text{Li}_2(\eta )+2 \pi ^2\right)^2 \nnb \\
	     && \hs{35}- 16 z^2 \left(40 z^3+30 z^2-6\ln (z)-1\right) q_4(z) -\frac{2}{3} \pi ^2 z (1649 z+9)\nnb \\
	     && \hs{35}+\frac{8}{3} z \left(120z^3+\!70 z^2-\!11 z-6\right) q_6(z) +\!6 (z-1) z \left(\ln ^2(z)+2 \text{Li}_2(1-z)\right) \nnb \\
	     && \hs{35}-2 z \left(240 z^3-330 z^2-971 z+9\right) q_7(z) \nnb \\
	     && \hs{35}+16  z^{3/2} \left(120 z^3-185 z^2-51 z-147\right) q_8(z) \nnb \\
	     && \hs{35}-4 z^2 \left(40 z^3-75 z^2+11\right) q_{10}(z) +42 z^2 q_{11}(z) -96 z^{3/2} q_{12}(z) \nnb \\
	     && \hs{35} -\frac{4}{3} z \left(2880 z^3-1140 z^2-823 z-54\right) \ln (z)\nnb \\
	     && \hs{35} -\frac{2}{9} z \left(17280 z^3-8280 z^2-2861 z+1908\right) \nnb \\
	     && +\,i\pi\bigg[ -48  z^2 \left(40 z^3+30 z^2-6 \ln
               (z)-1\right) \ln ^2(\eta )-192 z^2 q_1(z) \nnb \\
	     && \hs{29} -\frac{8}{3} z \left(720 z^3+480 z^2-37 z-183\right) 
\nnb \\
	     && \hs{29}-16 z \left(120 z^3+70 z^2-11 z-6\right) q_2(z)\bigg]\bigg\} \; ,
\eea
with $z\equiv m_c^2/m_b^2$. The variable $\eta$ and the functions 
$q_i(z)$ are defined in Appendix~\ref{ap:charmfunc}. One can check
that $\Delta V^{(2)}_{11} {}_{|_{m_c}}$ vanishes for $z\to 0$, and 
that for $z\to 1$ it approaches the corresponding contribution
of a bottom quark contained in~(\ref{eq:T211mc0}).

The case of $V_{12}^{(2)}$ is simpler since the corresponding 
one-loop expression vanishes due to the colour structure,
and therefore one also has
\bea
\Delta V^{(2)}_{12} {}_{\big|_{m_c}} &=&0.
\eea
Hence there is no need to distinguish between the massive and massless 
charm contribution. We obtain
\bea
V_{12}^{(2)} &=& \left\{ 18 L^2+156 L+3744 \zeta (5)-
936 \pi ^2 \zeta (3)-\frac{14833 \zeta (3)}{5}+72 \pi ^2 \ln (2)+
\frac{239 \pi ^4}{90} \right.\nnb \\
             && \hs{9}\left.+\frac{12487 \pi ^2}{12}+
\frac{5347}{60} +i\pi \left[ 18 L-16 \zeta(3)+\frac{47 \pi^2}{45}+
\frac{1333}{12}\right]\right\} \nnb \\
	     && + \,a_1^{M}\left\{ -33 L-19008 \zeta (5)+
4752 \pi^2 \zeta (3)+\frac{77157 \zeta (3)}{5}-24 \pi ^2\ln (2)-
\frac{181 \pi ^4}{10}\right.\nnb \\
	     && \hs{35}\left.-\frac{21807 \pi ^2}{4}+
\frac{4568}{15} +i\pi \left[18 L+\frac{23 \pi^2}{5}+
\frac{15}{4}\right]\right\} \nnb \\   
             && + \,a_2^{M}\left\{ \frac{63 L}{10}+
74304 \zeta (5)-18576 \pi ^2 \zeta (3)-\frac{2236872 \zeta (3)}{35}-
2064 \pi ^2 \ln (2)+\frac{797 \pi ^4}{10} \right.\nnb \\
	     && \hs{35}\left.+\frac{204218 \pi^2}{9}+
\frac{32369221}{12600} +i\pi \left[-\frac{18 \pi ^2}{35}-
\frac{173}{30}\right]\right\} \, .
\eea

The integrated two-loop kernels for the colour-suppressed tree amplitude 
follow from (\ref{eq:wronglinearright}), and read 
\bea     
V_{21}^{(2)} &=& -\frac{1}{3} \, V_{11}^{(2)} +\frac{4}{9} \, 
V_{12}^{(2)} - \frac{10}{3} \, n_l \, T_f - 
\frac{10}{3} \, T_f-\frac{49}{2} \, ,
\nonumber \\
V_{22}^{(2)} &=& 2 \, V_{11}^{(2)} +\frac{1}{3} \, V_{12}^{(2)} + 
6 \,  V_{11}^{(1)} \, .
\eea

Our results agree with those in~\cite{Bell:2007tv,Bell:2009nk}.
The comparison we performed is of analytic nature,
except for those contributions to the real part of the amplitudes
that stem from charm and bottom loop insertions into a gluon
propagator, for which the result has been obtained
in~\cite{Bell:2009nk} only in numerical form. For these terms 
the numerical agreement for the values of the quark masses covered in
Table~1 of~\cite{Bell:2009nk} with our analytic result is within
three permille.

\section{Phenomenological applications}
\label{sec:pheno}

\subsection{Input parameters}
\label{sec:input}

\renewcommand{\arraystretch}{1.3}
\begin{table}[t]
\begin{center}
\begin{tabular}{|cc|cc|}\hline
Parameter & Value/Range & Parameter & Value/Range \\
\hline
&&&\\[-0.5cm]
$\Lambda_{\overline{\mathrm{MS}}}^{(5)}$ & 0.225 &
$\mu_\mathrm{hc}$ & $1.5\pm 0.6$ \\
$m_c$ & $1.3\pm 0.2$  &
$f_{B_d}$ & $0.195 \pm 0.015$  \\
$m_s$(2 GeV) & $0.09\pm 0.02$ &
$f_{\pi}$ & 0.131  \\
$(m_u+m_d)/m_s$ & 0.0826 &
$f^{B\pi}_+(0)$ &  $0.25 \pm 0.05^\dagger$  \\
$m_b$ & 4.8 &
$f_{\rho}$ & 0.209  \\
$\bar m_b(\bar m_b)$ & 4.2 &
$A_0^{B\rho}(0)$ & $0.30\pm 0.05^\dagger$ \\
$|V_{cb}|$ & $0.0415\pm 0.0010$ &
$\lambda_B$(1 GeV) & $0.35 \pm 0.15^\dagger$  \\
$|V_{ub}/V_{cb}|$ &  $0.09\pm 0.02$  &
$\sigma_1$(1 GeV) &  $1.5 \pm 1$\\
$\gamma$ & $(70 \pm 10)^\circ$ &
$\sigma_2$(1 GeV) &  $3 \pm 2$\\
$\tau(B^-)$ & 1.64\,\mbox{ps} &
$a^{\pi}_2$(2 GeV) &  $0.2 \pm 0.15$ \\
$\tau(B_d)$ & 1.53\,\mbox{ps} &
$a^{\rho}_{2}$(2 GeV) &  $0.1 \pm 0.15$ \\
$\mu_b$ & $4.8^{+4.8}_{-2.4}$ &
$a^{\rho}_{2,\perp}$(2 GeV) &  $0.1 \pm 0.15$ \\[-0.4cm]
&&& \\
\hline
\end{tabular}

\vspace*{0.1cm}
\footnotesize ${}^\dagger\,$\parbox[t]{11.5cm}{
Value and range changed to 
$f^{B\pi}_+(0)=0.23\pm 0.03$, $A_0^{B\rho}(0)=0.28\pm 0.03$,
$\lambda_B(1\,\mbox{GeV})=(0.20^{+0.05}_{-0.00})\,$GeV 
in parameter set ``Theory II''. See text in 
Section~\ref{subsec:finalresults}.
}
\caption{List of input parameters. Dimensionful parameters are given
in units of 1 GeV.
\label{tab:inputs}}\vspace{0.2cm}
\end{center}
\end{table}

In this section we begin the discussion of the numerical
evaluation of the topological tree amplitudes $\alpha_1$, $\alpha_2$,
and the tree-dominated charmless $B$ decays into final states $\pi\pi$,
$\pi\rho$, and $\rho\rho$. The main result of this work, the
NNLO vertex correction, depends on very few input parameters:
the strong coupling $\alpha_s$ via $\Lambda_{\overline{\rm
MS}}^{(5)}$; the bottom pole mass $m_b$ or $\overline{\rm MS}$ mass
$\bar m_b(\bar m_b)$; the renormalization scale $\mu\equiv \mu_b$
(to distinguish it from the hard-collinear scale that appears in
spectator scattering); the Gegenbauer moments of the light-meson
light-cone distribution amplitude $a^{M}_n$,
the only hadronic parameters; and the electroweak-scale initial
conditions for the Wilson coefficients of the effective Hamiltonian.
Their values and uncertainties are specified in
Table~\ref{tab:inputs}, except for the electroweak-scale
parameters $M_W=80.4\,$GeV, $\bar m_t(\bar m_t)=167\,$GeV
and $\mbox{sin}^2\theta_W=0.23$. When we evaluate the full topological
tree amplitudes and branching fractions including penguin amplitudes
(using results from \cite{Beneke:2006mk}), we need further
Standard Model and hadronic parameters, which are also listed in
the Table. The values and uncertainties follow
\cite{Beneke:2006mk,Beneke:2006hg} with some minor adjustments
that account for recent reevaluations or progress in error
estimates of $\gamma$, $f_B$ \cite{Gamiz:2009ku,Bernard:2009wr},
$A_0^{B\rho}(0)$ \cite{Ball:2004rg}, and 
$a^\rho_{2,(\perp)}$ \cite{Ball:2007rt}.

Together with the NNLO non-leptonic matrix elements, we now use
three-loop evolution of the strong coupling, as well as the
next-to-next-to-leading-logarithmic (NNLL) approximation to the Wilson
coefficients appearing in the effective weak interaction Hamiltonian
in the CMM basis. This
approximation is constructed from the two-loop initial
conditions~\cite{Bobeth:1999mk,Gorbahn:2004my}, three-loop anomalous
dimension matrix~\cite{Gambino:2003zm,Gorbahn:2004my} and the 
NNLL solution to the renormalization group equation~\cite{Beneke:2001at}.
The penguin amplitudes, which are not yet known at NNLO, are evaluated
exactly as in \cite{Beneke:2006mk}, whenever needed,
except that we employ the same
three-loop evolved strong coupling in the entire numerical program.
The observables we discuss in this paper are chosen such that they
do not depend significantly on the penguin amplitudes.

\subsection{Topological tree amplitudes at NNLO}

\subsubsection{Two-loop vertex correction}

We first consider the loop expansion of the pieces of the topological
tree amplitudes not related to spectator scattering, as defined in
(\ref{alphaiexpansion}). To acquire an idea of the convergence of 
the perturbative expansion we provide a numerical representation of
\begin{equation}
\int_0^1 du\, T_j(u)\,\phi_M(u)
= V_{1j}^{(0)} + \sum_{l\geq 1}\left(\frac{\alpha_s}{4\pi}\right)^l
\frac{C_F}{2 N_c} V_{1j}^{(l)}
\end{equation}
up to ${\cal O}(\alpha_s^2)$, and similarly for $\tilde T_j(u)$ and
$V_{2j}^{(l)}$. Choosing $\mu=m_b$ we get
\bea
\int_0^1 du\, T_1(u)\,\phi_M(u)  &=& 
0+ \left[- 5 - 2.094 \, i
+\left(1.222 - 2.094 \, i\right) a_1^M  -0.233 \, a_2^M\right]
\left(\frac{\alpha_s}{4\pi}\right) \nnb\\
&& \hs{29} + \left[-38.510 - 75.389\, i +
\left( 142.014 - 24.416 \, i\right) a_1^M \right. \nnb \\
&& \hs{39}\left.+\left( -18.588 - 13.528 \, i\right) a_2^M + 
\Delta_{m_c}\right]
\left(\frac{\alpha_s}{4\pi}\right)^2 \, ,
\nonumber \\
\int_0^1 du\, T_2(u)\,\phi_M(u) &=& 1 +\left[71.598 + 71.320 \, i
+\left(-47.327 + 34.313 \, i\right) a_1^M \right. \nnb \\
&& \hs{39}\left.+\left(0.848 - 7.569 \, i\right) a_2^M\right]
\left(\frac{\alpha_s}{4\pi}\right)^2 \, ,
\nonumber \\
\int_0^1 du\, \widetilde T_1(u)\,\phi_M(u)  &=& 
\frac{4}{9}+\left[ 0.333 + 0.698\, i
+\left(-0.407 + 0.698 \, i\right) a_1^M +0.078 \, a_2^M\right]
\left(\frac{\alpha_s}{4\pi}\right) \nnb\\
&& \hs{29}+ \bigg[37.362 + 56.828 \, i +
\left(-68.372 + 23.389 \, i\right) a_1^M \nnb \\
&& \hs{38}\left.+\left( 6.573 + 1.145 \, i\right) a_2^M - 
\frac{1}{3}\,\Delta_{m_c}\right]
\left(\frac{\alpha_s}{4\pi}\right)^2 \, ,
\nonumber \\
\int_0^1 du\, \widetilde T_2(u)\,\phi_M(u) &=& 
\frac{1}{3} +\left[ -10 - 4.189 \, i
+\left(2.444 - 4.189 \, i\right) a_1^M -0.467 \, a_2^M\right]
\left(\frac{\alpha_s}{4\pi}\right) \nnb\\
&& \hs{29}+ \left[-83.154 - 139.571 \, i +
\left(275.585 - 49.960 \, i\right) a_1^M \right. \nnb \\
&& \hs{38}\left.+\left(-38.293 - 29.580 \, i\right) a_2^M+ 
2 \,\Delta m_c\right]
\left(\frac{\alpha_s}{4\pi}\right)^2\, .\label{eq:convT2ti}
\eea
We observe that the NNLO terms have relatively large coefficients, 
but since $\alpha_s(m_b)\approx 0.22$, the perturbative expansion 
is still under control for most of the terms. 
The coefficients of the second Gegenbauer 
moment are generally rather small, contrary to those of the 
first one, as we will discuss in more detail at the end of this subsection.

The numbers in (\ref{eq:convT2ti}) apply when we treat the charm quark 
as massless. The charm mass effect, the difference between the 
contribution of a quark of mass $m_c$ and that of a massless quark, 
is parameterized by $\Delta_{m_c}$, for which we find the 
fitting formula 
\bea
\Delta_{m_c}&=& -0.9661 -2.0849\, i -
(11.4437 +12.9458\, i) (z - 0.075)  \nnb \\
&& \hs{12} + \,(28.0366 + 64.7300\, i)  (z - 0.075)^2 \nnb \\
&& \hs{0}+ \left[ 4.5919 -1.9379\, i + 
(14.4270 -12.2979\,i) (z - 0.075)\right. \nnb \\
&& \hs{12}\left. -\,(96.8420  - 59.5511\,i) (z - 0.075)^2\right]a_1^M \nnb \\
&& \hs{0}+ \left[-0.3599 -0.3833\,i -
(2.2574 +0.5738 \,i) (z - 0.075) \right. \nnb \\
&& \hs{12}\left.+ \,(14.1412 + 7.0945\,i) 
(z - 0.075)^2\right]a_2^M 
\eea
in the region of $z=m_c^2/m_b^2$ of interest. The second-order 
polynomials in $z$ were obtained from a least-squares fit
to the exact functions, which works better than $0.4\%$ for 
$z \in [0.05,0.10]$. The size of the charm-mass dependent terms is 
moderate. Most of the numbers
in~(\ref{eq:convT2ti}) receive small corrections, but the shifts
induced by the $z$-dependent terms can be up to $11\%$ for single entries.

\begin{figure}[t]
\centerline{\parbox{10cm}{
\centerline{\includegraphics[width=7cm]{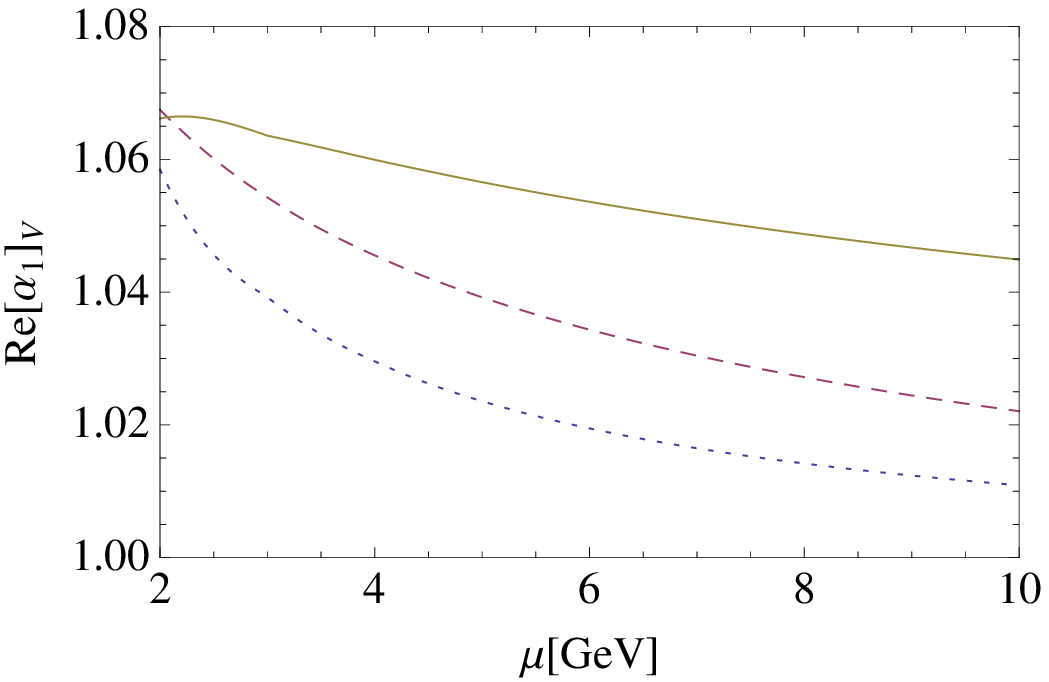}
\hspace{10mm}
\includegraphics[width=7cm]{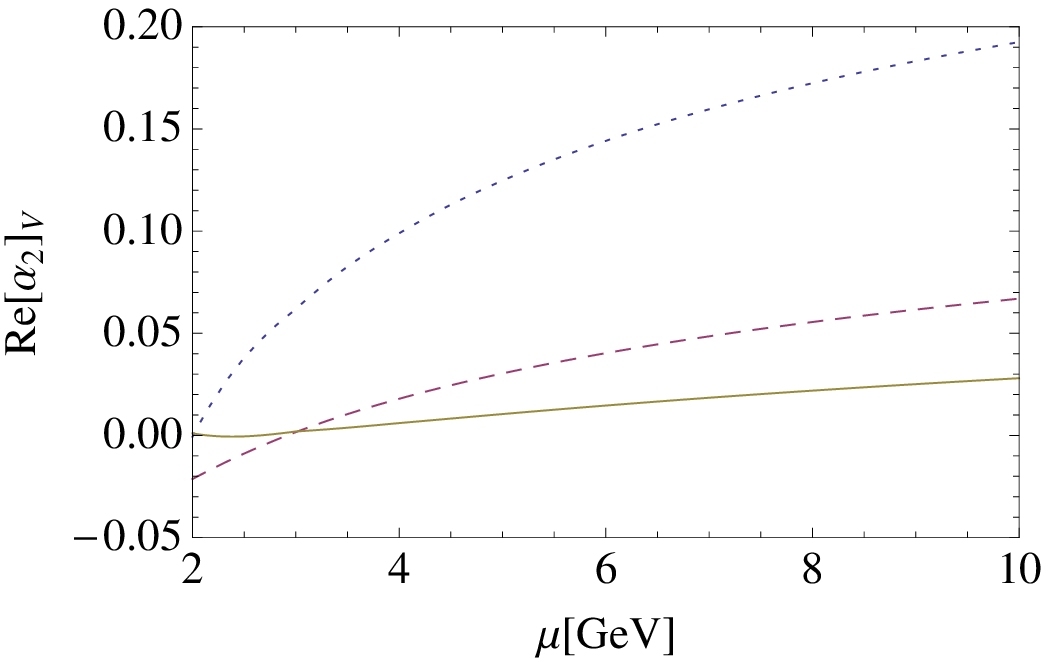}}
\vspace{10mm}
\centerline{\includegraphics[width=7cm]{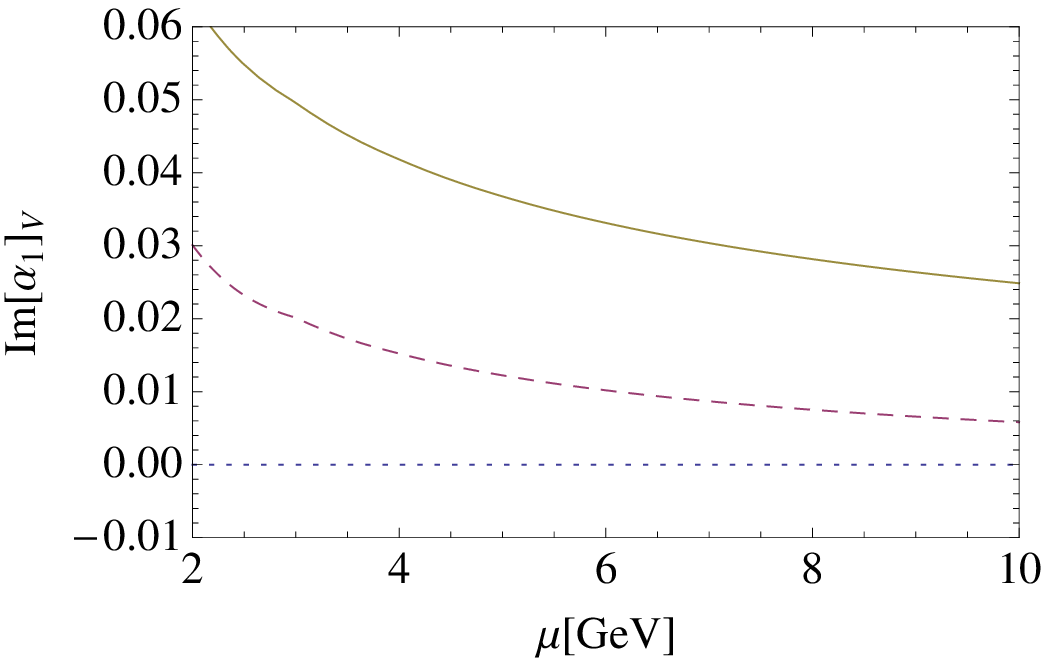}
\hspace{10mm}
\includegraphics[width=7cm]{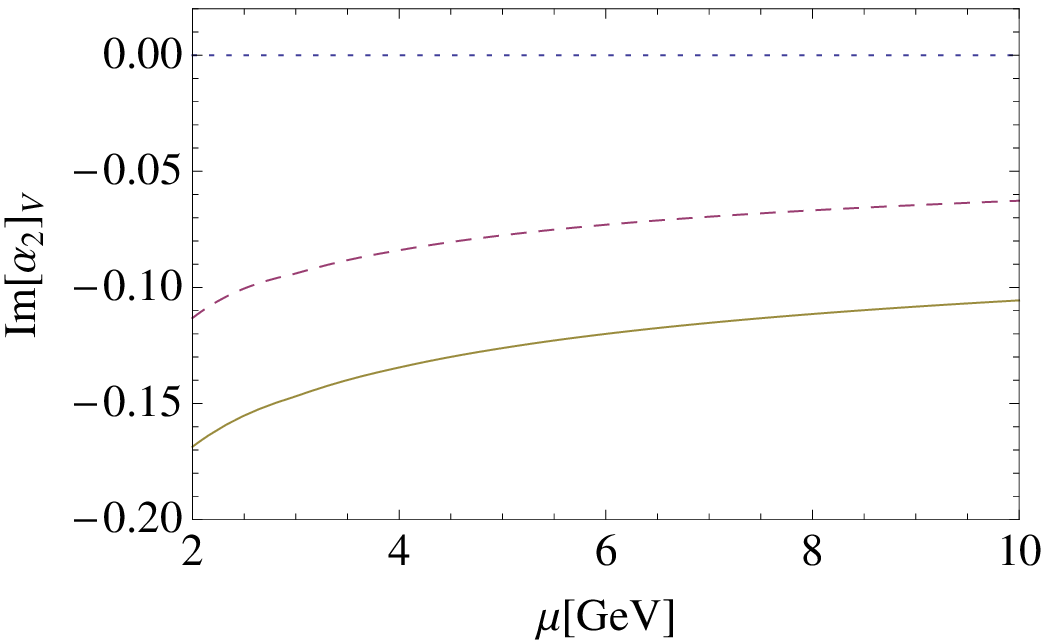}}}}
\vspace{5mm}
\centerline{\parbox{16cm}{\caption{\label{fig:scale} Dependence of the
topological tree amplitudes $\alpha_1(\pi\pi)$ and $\alpha_2(\pi\pi)$
on the hard scale $\mu$ (vertex corrections only). The dotted, dashed and solid
lines refer to the theoretical predictions at LO, NLO and NNLO, 
respectively.}}}
\end{figure}

The successive LO, NLO and NNLO 
approximations\footnote{The LO and NLO result is computed in the traditional 
operator basis for the effective weak Hamiltonian, so that it corresponds 
to the expressions used in \cite{Beneke:1999br,Beneke:2001ev}, but is 
slightly different from the NLO truncations of the NNLO results 
of the present calculation.}
to the topological tree amplitude parameters $\alpha_1(\pi\pi)$ and
$\alpha_2(\pi\pi)$ (vertex correction only, i.e.~no spectator scattering) 
are shown in Figure~\ref{fig:scale}, depending on the
renormalization scale $\mu$. Relative to the NLO correction the NNLO term
turns out to be sizeable. For the colour-allowed amplitude
$\alpha_1(\pi\pi)$ this can be explained by the observation that
the NLO correction is colour-suppressed, while the NNLO correction
is no longer colour-suppressed for any of the two amplitudes.
As in any perturbative QCD calculation the
higher-order corrections are expected to reduce the scale uncertainties
compared to lower-order approximations.  We observe that this is indeed
the case after taking into account the NNLO corrections for the
real parts of the amplitudes, but the reduction is absent for the
imaginary parts. We attribute this to the fact that the two-loop
contribution to the imaginary part is really a next-to-leading order
effect since the imaginary part is zero at tree-level, and to the large
size of the two-loop correction.

Besides the renormalization scale the vertex correction to the 
topological tree amplitudes depends only on the Gegenbauer moments.
At leading order there is no dependence on these moments,
whereas at NLO this dependence is given by 
\begin{eqnarray}
{[\alpha_1]}_V &=& 1.040 + 0.013 i - 
(0.007-0.013 i)\,a_1^{M_2} + 0.001 \,a_2^{M_2}, 
\nonumber \\
{[\alpha_2]}_V &=& 0.029 - 0.079 i + 
(0.046-0.079 i)\,a_1^{M_2} - 0.009 \,a_2^{M_2}, 
\end{eqnarray}
where we choose again $\mu=m_b$. At NNLO we obtain
\begin{eqnarray}
{[\alpha_1]}_V &=& 1.057 + 0.038 i - 
(0.032-0.022 i)\,a_1^{M_2} + (0.003-0.001 i)  \,a_2^{M_2},
\nonumber \\
{[\alpha_2]}_V &=& 0.013 -0.126 i + 
(0.139 - 0.096 i)\,a_1^{M_2} - (0.021 +0.009 i)  \,a_2^{M_2}.
\end{eqnarray}
Since the Gegenbauer moments are typically smaller than 0.3, we conclude 
that the dependence on the second moment $a_2^{M_2}$ is very small for 
both tree amplitudes. This implies that there is practically no 
dependence on the shape of the pion or rho meson light-cone
distribution amplitude in the vertex term of the factorization
formula~(\ref{factformula}). This is no longer true once
spectator scattering is included, especially for the colour-suppressed
tree amplitude  $\alpha_2$, as discussed in~\cite{Beneke:2005vv}. 
The first Gegenbauer moment is more important and has a significant 
effect on the value of the colour-suppressed amplitude 
${[\alpha_2]_V}$, which is furthermore enhanced at NNLO. This is a 
source of non-factorizable SU(3) flavour symmetry breaking (not related 
to decay constants and heavy-to-light form factors) in the 
relation of amplitudes with pions and kaons.


\subsubsection{Complete NNLO amplitude including spectator scattering}

We now proceed with our investigation of the topological tree amplitudes
by adding the corrections from spectator scattering on top of those from
the vertex corrections. Our numerical results for the $\pi\pi$ final
states read
\bea
 \alpha_1(\pi\pi) &=&   1.009 + \left[ 0.023+ \, 0.010 \, i\right]_{\rm NLO}
+ \left[0.026+ \, 0.028 \, i\right]_{\rm NNLO}
\nonumber \\
&& -\, \left[\frac{r_{\rm sp}}{0.445}\right]
\left\{\left[0.014\right]_{\rm LOsp}+
\left[0.034 +\,0.027 i\,\right]_{\rm NLOsp}+
\left[0.008\right]_{\rm tw3}\right\}
\nonumber\\
&=& 1.000^{+0.029}_{-0.069} + (0.011^{+0.023}_{-0.050})i \, ,
\\[0.3cm]
\alpha_2(\pi\pi) &=&   0.220 - \left[ 0.179+ \, 0.077 \, i\right]_{\rm NLO}
- \left[0.031+ \, 0.050 \, i\right]_{\rm NNLO}
\nonumber\\
&& +\, \left[\frac{r_{\rm sp}}{0.445}\right]
\left\{\left[0.114\right]_{\rm LOsp}+
\left[0.049 +\,0.051 i\,\right]_{\rm NLOsp}+
\left[0.067\right]_{\rm tw3}\right\}\nnb \\
&=& 0.240^{+0.217}_{-0.125} + (-0.077^{+0.115}_{-0.078})i \, .
\label{a2num}
\eea
Here the ``NNLO'' term in the first line of each of the two
equations corresponds to the numerical evaluation of the calculation of
this paper, while the second line accounts for spectator scattering
at ${\cal O}(\alpha_s)$ (``LOsp'') and ${\cal O}(\alpha_s^2)$ 
(``NLOsp'') and a certain power correction (``tw3''). The theoretical 
calculation for spectator scattering is taken from~\cite{Beneke:2005vv}, 
but the numerical values differ, since we now use a different operator
basis, which redistributes the perturbative expansion, and slightly
different input parameters.
The third line sums all contributions and provides an estimate of
the theoretical uncertainties from the parameter variations
detailed in Table~\ref{tab:inputs}.
By comparing the theoretical uncertainty in the full expressions above 
to the one from the vertex correction alone, see Figure~\ref{fig:scale},
we see that it arises primarily from spectator scattering. The 
main contributors to the uncertainty are the parameter combination
\begin{equation}
r_{\rm sp} = \frac{9 f_\pi \hat f_B}{m_b f_+^{B\pi}(0) \lambda_B},
\label{defrsp}
\end{equation}
which appears as an overall normalization 
factor of the spectator-scattering term, 
the second Gegenbauer moment $a^{\pi}_2(2\,\mbox{GeV})$, 
and ``tw3'' power correction.

Relative to the full amplitude the two-loop vertex correction is only
a few percent for the colour-allowed amplitude $\alpha_1$, 
but it is quite significant 
for $\alpha_2$. The negative $(10-15)\%$ correction to the real part 
decreases the branching fractions of the colour-suppressed decays 
by about $25\%$ relative to previous results that included the 
${\cal O}(\alpha_s^2)$ correction only in spectator scattering. 
The magnitude of the correction to the imaginary part of $\alpha_2$ 
at two loops still reaches $25\%$ of the leading-order real part.
However, we also
observe that the two  ${\cal O}(\alpha_s^2)$ corrections, ``NNLO''
in the vertex term and ``NLOsp'' in spectator scattering, cancel to a
large extent in both, their real and imaginary parts. This is
somewhat unfortunate, since an enhancement rather than a cancellation
in the colour-suppressed tree amplitude might have been welcome
in view of the trend indicated by experimental data, as will be seen
below. 

The colour-suppressed tree amplitude $\alpha_2(\pi\pi)$
exhibits an interesting structure. It starts out with a positive
real value 0.220 that corresponds to the naive factorization
approximation. After adding the perturbative corrections to the
four-quark vertex (the first line of (\ref{a2num})), it is
found to be almost purely imaginary, $0.01 - 0.13 i$. However,
the spectator-scattering mechanism regenerates a real part of
roughly the original size and cancels part of the strong phase.
The net result of this is that the colour-suppressed tree amplitude
can become sizeable in QCD factorization when $r_{\rm sp}$ is
large, but since this enhances the cancellation of the imaginary
part, one cannot have both, a large magnitude and a large strong phase.
In comparison, the colour-allowed tree amplitude $\alpha_1(\pi\pi)$
is rather stable against radiative corrections, and never
deviates by a large amount from its naive-factorization estimate.

\subsection{Parameter dependence of branching fraction ratios}

\subsubsection{Factorization test}
\label{sec:facttest}
In this subsection we consider a non-leptonic to semi-leptonic decay
ratio that provides direct access to the magnitude of the topological
tree amplitudes. This analysis has first been performed
in~\cite{Beneke:2003zv} at NLO and can now be repeated at NNLO,
and with improved experimental data (see also \cite{Bell:2009fm}).

Within the approximation that all electroweak contributions are
neglected since they are never CKM-enhanced and formally of order
${\cal O}(\alpha_{\rm em})$, the process $B^-\to\pi^-\pi^0$ is a
pure tree decay. The
corresponding decay amplitude in the SM can be written
as~\cite{Beneke:2003zv}
\begin{equation}
\begin{aligned}
   \sqrt2\,{\cal A}_{B^-\to\pi^-\pi^0}
   &= A_{\pi\pi}\,\sum_{p=u,c} \lambda_{p}^d\,\delta_{pu}\,
   (\alpha_1 + \alpha_2)\,,
\end{aligned}
\end{equation}
with $A_{\pi\pi}= i\,\frac{G_F}{\sqrt2}\,m_B^2 f_+^{B\pi}(0)
f_{\pi}$, and $\lambda_{p}^d=V_{pb}\,V_{pd}^{\ast}$.
A substantial uncertainty in the prediction of the branching ratio arises
from the overall normalization due to $|V_{ub}|$ and
$f_+^{B\pi}(0)$. These two quantities cancel out in the
ratio~\cite{Beneke:2003zv,Bjorken:kk}
\begin{equation}
\label{R_ratio}
   R_\pi\equiv\frac{\Gamma(B^-\to\pi^-\pi^0)}
        {d\Gamma(\bar B^0\to\pi^+ l^-\bar\nu)/dq^2\big|_{q^2=0}}
   = 3\pi^2 f_\pi^2\,|V_{ud}|^2\,|\alpha_1(\pi\pi)+\alpha_2(\pi\pi)|^2 \,,
\end{equation}
and a direct measurement of the tree amplitude
coefficients can be performed
if the differential semi-leptonic $B\to\pi l\,\nu$ rate is
measured near $q^2=0$. The ratio $R_\pi$ in~(\ref{R_ratio}) therefore
constitutes a test of QCD factorization, and furthermore
underlines the importance of measuring the semi-leptonic decay spectrum
for understanding the pattern of non-leptonic $B$ decays. Replacing 
pions by longitudinally polarized rho mesons in (\ref{R_ratio}), 
we define an analogous ratio $R_\rho$, which allows us to test 
factorization in $B\to VV$ decays. 

This test can currently be carried out for pions. 
The latest experimental data on the $B^- \to \pi^-\pi^0$ branching fraction 
reads~\cite{Barberio:2008fa,Aubert:2007hh,Abe:2006qx,Bornheim:2003bv}
\begin{eqnarray}\label{exp_pipi}
\mbox{Br}\,(B^{-}\to\pi^{-}\pi^0)&=&(5.59^{+0.41}_{-0.40})\cdot 10^{-6} \, .
\end{eqnarray}
Since the theoretical expression for the differential
semi-leptonic spectrum at the end-point assumes the form
\be{d\Gamma(\bar B^0\to\pi^+
l^-\bar\nu)/dq^2\big|_{q^2=0}}=\frac{G_F^2\,
m_B^3}{192\,\pi^3}\,\Big[|V_{ub}|\,f_+^{B\pi}(0)\Big]^2 \, ,
\ee
the measurement of the left-hand side is equivalent to 
the experimental determination of $|V_{ub}|\,f_{+}^{B\pi}(0)$ from 
the full lepton invariant mass spectrum in $B\to\pi l\,\nu$ decay 
together with an extrapolation to $q^2=0$ based on a form-factor 
parameterization. Several extractions of  
$|V_{ub}|\,f_{+}^{B\pi}(0)$~\cite{Ball:2006jz,Becher:2005bg,Aubert:2006px}
employing  different parameterizations of the transition form
factor~\cite{BK,BZ04,flynn,disper}
can be found in the literature, which all yield similar results.
We shall adopt the value
\begin{eqnarray}
\label{Ball}
|V_{ub}|f_+^{B\pi}(0) &= & (9.1\pm 0.7)\cdot 10^{-4}\,.
\end{eqnarray}
Combining this and the experimental data on $\mbox{Br}(B^{-}\to\pi^{-}\pi^0)$
given by (\ref{exp_pipi}) allows us to extract the following
experimental constraint on $R_\pi$,
\be\label{exp_R}
[R_\pi]_{\rm exp}= 0.81 \pm 0.14 \, ,
\ee
which can be translated into a constraint on the tree amplitudes,
\be\label{exp_a1a2}
   \left[\,|\alpha_1(\pi\pi)+\alpha_2(\pi\pi)|\,\right]_{\rm exp}
=  1.29 \pm 0.11 \, .
\ee
On the other hand, varying the parameters in the range specified in
Table~\ref{tab:inputs}
and adding the errors in quadrature,
our theoretical prediction at NNLO is given by
\begin{equation}\label{th_a1a2}
   |\alpha_1(\pi\pi)+\alpha_2(\pi\pi)|=
    1.24^{+0.16}_{-0.10} \, , 
\end{equation}
which is in good agreement with the experimental
data and hence provides support for the factorization assumption.
\begin{figure}[t]
\centerline{\includegraphics[width=8cm]{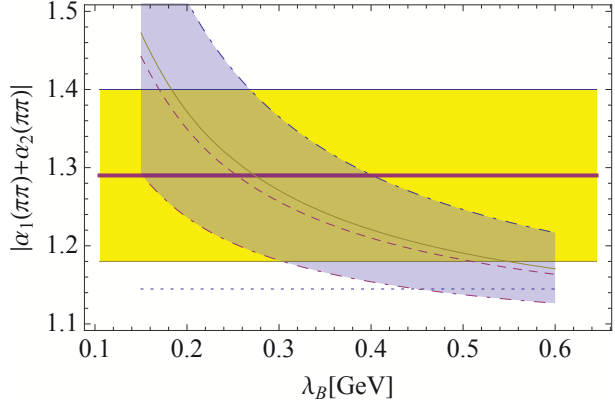}}
\vspace{5mm}
\centerline{\parbox{16cm}{\caption{\label{fig:lambdaB}
Dependence of the theoretical prediction of
$|\alpha_1(\pi\pi)+\alpha_2(\pi\pi)|$ (equivalent to $R_\pi$) on
$\lambda_B$ and comparison to
$[\,|\alpha_1(\pi\pi)+\alpha_2(\pi\pi)|\,]_{\rm exp}=1.29 \pm 0.11$.
The dotted, dashed and solid lines denote the LO, NLO and NNLO
predictions, respectively.
The dot-dashed lines give the error band upon varying the input parameters
within the ranges specified in Table~\ref{tab:inputs}, excluding
the parameters contained in $r_{\rm sp}$. The experimental data within
$1\sigma$ error is also shown (horizontal band).}}}
\end{figure}
The theoretical uncertainty is still large, but much of it arises from
$\lambda_B$, $\hat f_B$ and $f_+^{B\pi}(0)$, on which the theoretical
result depends only through the overall factor $r_{\rm sp}$ in (\ref{defrsp}).
In Figure~\ref{fig:lambdaB} we therefore plot the dependence of
$|\alpha_1(\pi\pi)+\alpha_2(\pi\pi)|$ on the first inverse moment of
the $B$-meson distribution amplitude,
$\lambda_B$, and the remaining theoretical error (grey band).
For comparison we show the LO (flat, dotted line), NLO (dashed)
and NNLO calculation, which makes it clear that the combined NNLO
correction is rather small. The overlaid horizontal band
refers to the experimental value~(\ref{exp_a1a2}), displaying
again the good agreement with the theoretical calculation. The
errors are too large to turn this into a determination of the
parameter $\lambda_B$, which may then serve as an input for the
remaining non-leptonic observables. The central experimental value
(thick horizontal line)
alone would allow the range $\lambda_B \in [150,400]\,$MeV,
which is on the lower side of theoretical expectations~\cite{Beneke:1999br,Beneke:2001ev,Ball:2003fq,Braun:2003wx,Kawamura:2008vq}.

In the future it should be of interest to extend the factorization test 
to other final states. In Table~\ref{tab:resultsR} we summarize our 
predictions for pions and longitudinal rho mesons. Here ``Theory I'' 
corresponds to the input parameters of Table~\ref{tab:inputs}. As 
will become clear below, the analysis of non-leptonic decay data 
suggests the hypothesis of smaller form factors and $\lambda_B$. 
This leads us to choose a second parameter set, to be discussed in  
Section~\ref{subsec:finalresults} and labelled ``Theory II''. We 
consider the corresponding entries in Table~\ref{tab:resultsR} to be 
our reference predictions.

\tabcolsep0.5cm
\begin{table}[b]
  \centering
  \let\oldarraystretch=\arraystretch
  \renewcommand*{\arraystretch}{1.3}
\begin{tabular}{lccc}
    \toprule
        & \phantom{0}Theory I& \phantom{0}Theory II
        & Experiment
    \\
    \midrule

$R_{\pi}$
 & $0.75^{+0.20}_{-0.11}$
 & $0.94^{+0.23}_{-0.22}$
 & $0.81 \pm 0.14$ \\
$R_{\rho}$
 & $1.75^{+0.37}_{-0.24}$
 & $2.08^{+0.50}_{-0.46}$
 & --- \\ \addlinespace
\bottomrule
\end{tabular}
\let\arraystretch=\oldarraystretch
\caption{Theoretical results for the factorization test ratios 
(\ref{R_ratio}) for pions and longitudinal rho mesons. ``Theory II'' 
refers to our preferred input parameter set, see text.
  \label{tab:resultsR}}
\end{table}

\subsubsection{Ratios involving the colour-suppressed tree amplitude}
\label{sec:coloursuppressed}

Next we consider a number of branching fraction ratios to further
highlight the importance of hard spectator-scattering, and thereby
the dependence on $\lambda_B$ and $r_{\rm sp}$.
Some of these ratios have been considered before, for instance 
in~\cite{Beneke:2003zv,pipipuz,Bell:2009fm}, for a different purpose.
We define
\begin{eqnarray}
\begin{aligned}
 R_{+-}^{\pi\pi} &\equiv 2\,\frac{\Gamma(B^-\to\pi^-\pi^0)}
 {\Gamma(B^{0}\to\pi^{+}\pi^{-})}\,,\qquad\quad
 &R_{00}^{\pi\pi} &\equiv 2\, \frac{\Gamma(B^{0}\to\pi^0\pi^0)}
 {\Gamma(B^{0}\to\pi^{+}\pi^{-})}\,, \\
 R_{+-}^{\rho\rho} &\equiv 2\,\frac{\Gamma(B^-\to\rho_L^-\rho_{L}^0)}
 {\Gamma(B^{0}\to\rho_L^{+}\rho_L^{-})}\,,\qquad\quad
 &R_{00}^{\rho\rho} &\equiv 2\, \frac{\Gamma(B^{0}\to\rho_L^0\rho_L^0)}
 {\Gamma(B^{0}\to\rho_L^{+}\rho_L^{-})}\,, \\
 R_{00}^{\pi\rho} &\equiv \frac{2\,\Gamma(B^0\to \pi^0\rho^0)}
                    {\Gamma(B^0\to \pi^+\rho^-)+
                     \Gamma(B^0\to \pi^-\rho^+)} \,. &&&
\label{firstratios}
\end{aligned}
\end{eqnarray}
Here, and throughout the paper, $\Gamma$ denotes the CP-averaged decay rate, 
i.e.~$\Gamma(B\to f)$ is in fact the average of $\Gamma(B\to f)$ and 
the decay rate ~$\Gamma(\bar B\to \bar f\,)$ of the CP-conjugate mode.
The $R_{00}$ ratios are approximately proportional to the square of the
ratio of the colour-suppressed to the colour-allowed tree amplitude. This
approximation is very good for the modes involving the rho meson
final states, where the penguin contribution is known to be
small.\footnote{For the vector-vector final state, we consider the
longitudinally polarized final state. Accordingly, the experimental
unpolarized 
branching fraction is multiplied by the measured longitudinal polarization
fraction.} The $R_{+-}$ ratios involve $|\alpha_1+\alpha_2|$, similar
to $R_\pi$ and $R_\rho$ in the previous subsection, but divided by $|\alpha_1|$ and
squared. These ratios also display some sensitivity to the CKM angle
$\gamma$ for $\pi\pi$, but we do not discuss this here, since it arises
through interference with the penguin amplitudes.

\begin{figure}[p]
\centerline{
\includegraphics[width=7cm]{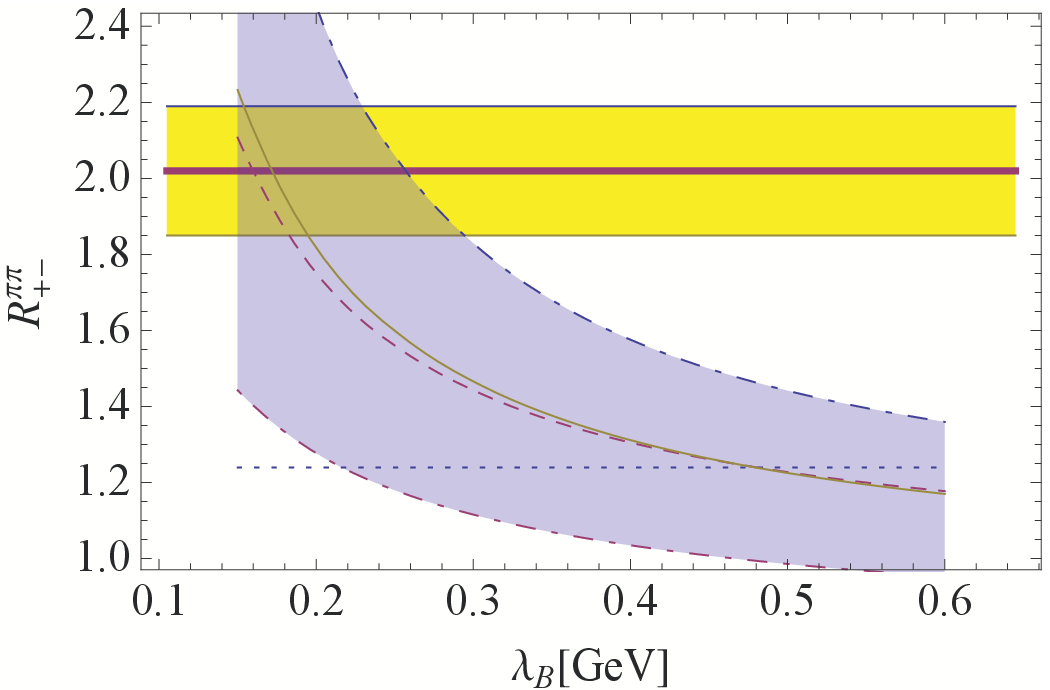}
\hspace{10mm}
\includegraphics[width=7cm]{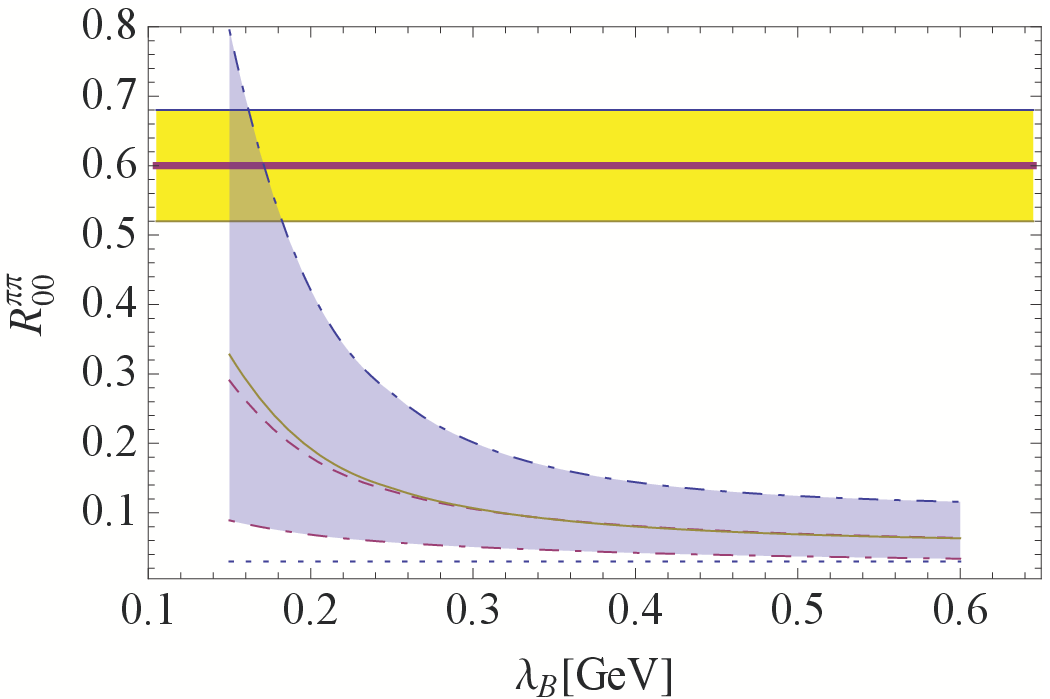}}
\vspace{10mm}
\centerline{
\includegraphics[width=7cm]{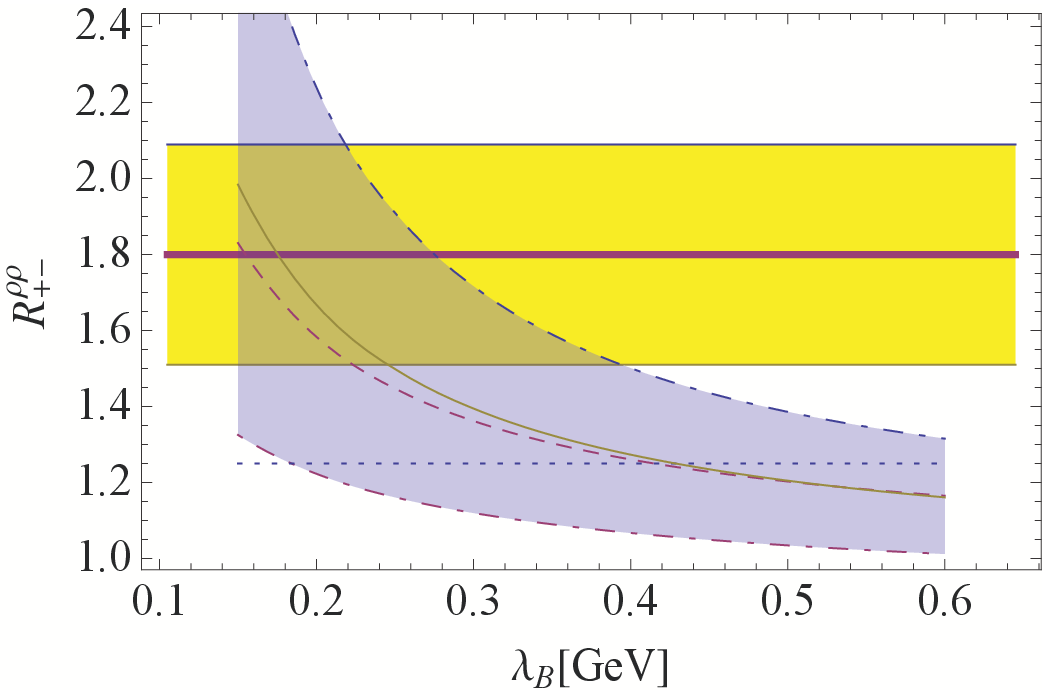}
\hspace{10mm}
\includegraphics[width=7cm]{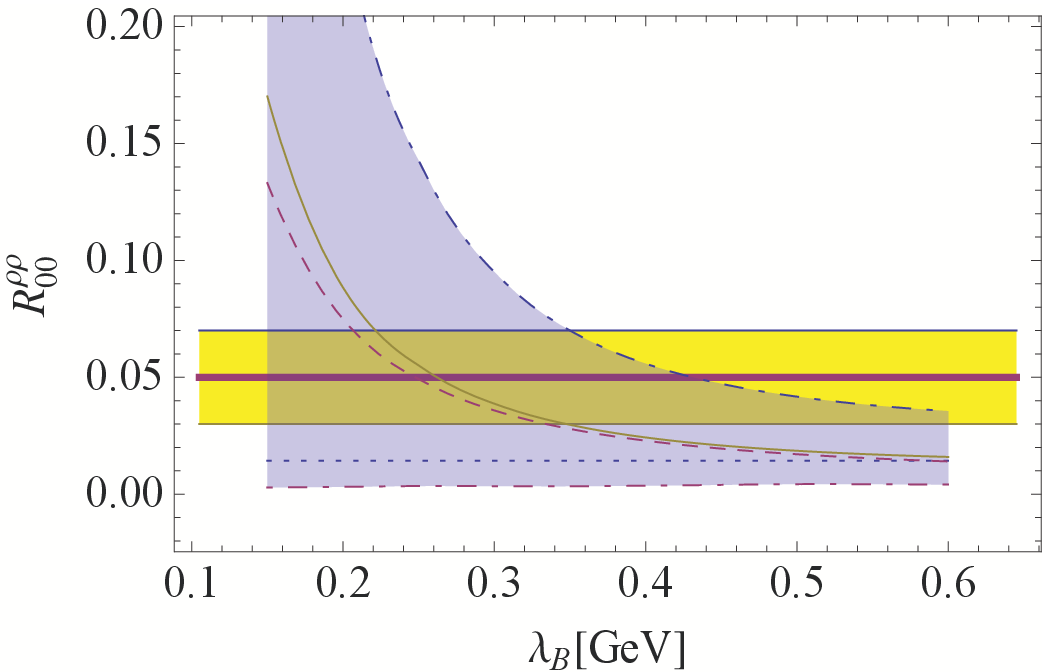}}
\vspace{10mm}
\centerline{
\includegraphics[width=7cm]{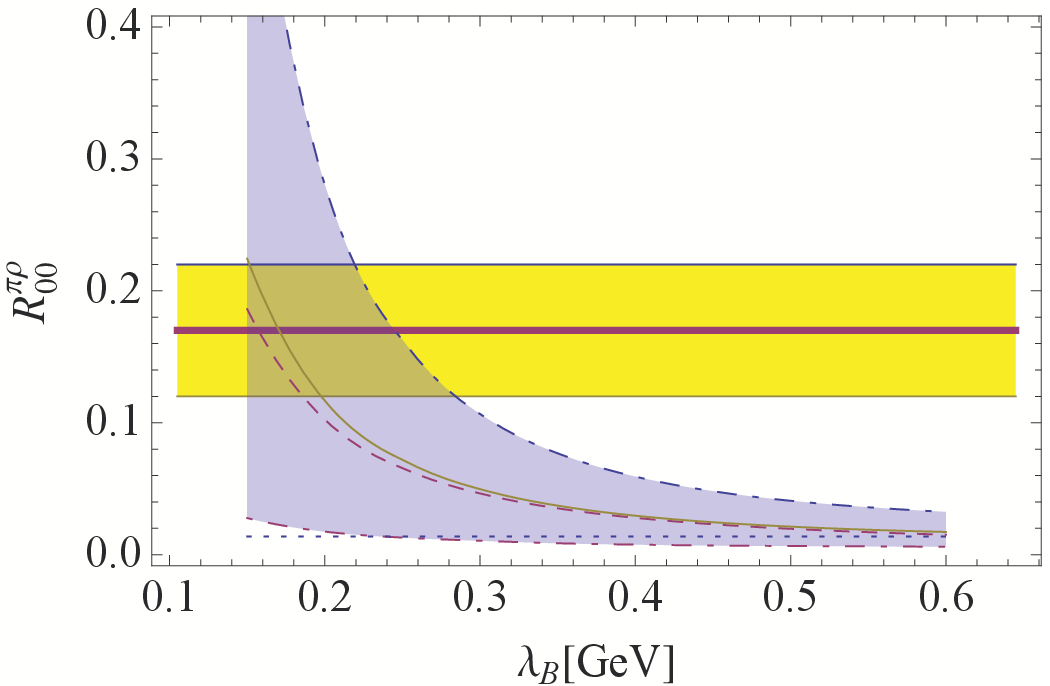}}
\vspace{10mm}
\centerline{\parbox{16cm}{\caption{\label{fig:pipiratio}
Constraints on $\lambda_B$ from the ratios (\ref{firstratios}) 
of $B\to \pi\pi$,
$B\to \pi\rho$, and $B\to \rho_L\rho_L$ decays.
Line styles have the same meaning as in Figure~\ref{fig:lambdaB}.}}}
\end{figure}

Our results for these ratios are displayed in Figure~\ref{fig:pipiratio}.
Again the grey band refers to the theoretical prediction (with uncertainty)
for the value of $\lambda_B$ on the horizontal axis, while the horizontal
band gives the current
measurement~\cite{Barberio:2008fa,Aubert:2007hh,Abe:2006qx,Bornheim:2003bv,
Aubert:2006fha,Morello:2006pv,Aubert:2008sb,Abe:2006cx,Aubert:2009it,
Zhang:2003up,Aubert:2007nua,Somov:2006sg,:2008iha,:2008et,Godang:2001sg,
Aubert:2003fm,:2007mj,Jessop:2000bv,Aubert:2003wr}.
It is clear that naive factorization (the dotted line) fails to describe 
the data. In general, QCD factorization provides a good description of 
data, especially if $\lambda_B$ is around $200\,$MeV, where, however, 
the theoretical uncertainty becomes large. In the QCD factorization 
approach small $\lambda_B$ and a large colour-suppressed amplitude is 
connected with an important role of spectator scattering in the dynamics 
underlying this amplitude as already noted
in~\cite{Beneke:2003zv}. An exception to the good agreement between 
theory and data is $R_{00}^{\pi\pi}$ which can be accommodated even
for $\lambda_B\simeq 200\,$MeV only marginally within errors. Thus, it 
is plausible or even likely that power-suppressed corrections are 
also important for the colour-suppressed amplitude, especially if 
it is accompanied by a large phase, an issue that we do not discuss in 
the present paper, since it requires the consideration of penguin 
amplitudes. Explaining why this effect should be large for pions but 
small for rho mesons (rather than fitting it with new 
parameters~\cite{Li:2009wba,Cheng:2009eg}) appears to be a theoretical 
challenge.

\subsubsection{Ratios involving other $\pi\rho$ final states}

\begin{figure}[t]
\centerline{
\includegraphics[width=7cm]{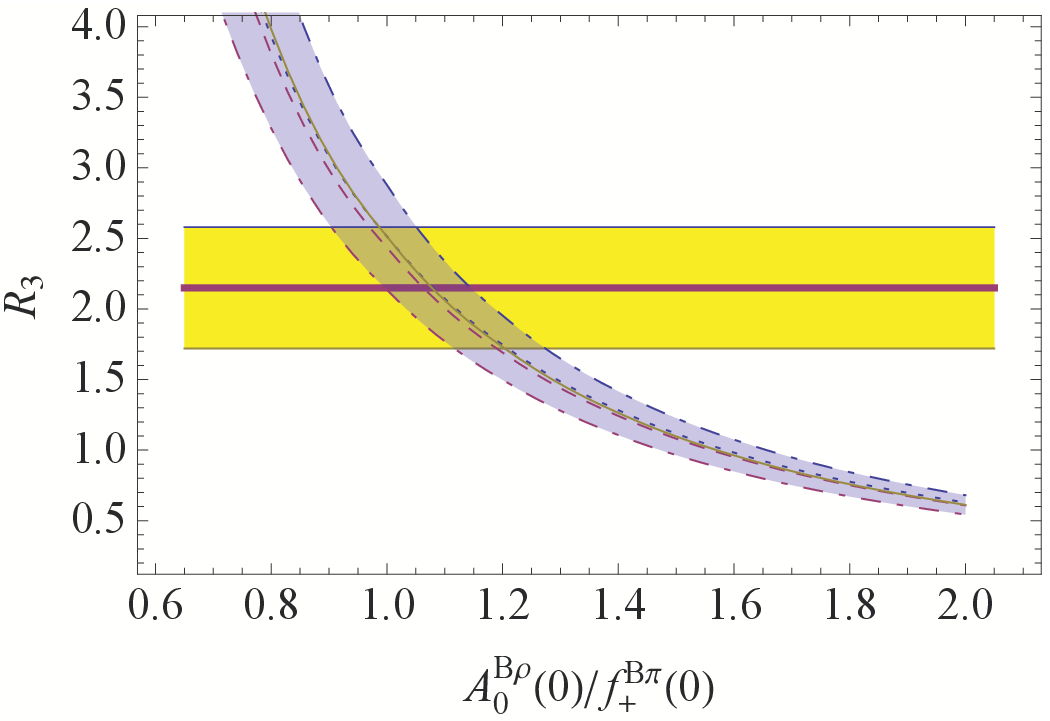}
\hspace{10mm}
\includegraphics[width=7cm]{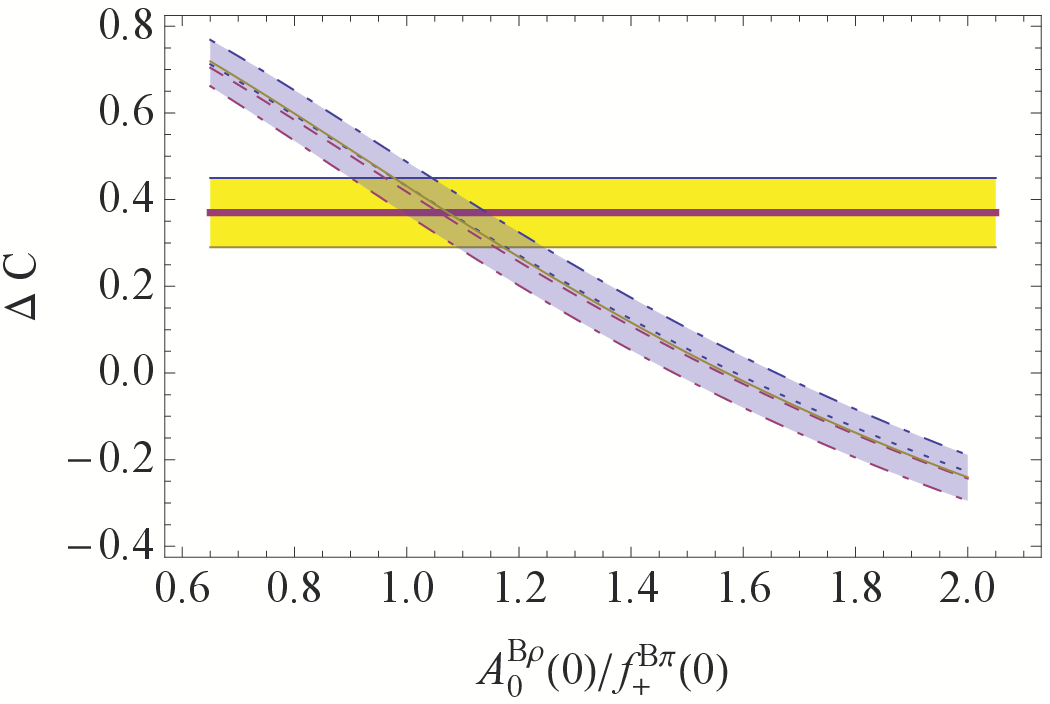}}
\vspace{5mm}
\centerline{\parbox{16cm}{\caption{\label{fig:R3PiRho}
Dependence of $R_3$ and $\Delta C$ on the ratio 
$A_0^{B\rho}(0)/f_+^{B\pi}(0)$ of $B$-meson form factors
for fixed $f_+^{B\pi}(0)=0.25$.
Our default input value for the form factor ratio
is 1.20. Lines and bands have the same meaning as in
Figure~\ref{fig:lambdaB}.}}}
\end{figure}

Next we discuss two ratios that involve only the charged
$\pi^\pm \rho^\mp$ final states. These final states are dominated
by the two colour-allowed tree amplitudes
$\alpha_1(\pi\rho)$ and $\alpha_1(\rho\pi)$. The ratios
we consider have few theoretical uncertainties other than the
one from the heavy-to-light form factors~\cite{Beneke:2003zv}.
Specifically, we define
\bea
\begin{aligned}
R_3 &\equiv \frac{\Gamma(\bar B^0\to \pi^+\rho^-)}
               {\Gamma(\bar B^0\to \pi^-\rho^+)} \,,
    \qquad\quad
&\Delta C &\equiv
\frac{1}{2} \left[C(\pi^-\rho^+)-C(\pi^+\rho^-)\right] \,, &&
\end{aligned}
\label{secondratios}
\eea
where $C(f)$ denotes the coefficient of the $\cos(\Delta m_B t)$
term in the time-dependent decay amplitude into the final state $f$,
see~\cite{Beneke:2003zv} for the relevant definitions. The dependence
of the theoretical calculation on the form factor ratio
$F_{\pi\rho}=A_0^{B\rho}(0)/f_+^{B\pi}(0)$ is shown in
Figure~\ref{fig:R3PiRho} together with the present experimental
data~\cite{Barberio:2008fa,Aubert:2003wr,:2007mj,Jessop:2000bv,Aubert:2007jn,
Kusaka:2007dv}.
The agreement is excellent for $F_{\pi\rho} \in [1.0, 1.2]$, which
is consistent with the input value 1.2 that follows from
Table~\ref{tab:inputs}.

\subsection{Final results}
\label{subsec:finalresults}

In this subsection we give our final theoretical predictions for the 
CP-averaged branching fractions and a number of ratios of CP-averaged 
$\pi\pi$, $\pi\rho$ and $\rho_L\rho_L$ decay rates.
In addition to the ones already discussed in the previous subsections, 
we further consider the ratios
\begin{eqnarray}
\begin{aligned}
   R_1 &\equiv \frac{\Gamma(\bar B^0\to \pi^+\rho^-)}
                    {\Gamma(\bar B^0\to \pi^+\pi^-)} \,,
    \qquad\quad
  &R_2 &\equiv \frac{\Gamma(\bar B^0\to \pi^+\rho^-)+
                     \Gamma(\bar B^0\to \pi^-\rho^+)}
                    {2\,\Gamma(\bar B^0\to \pi^+\pi^-)} \,, \\
  R_4 &\equiv \frac{2\,\Gamma(B^-\to \pi^-\rho^0)}
                    {\Gamma(\bar B^0\to \pi^-\rho^+)} - 1 \,,  \qquad\quad
  & & 
  && \\
   R_5 &\equiv \frac{2\,\Gamma(B^-\to \pi^0\rho^-)}
                    {\Gamma(\bar B^0\to \pi^+\rho^-)} - 1 \,,
    \qquad\quad
  &R_6 &\equiv \frac{\Gamma(\bar B^0\to \pi^+\rho^-)+
                     \Gamma(\bar B^0\to \pi^-\rho^+)}
                    {2\,\Gamma(\bar B^0\to \rho_L^{+}\rho_L^{-})} \, .
\end{aligned}
\end{eqnarray}
$R_3$ has already been defined in (\ref{secondratios}).
Moreover, we define the following ratios of the 
colour-suppressed decays,
\be
R_C^{\pi} = \frac{\Gamma(\bar B^0\to\pi^0\pi^0)}
 {\Gamma(\bar B^{0}\to\pi^0\rho^0)}\,,
 \qquad
R_{C}^{\rho} =  \frac{\Gamma(\bar B^{0}\to\rho_L^0\rho_L^0)}
{\Gamma(\bar B^0\to\pi^0\rho^0)}\,,
\ee
to eliminate the input parameter correlations among the three decay 
modes. 

\tabcolsep0.5cm
\begin{table}
  \centering
  \let\oldarraystretch=\arraystretch
  \renewcommand*{\arraystretch}{1.3}
\begin{tabular}{llll}
    \toprule
        & \phantom{0}Theory I& \phantom{0}Theory II
        & Experiment
    \\
    \midrule
$B^-\to\pi^-\pi^0$
 & $\phantom{0}5.43_{\,-0.06\,-0.84}^{\,+0.06\,+1.45}\;\;(\star)$
 & $\phantom{0}5.82_{\,-0.06\,-1.35}^{\,+0.07\,+1.42}\;\;(\star)$
 & $\phantom{0}5.59^{+0.41}_{-0.40}$ \\
$\bar B_d^0\to\pi^+\pi^-$
 & $\phantom{0}7.37_{\,-0.69\,-0.97}^{\,+0.86\,+1.22}\;\;(\star)$
 & $\phantom{0}5.70_{\,-0.55\,-0.97}^{\,+0.70\,+1.16}\;\;(\star)$
 & $\phantom{0}5.16 \pm 0.22$ \\
$\bar B_d^0\to\pi^0\pi^0$
 & $\phantom{0}0.33_{\,-0.08\,-0.17}^{\,+0.11\,+0.42}$
 & $\phantom{0}0.63_{\,-0.10\,-0.42}^{\,+0.12\,+0.64}$
 & $\phantom{0}1.55 \pm 0.19$ \\ \addlinespace
$B^-\to\pi^-\rho^0$
 & $\phantom{0}8.68_{\,-0.41\,-1.56}^{\,+0.42\,+2.71}\;\;(\star\star)$
 & $\phantom{0}9.84_{\,-0.40\,-2.52}^{\,+0.41\,+2.54}\;\;(\star\star)$
 & $\phantom{0}8.3^{+1.2}_{-1.3}$ \\
$B^-\to\pi^0\rho^-$
 & $12.38_{\,-0.77\,-1.41}^{\,+0.90\,+2.18}\;\;(\star)$
 & $12.13_{\,-0.73\,-2.17}^{\,+0.85\,+2.23}\;\;(\star)$
 & $10.9^{+1.4}_{-1.5}$ \\
$\bar B^0\to\pi^+\rho^-$
 & $17.80_{\,-0.56\,-2.10}^{\,+0.62\,+1.76}\;\;(\star)$
 & $13.76_{\,-0.44\,-2.18}^{\,+0.49\,+1.77}\;\;(\star)$
 & $15.7 \pm 1.8$ \\
$\bar B^0\to\pi^-\rho^+$
 & $          10.28_{\,-0.39\,-1.42}^{\,+0.39\,+1.37}\;\;(\star\star)$
 & $\phantom{0}8.14_{\,-0.33\,-1.49}^{\,+0.34\,+1.35}\;\;(\star\star)$
 & $\phantom{0}7.3 \pm 1.2$ \\
$\bar B^0\to\pi^\pm\rho^\mp$
 & $28.08_{\,-0.19\,-3.50}^{\,+0.27\,+3.82}\;\;(\dagger)$
 & $21.90_{\,-0.12\,-3.55}^{\,+0.20\,+3.06}\;\;(\dagger)$
 & $23.0 \pm 2.3$ \\
$\bar B^0\to\pi^0\rho^0$
 & $\phantom{0}0.52_{\,-0.03\,-0.43}^{\,+0.04\,+1.11}$
 & $\phantom{0}1.49_{\,-0.07\,-1.29}^{\,+0.07\,+1.77}$
 & $\phantom{0}2.0 \pm 0.5$ \\ \addlinespace
$B^-\to\rho^-_L\rho^0_L$
 & $18.42^{+0.23}_{-0.21}{}^{+3.92}_{-2.55}\;\;(\star\star)$
 & $19.06^{+0.24}_{-0.22}{}^{+4.59}_{-4.22}\;\;(\star\star)$
 & $22.8^{+1.8}_{-1.9}$ \\
$\bar B_d^0\to\rho^+_L\rho^-_L$
 & $25.98^{+0.85}_{-0.77}{}^{+2.93}_{-3.43}\;\;(\star\star)$
 & $20.66^{+0.68}_{-0.62}{}^{+2.99}_{-3.75}\;\;(\star\star)$
 & $23.7^{+3.1}_{-3.2}$ \\
$\bar B_d^0\to\rho^0_L\rho^0_L$
 & $\phantom{0}0.39^{+0.03}_{-0.03}{}^{+0.83}_{-0.36}$
 & $\phantom{0}1.05^{+0.05}_{-0.04}{}^{+1.62}_{-1.04}$
 & $\phantom{0}0.55^{+0.22}_{-0.24}$
        \\
    \midrule
$R_{+-}^{\pi\pi}$
 & $1.38^{+0.12}_{-0.13}{}^{+0.53}_{-0.32}$
 & $1.91^{+0.18}_{-0.20}{}^{+0.72}_{-0.64}$
 & $2.02 \pm 0.17$ \\
$R_{00}^{\pi\pi}$
 & $0.09^{+0.03}_{-0.02}{}^{+0.12}_{-0.04}$
 & $0.22^{+0.06}_{-0.05}{}^{+0.28}_{-0.16}$
 & $0.60 \pm 0.08$ \\ \addlinespace
$R_{+-}^{\rho\rho}$
 & $1.32^{+0.02}_{-0.03}{}^{+0.44}_{-0.27}$
 & $1.72^{+0.03}_{-0.03}{}^{+0.64}_{-0.53}$
 & $1.80^{+0.28}_{-0.29}$ \\
$R_{00}^{\rho\rho}$
 & $0.03^{+0.00}_{-0.00}{}^{+0.07}_{-0.03}$
 & $0.10^{+0.01}_{-0.01}{}^{+0.19}_{-0.11}$
 & $0.05 \pm 0.02$ \\ \addlinespace
$R_{00}^{\pi\rho}$
 & $0.04^{+0.00}_{-0.00}{}^{+0.09}_{-0.03}$
 & $0.14^{+0.01}_{-0.01}{}^{+0.20}_{-0.13}$
 & $0.17 \pm 0.05$ \\ \addlinespace
$R_1$
 & $2.41^{+0.16}_{-0.18}{}^{+0.32}_{-0.37}$
 & $2.41^{+0.17}_{-0.20}{}^{+0.37}_{-0.43}$
 & $3.04 \pm 0.37$ \\
$R_2$
 & $1.90^{+0.18}_{-0.19}{}^{+0.53}_{-0.41}$
 & $1.92^{+0.19}_{-0.20}{}^{+0.42}_{-0.40}$
 & $2.23 \pm 0.24$ \\
$R_3$
 & $1.73^{+0.13}_{-0.12}{}^{+1.12}_{-0.82}$
 & $1.69^{+0.13}_{-0.12}{}^{+0.72}_{-0.59}$
 & $2.15 \pm 0.43$ \\
$R_4$
 & $0.58^{+0.02}_{-0.02}{}^{+0.67}_{-0.35}$
 & $1.26^{+0.00}_{-0.00}{}^{+0.84}_{-0.75}$
 & $1.12^{+0.46}_{-0.48}$ \\
$R_5$
 & $0.30^{+0.05}_{-0.04}{}^{+0.36}_{-0.20}$
 & $0.64^{+0.06}_{-0.05}{}^{+0.50}_{-0.41}$
 & $0.30^{+0.22}_{-0.23}$ \\
$R_6$
 & $0.54^{+0.01}_{-0.01}{}^{+0.23}_{-0.17}$
 & $0.53^{+0.01}_{-0.01}{}^{+0.16}_{-0.13}$
 & $0.49 \pm 0.08$ \\ \addlinespace
$R_C^\pi$
 & $0.64^{+0.22}_{-0.17}{}^{+0.64}_{-0.37}$
 & $0.42^{+0.09}_{-0.08}{}^{+0.28}_{-0.16}$
 & $0.78 \pm 0.22$ \\
$R_C^\rho$
 & $0.74^{+0.10}_{-0.09}{}^{+0.58}_{-0.46}$
 & $0.70^{+0.06}_{-0.06}{}^{+0.46}_{-0.39}$
 & $0.27^{+0.13}_{-0.14}$ \\ \addlinespace
\bottomrule
\end{tabular}
\let\arraystretch=\oldarraystretch
\caption{
CP-averaged branching fractions in units of
$10^{-6}$~\cite{Barberio:2008fa,Aubert:2007hh,Abe:2006qx,Bornheim:2003bv,
Aubert:2006fha,Morello:2006pv,Aubert:2008sb,Abe:2006cx,Aubert:2009it,
Zhang:2003up,Aubert:2007nua,Somov:2006sg,:2008iha,:2008et,Godang:2001sg,
Aubert:2003fm,:2007mj,Jessop:2000bv,Aubert:2003wr,:2009az,Gordon:2002yt,
Aubert:2007py,Zhang:2004wza} and various ratios
of tree-dominated $B\to \pi\pi$, $\pi\rho$ and $\rho_L\rho_L$ decays. 
The first error on a quantity comes from the CKM parameters, while the 
second one stems from all other parameters added in 
quadrature~\cite{Beneke:2006hg}.  We consider ``Theory II'' as our 
reference values, see text.}
\label{tab:results}
\end{table}

The theoretical results and experimental measurements of the 
CP-averaged branching 
fractions of the twelve final states composed of pions and rho mesons, 
as well their ratios as defined above are summarized in 
Table~\ref{tab:results}. The column
labelled ``Theory I'', on which we focus first, uses the input values and
uncertainties defined in Table~\ref{tab:inputs}. The first error 
comprises the uncertainties from CKM parameters ($\gamma$, $|V_{cb}|$, 
$|V_{ub}|$ -- see below), the second combines all other 
uncertainties (scale, hadronic parameters, power correction
parameters) in quadrature. For the first part of the table 
containing the branching fractions, we apply the following modified 
procedure. Since the tree-dominated modes
are dominated by amplitudes proportional to $V_{ub}$, there is a 
large normalization uncertainty from $|V_{ub}|^2$, which we do 
{\em not} include in the table. Instead, we calculate the theoretical 
uncertainty of the quantity 
$\mbox{BrAv}(\bar B \to f)/|V_{ub}|^2$.\footnote{BrAv denotes the 
CP-averaged branching fraction.} 
The error from $|V_{ub}|$ can easily be restored by assuming 
that $\mbox{BrAv}(\bar B \to f) \propto |V_{ub}|^2$. In the same 
way, one can also rescale the branching fractions to account for 
a value of $|V_{ub}/V_{cb}|$ different from 0.09. We handle the 
dependence on the heavy-to-light form factors in a similar way. 
In this case one notes that the colour-suppressed final states 
$\pi^0\pi^0$, $\pi^0\rho^0$, $\rho^0\rho^0$ are almost independent 
of the form factors, since they are dominated by 
spectator scattering. The other modes, however, are nearly proportional 
to the square of the form factor. To remove the trivial dependence 
on the form factor uncertainty, which might be reducible in the future, 
we calculate  the theoretical 
uncertainty of the quantity $\mbox{BrAv}(\bar B \to f)/f_+^{B\pi}(0)^2$ 
for the modes marked in the table with $(\star)$, 
of $\mbox{BrAv}(\bar B \to f)/A_0^{B\rho}(0)^2$ for those marked 
with  $(\star\star)$, and of  
$\mbox{BrAv}(\bar B \to f)/(f_+^{B\pi}(0)A_0^{B\rho}(0))$ 
for the mode $ \bar B \to \pi^\pm\rho^\mp $ marked with $(\dagger)$. 
Once again, the full form-factor error can be restored by 
assuming the dependence on the form factor as divided out above, 
and the branching fractions can be approximately rescaled to other 
form-factor values by multiplying the appropriate factor.

Scanning the numbers in the ``Theory I'' column of 
Table~\ref{tab:results}, we notice that the $R_{+-}$ and 
$R_{00}$ ratios are systematically below the data, while the 
absolute branching fractions for the colour-allowed modes with 
only charged particles in the final state are consistently 
above. A plausible interpretation of this trend is that $\lambda_B$ is
in fact smaller than the default value $\lambda_B=350\,$MeV (a smaller 
value enhancing the colour-suppressed amplitude), and that the 
heavy-to-light form factors are about 10\% smaller than the QCD sum 
rule central values that we assumed in 
Table~\ref{tab:inputs} and used 
for ``Theory I''.\footnote{Alternatively, a smaller value of 
$|V_{ub}|$ might be considered. Indeed, our central values imply
$|V_{ub}| f_+^{B\pi}(0) = 9.34 \cdot 10^{-4}$,
which is slightly larger than~(\ref{Ball}).} Since we have already 
seen in Section~\ref{sec:coloursuppressed} that a large value of 
$\lambda_B$ is in conflict with data, we adopt the 
``small $\lambda_B$ and form-factor hypothesis'' and recalculate 
the theoretical prediction with the modified parameter 
values and ranges $f^{B\pi}_+(0)=0.23\pm 0.03$, $A_0^{B\rho}(0)=0.28\pm 0.03$,
$\lambda_B(1\,\mbox{GeV})=(0.20^{+0.05}_{-0.00})\,$GeV. The 
result is column ``Theory II'' in Table~\ref{tab:results}. We 
consider these as our reference predictions to be tested against 
future more accurate experimental results. 

Comparing now ``Theory II'' with measurements, we find good agreement 
for the branching fractions, except for $\pi^0\pi^0$ as already 
discussed. However, the theoretical uncertainty is also large for 
this mode. The same feature is reflected in the ratios, where 
$R_{00}^{\pi\pi}$ stands out as too small. The ratios $R_C^\pi$, 
$R_C^\rho$ of colour-suppressed modes do not fit very well either, 
reflecting the fact that factorization likes to have 
$[\rho^0\rho^0]_L$ to be larger and $\pi^0\pi^0$ to be smaller than 
data. Nevertheless, uncertainties taken face value, there is no 
disagreement. The $\pi\rho$ ratios $R_{1-6}$ are in good agreement 
with data, though the central value of $R_5$ shifts away from 
the data in scenario ``Theory II''. Since $R_4$ and $R_5$ 
provide access to the real part of $\alpha_2(\pi\rho)$ and 
 $\alpha_2(\rho\pi)$, respectively~\cite{Beneke:2003zv}, precise 
measurements of the $\pi\rho$ final states should provide further 
insight into the mechanism that generates the colour-suppressed 
tree amplitude.

\section{Conclusion}
\label{sec:conclude}

We computed the two-loop vertex corrections to the colour-allowed and
colour-suppressed tree amplitude in QCD factorization, completing 
the calculation of these amplitudes at NNLO. Technically, 
the calculation amounts to a matching calculation from QCD to SCET, 
involving two-loop renormalization and infrared subtractions 
of evanescent operators, and massive two-loop vertex integrals 
that depend on one dimensionless parameters, and the charm quark 
mass in some cases. We obtain fully analytic expressions for 
both amplitudes after integration over the Gegenbauer expansion,
including the exact dependence on the charm quark mass. The massless 
result is in complete agreement with a recent independent 
calculation~\cite{Bell:2009nk}, as well as the charm mass dependence
of the imaginary part~\cite{Bell:2007tv}, while the charm mass dependence 
of the real part agrees numerically.

The NNLO vertex correction to the colour-suppressed tree amplitude is 
sizable, ranging from 10\% to 25\% for the real and imaginary part, 
respectively, and is a few percent for the colour-allowed amplitude. 
When combined with the ${\cal O}(\alpha_s^2)$ 
correction to spectator scattering, already known 
from~\cite{Beneke:2005vv,Pilipp:2007mg,Kivel:2006ki}, 
we find a large cancellation, both in the real and imaginary parts, so
that the overall NNLO correction to the topological 
tree amplitudes is small. This is 
somewhat unfortunate, since an enhancement rather than a cancellation 
in the colour-suppressed tree amplitude might have helped to
cure the large discrepancy with experimental data in 
the $\pi^0\pi^0$ channel.

A dedicated phenomenological analysis of tree-dominated $B$~decays 
to (quasi) two-body final states with pions and rho mesons shows
that the QCD factorization approach obtains strong support from the 
factorization test performed in Section~\ref{sec:facttest}. Overall, 
the data is described very well within theoretical and experimental 
uncertainties, especially for low values of $\lambda_B \simeq 200$~MeV
and smaller form factors. The most problematic observables remain 
those related to the $\pi^0\pi^0$
branching fraction, which is predicted too low. Since the 
combined NNLO correction is rather small, the generic features 
of factorization are unchanged compared to the NLO analysis 
of~\cite{Beneke:2003zv}. Numerical differences arise primarily from 
modified parameter choices. Our final results are the columns 
labelled ``Theory II'' contained in 
Tables~\ref{tab:resultsR} and \ref{tab:results}. 

\vspace*{0.5em}
\noindent
\subsubsection*{Acknowledgement}
We would like to thank Bernd Jantzen for useful discussions, 
and Oleg Tarasov for collaboration
at an early stage of this work.
This work is supported in part by the 
DFG Sonder\-forschungsbereich/Transregio~9 
``Computergest\"utzte Theoretische Teilchenphysik''. 
X.-Q.~Li acknowledges support from the Alexander-von-Humboldt Stiftung.
T.H. acknowledges support from the German Federal Ministry of
Education and Research (BMBF).
M.B. and T.H. acknowledge hospitality from the CERN
theory group, where part of this work was performed.

\appendix

\section{Master integrals}
\label{ap:masters}

\begin{figure}[t]
\hspace*{1.2cm}\includegraphics[scale=0.5]{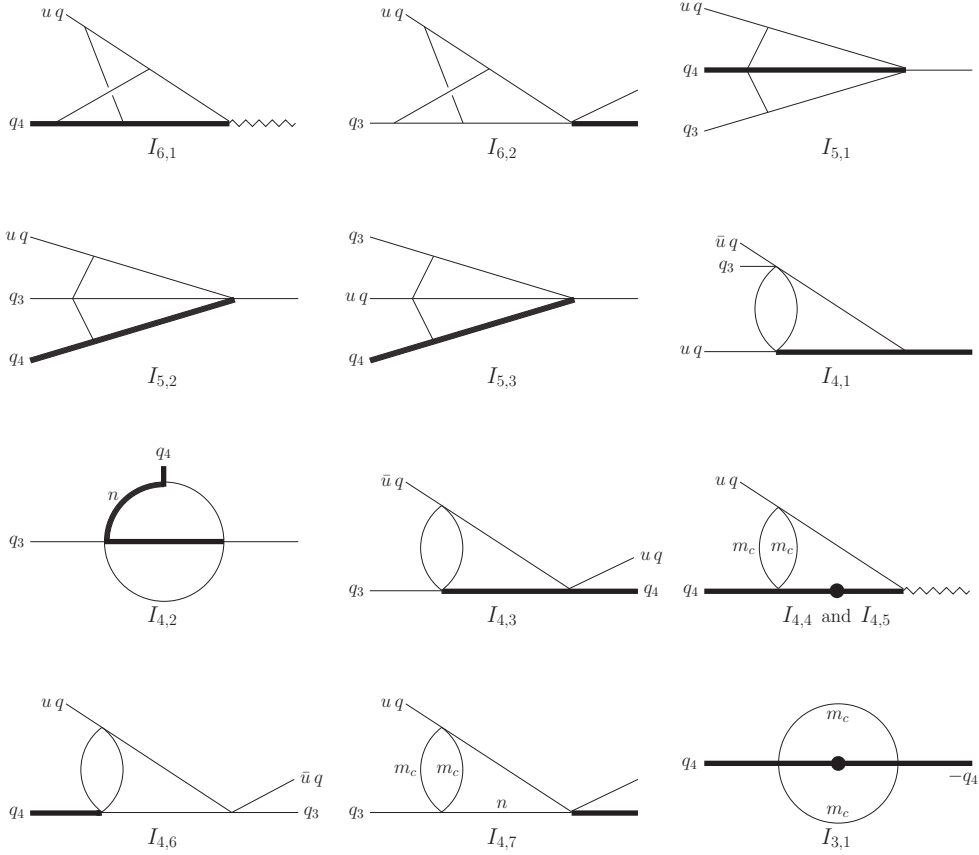}
\vskip0.5cm
\caption{Selected master integrals. All momenta indicated are assumed 
to be incoming. Bold lines denote lines of mass $m_b$, and thin
lines are massless unless otherwise indicated. Zig-zag lines have 
virtuality $\bar u m_b^2$. Dots
on lines denote squared propagators, and the diagrams in question 
stand for both the undotted and the dotted integral. $n$ represents a
generalized propagator power.}\label{fig:masters}
\end{figure}

In this Appendix we list a few master integrals on which our results 
add to the ones given in~\cite{Bell:2006tz},
for instance by giving closed forms valid to all orders in $\eps=(4-D)/2$. 
They are depicted in Figure~\ref{fig:masters}. The expansions
in $\eps$ are conveniently done with the package 
{\tt HypExp}~\cite{Huber:2005yg,Huber:2007dx} and agree with 
the results given in~\cite{Bell:2006tz}.
Our notation for the integration measure is
\be
\int \! \left[dk\right] \equiv \loopint{k} \; ,
\ee
and we define the prefactor
\be
\ESGamma \equiv \frac{1}{\left(4\pi\right)^{D/2} \Gamma(1-\eps)}\; .
\ee

\vspace*{5pt}

\noindent All momenta are incoming, and the kinematics is such 
that $q+q_3+q_4=0$ with $q^2=q_3^2=0$ and $q_4^2=m_b^2$.
We tacitly assume that all propagators in the integrals below contain an infinitesimal $+i\eta$. We then have
\bea
I_{5,1} &=& \int \! \left[dk_1\right] \!\int \! \left[dk_2\right] \frac{1}{\left[(k_1+k_2+q_4)^2-m_b^2\right] (k_1-uq)^2 (k_2-q_3)^2 \, k_1^2 \, k_2^2} \nnb\\
&=& -\ESGamma^2 \; (m_b^2)^{-1-2\eps} \; \frac{\Gamma^2(1-\eps) \Gamma(1+2\eps)\Gamma(2+\eps)\Gamma^2(-\eps)}{\Gamma(2-\eps)} \nnb\\
&& \hspace*{10pt}\times \left\{ \frac{1-u}{u} \; \pFq{4}{3}{1,1,1+2\eps,2+\eps}{2,2,2-\eps}{1-u}\right. \nnb\\
&&\hspace*{30pt} \left. -\frac{1}{u} \; \pFq{4}{3}{1,1,1+2\eps,2+\eps}{2,2,2-\eps}{1}\right\} \, . \label{eq:I51} \\
I_{5,2} &=& \int \! \left[dk_1\right] \!\int \! \left[dk_2\right] \frac{1}{\left[(k_2-q_4)^2-m_b^2\right] (k_1+k_2+q_3)^2 (k_1-u q)^2 \, k_1^2 \, k_2^2} \nnb\\
&=& -\ESGamma^2 \; (m_b^2)^{-1-2\eps} \; \Gamma^2(1-\eps)\Gamma(-\eps)\Gamma(1+\eps) \nnb \\
&& \hspace*{10pt}\times \bigg\{ -u^{-3\eps} \; \Gamma(3\eps-1)\Gamma(-\eps) \; e^{3\pi i\eps}\; \pFq{2}{1}{1-3\eps,1-2\eps}{2-3\eps}{u} \nnb\\
&&\hspace*{30pt} \left. +u^{-2\eps} \; \frac{\Gamma(1-\eps)\Gamma(2\eps-1)\Gamma(-\eps)}{\Gamma(1-3\eps)\Gamma(1+\eps)} \; e^{2\pi i\eps}\; \pFq{3}{2}{1,1-2\eps,1-\eps}{2-2\eps,1+\eps}{u}\right. \nnb\\
&&\hspace*{30pt} \left. - \frac{\Gamma(2\eps)}{3\eps} \; \pFq{3}{2}{1,1,1+\eps}{2,1+3\eps}{u}\right\} \; .
\label{eq:I52}
\eea
For the next integral it is convenient to split up the integration over $y$ into the three sub-intervals $0\ldots u/(u+1) \ldots u \ldots 1$. Alternatively, one can compute this master integral by using the
method of differential equations.
\bea
I_{5,3} &=& \int \! \left[dk_1\right] \!\int \! \left[dk_2\right] \frac{1}{\left[(k_2-q_4)^2-m_b^2\right] (k_1+k_2+u q)^2 (k_1-q_3)^2 \, k_1^2 \, k_2^2} \nnb\\
&=& \ESGamma^2 \; (m_b^2)^{-1-2\eps} \; \frac{\Gamma^2(1-\eps)\Gamma(1+2\eps)\Gamma(1+\eps)\Gamma(-\eps)}{2\eps^2} \int\limits_0^1 \! dy \, \frac{y^{-4\eps}}{u-y}\nnb \\
&& \hspace*{10pt}\times \left[ \pFq{2}{1}{\eps,2\eps}{1-\eps}{\frac{\bar y (u+i\eta-y)}{y^2}}-\pFq{2}{1}{\eps,2\eps}{1-\eps}{\frac{(y-u+i\eta)}{y}}\right] \nnb\\
&=& -\ESGamma^2 \; (m_b^2)^{-1-2\eps} \left\{\frac{1}{\eps}\left[-\frac{1}{6}\ln ^3(u)+\frac{1}{2} \ln (1-u) \ln ^2(u)-\frac{1}{6} \pi ^2 \ln (u)+\text{Li}_3(1-u)\right.\right. \nnb \\
&& \hspace*{100pt}\left.+\text{Li}_3(u)+i\pi \left(\frac{1}{2}  \ln ^2(u)+\text{Li}_2(1-u)+\frac{\pi^2}{6}\right)\right] \nnb \\
&& +\frac{3 \ln ^4(u)}{8}-\frac{5}{6} \ln (1-u) \ln ^3(u) -4 \ln ^2(1-u) \ln ^2(u)+\frac{5}{2} \text{Li}_2(u) \ln ^2(u)-\frac{1}{3} \pi ^2 \ln ^2(u)\nnb \\
&&+\frac{19}{6} \pi ^2 \ln (1-u) \ln (u)-5 \ln (1-u) \text{Li}_2(u) \ln (u)-11\text{Li}_3(1-u) \ln (u)-\frac{5}{2} \text{Li}^2_2(u)\nnb \\
&& -11 \text{Li}_3(u) \ln (u)+\text{Li}_4(1-u)+13 \text{Li}_4(u)-6 S_{2,2}(u)+6 \ln (1-u) \zeta(3)-\frac{209 \pi ^4}{360}\nnb \\
&& +\frac{13}{6} \pi ^2 \text{Li}_2(u)-6 \ln (1-u) \text{Li}_3(u)+i\pi\left(\frac{5}{2} \ln (1-u) \ln ^2(u)- \text{Li}_2(u) \ln (u)\right.\nnb \\
&& \left.\left.-\frac{5}{3} \pi^2 \ln(u)+2 \text{Li}_3(1-u)+7 \text{Li}_3(u) -\frac{3}{2} \ln ^3(u)\right) +{\cal O}(\eps)\right\}\; .
\label{eq:I53} \\
I_{4,1} &=& \int \! \left[dk_1\right] \!\int \! \left[dk_2\right] \frac{1}{\left[(k_1-u q)^2-m_b^2\right] (k_1+q_3+\bar u q)^2 (k_1+k_2)^2 \, k_2^2} \nnb\\
&=& -\ESGamma^2 \; (m_b^2)^{-2\eps} \; \frac{\Gamma^5(1-\eps)}{\Gamma(2-2\eps)} \nnb \\
&& \hspace*{10pt}\times \left\{ -\bar u^{1-3\eps} \; \frac{\Gamma(3\eps-1)\Gamma(1-2\eps)}{\Gamma(2-2\eps)} \; e^{3\pi i\eps}\; \pFq{2}{1}{1-\eps,1-2\eps}{2-2\eps}{u}\right. \nnb\\
&&\hspace*{30pt} \left. +\frac{\Gamma(1-3\eps) \Gamma(3\eps)}{\Gamma^2(1-\eps)\Gamma(2-3\eps)} \; \MeijerG{23}{33}{\bar u}{0,1-\eps,1-2\eps}{}{0,1-2\eps}{1-3\eps} \right\} \, .
\label{eq:I41}
\eea
The Meijer-G functions can be written as a linear combination of hypergeometric functions~\cite{thebook}. The latter are then expanded in $\eps$ with {\tt HypExp}. The next integral is a master only for
$n=1$ but for convenience we generalize one of the propagator powers.
\bea
I_{4,2} &=& \int \! \left[dk_1\right] \!\int \! \left[dk_2\right] \frac{1}{\left[(k_2+q_4)^2-m_b^2\right]^n \left[(k_1+k_2+q_4)^2-m_b^2\right] (k_1-q_3)^2 \, k_2^2} \nnb\\
&=& \ESGamma^2 \; (-1)^{-n} \; (m_b^2)^{1-n-2\eps} \; \frac{\Gamma^2(1-\eps)}{\Gamma(n)} \;
\MeijerG{24}{44}{1}{1-\eps-n,1-n,\eps,2-n-2\eps}{}{0,1-\eps-n}{-n,\eps-1} \, . \nnb \\
&&  \label{eq:I42} \\
I_{4,3} &=& \int \! \left[dk_1\right] \!\int \! \left[dk_2\right] \frac{1}{\left[(k_2+q_4)^2-m_b^2\right] \, k_1^2 \, (k_1+k_2-q)^2 \, (k_2-u q)^2} \nnb\\
&=& -\ESGamma^2 \; (m_b^2)^{-2\eps} \; \frac{\Gamma^4(1-\eps)\Gamma(\eps)\Gamma(2\eps)\Gamma(1-2\eps)}{\Gamma(2-2\eps)\Gamma(2-\eps)}
\; \pFq{2}{1}{1,2\eps}{2-\eps}{\bar u} \, . \label{eq:I4,3} \\
I_{4,4} &=& \int \! \left[dk_1\right] \!\int \! \left[dk_2\right] \frac{1}{\left[(k_1+q_4)^2-m_b^2\right] (k_1-u q)^2 \left[k_2^2-m_c^2\right] \, \left[\left(k_1+k_2\right)^2-m_c^2\right]} \nnb\\
&=& -\ESGamma^2 \; (m_b^2)^{-2\eps} \; \Bigg\{\frac{1}{2\eps^2}+\frac{1}{\eps}\left(\frac{5}{2}-\frac{u \ln (u)}{u-1}\right) + \frac{2 u \text{Li}_2(1-u)}{u-1}+\frac{2 u \ln ^2(u)}{u-1}\nnb\\
&&  \left.-\frac{5 u \ln (u)}{u-1}+\frac{\pi ^2}{2}+\frac{19}{2} +f_{44}(z,1-u)+{\cal O}(\eps)\right\} \, ,
\label{eq:I44}
\eea
with
\bea
f_{44}(z,x) &\equiv& -\MB{c_1}{w_1} \!\! \MB{c_2}{w_2} \nnb\\
 && \times \, \frac{2^{2 w_1-1} \pi ^{5/2} (1-x)^{w_2} z^{w_1} \csc (\pi w_1) \csc (\pi  w_2) \Gamma (-2 w_1-w_2+1) \Gamma(w_1+w_2)}{(w_1-1) w_1 \Gamma\left(\frac{3}{2}-w_1\right)} \nnb \\
\label{eq:f44}
\eea
and $z=m_c^2/m_b^2$. The two-fold Mellin-Barnes integration is along straight lines parallel to the imaginary axis, hence the real parts along the curves are constant. They read, respectively, $c_1=2/3$ and $c_2=-1/2$.
The form (\ref{eq:I44}) of $I_{4,4}$ enables us to obtain fully analytic results for the amplitudes $\alpha_1$ and $\alpha_2$, since we can interchange the integration over $u$ with the Mellin-Barnes integrations.
\bea
I_{4,5} &=& \int \! \left[dk_1\right] \!\int \! \left[dk_2\right] \frac{1}{\left[(k_1+q_4)^2-m_b^2\right]^2 (k_1-u q)^2 \left[k_2^2-m_c^2\right] \, \left[\left(k_1+k_2\right)^2-m_c^2\right]} \nnb\\
&& \hspace*{-1cm}
= -\, \ESGamma^2 \; (m_b^2)^{-1-2\eps} \; \bigg\{\!\!-\frac{1}{\eps}\,\frac{\ln (u)}{(u-1)}+\frac{\ln (u) \ln (z u)}{u-1}+\frac{\text{Li}_2(1-u)}{u-1}  + f_{45}(z,1-u) +{\cal O}(\eps)\bigg\}\; ,\nnb\\
\label{eq:I45}
\eea
with
\bea
f_{45}(z,x) &\equiv& -\MB{c_1}{w_1} \!\! \MB{c_2}{w_2} \nnb \\
&& \times \, \frac{2^{2 w_1-1} \pi ^{5/2} (1-x)^{w_2} z^{w_1} \csc(\pi w_1) \csc (\pi  w_2) \Gamma (-2 w_1-w_2) \Gamma(w_1+w_2+1)}{w_1 \Gamma \left(\frac{3}{2}-w_1\right)}\nnb \\
\label{eq:f45}
\eea
and $c_1=-1/24$ and $c_2=-5/6$.
\bea
I_{4,6} &=& \int \! \left[dk_1\right] \!\int \! \left[dk_2\right] \frac{1}{k_1^2 \, (k_1+k_2+q)^2 \, (k_2-q_3)^2 \, (k_2+\bar u q)^2} \nnb\\
&=& -\ESGamma^2 \; (m_b^2)^{-2\eps} \; \frac{\Gamma^5(1-\eps)\Gamma(\eps)\Gamma(2\eps)\Gamma(1-2\eps)}{\Gamma(2-2\eps)\Gamma(2-3\eps)\Gamma(1+\eps)}
\; e^{2\pi i\eps} \; \pFq{2}{1}{1,2\eps}{1+\eps}{u} \, .
\label{eq:I4,6}
\eea
In the next integral we again generalize one of the propagator powers.
\bea
I_{4,7} &=& \int \! \left[dk_1\right] \!\int \! \left[dk_2\right] \frac{1}{\left[(k_1+q_3)^2\right]^n (k_1-u q)^2 \left[k_2^2-m_c^2\right] \, \left[\left(k_1+k_2\right)^2-m_c^2\right]} \nnb\\
&=& \ESGamma^2 \; (-1)^{-n} \; (m_c^2)^{1-n-2\eps} \; \frac{\Gamma^2(1-\eps)}{\Gamma(n)} \, \sqrt{\pi} \, 2^{1-2n-2\eps}\nnb \\
&& \times \MeijerG{24}{44}{-\frac{u}{4z}-i\eta}{1-\eps-n,1-n,0,2-n-2\eps}{}{0,1-\eps-n}{\frac{1}{2}-n-\eps,\eps-1} \, .
\label{eq:I4,7}
\eea
The remaining integrals from Figure~\ref{fig:masters} can be 
found in the literature and shall not be given explicitly here: 
The integral $I_{6,1}$ through order ${\cal O}(\eps^0)$ can be 
found in~\cite{Huber:2009se}, whereas an all-order
result for $I_{6,2}$ is contained in~\cite{Gehrmann:2005pd}. 
The three-line master integral $I_{3,1}$ can be found in various 
papers~\cite{Berends:1997vk,Argeri:2002wz,Huber:2008mz,Bekavac:2009gz}.


\section{Auxiliary functions}
\label{ap:auxfunc}

\subsection{Hard-scattering kernels}
\label{ap:hsk}
In this Appendix we list the explicit expressions for the 
right-insertion hard-scattering kernels. The wrong-insertion
kernels can be obtained from linear combinations of 
the right-insertion ones, see~(\ref{eq:wronglinearright}).
The one-loop hard-scattering kernels 
in the CMM operator basis read
\bea 
 T_1^{(1)}  &=&\frac{C_F}{2N_c}
\bigg(-6 L+ \ln ^2(1-u)+2 \ln (u) \ln (1-u)-
\frac{2 \ln (1-u)}{u} \nnb \\[-0.1cm]
&& \hs{32} +\,3 \ln (1-u) -\ln ^2(u) +\frac{\ln (u)}{u-1}+3\ln (u)+
4 \text{Li}_2(u)-\frac{\pi ^2}{3}-22 \nnb \\
  && \hs{32} + \,i\pi \,\Big[ 2 \ln (1-u)-2 \ln (u) -3\Big]
\bigg) \, , \nonumber \\
 T_2^{(1)}  &=&0 \, ,
\eea
with
\be
L \equiv \ln\left(\frac{\mu^2}{m_b^2}\right) \; .
\ee

\begin{figure}[t]
\hspace*{60pt}\includegraphics[width=0.75\textwidth]{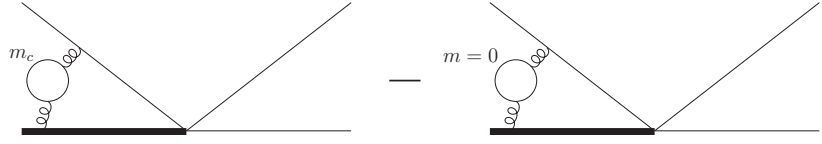}

\vs{-6}

\caption{Diagrammatic representation of the charm-mass dependent term. 
Three further diagram differences are not
shown.}\label{fig:charmdifference}
\end{figure}

At two loops, the kernels are conveniently subdivided into
the following building blocks
\bea
T_1^{(2)} &=&  T_1^{(2),re} + i \pi \, T_1^{(2),im} + 
n_l \, T_f \, ( T_{1,n_l}^{(2),re} + i \pi \, T_{1,n_l}^{(2),im}) \nnb \\
          && + T_f \, ( T_{1,T_f}^{(2),re} + 
i \pi \, T_{1,T_f}^{(2),im}) + 
\, T_f \, ( T_{1,c}^{(2),re} + i \pi \, T_{1,c}^{(2),im}) \, , 
\nonumber \\
T_2^{(2)} &=&  T_2^{(2),re} + i \pi \, T_2^{(2),im} \, . 
\label{eq:hskT22}
\eea
The charm-quark contribution $T_f \, ( T_{1,c}^{(2),re} + 
i \pi \, T_{1,c}^{(2),im})$ to $T_1^{(2)}$  arises from charm-loop 
insertions into the gluon
propagator of the non-factorizable one-loop diagrams. It is 
designed such as to give the difference between the contribution of a
quark of mass $m_c$ and a massless quark, see 
Figure~\ref{fig:charmdifference}. This construction ensures that 
we set $n_l=4$ irrespective
of whether we treat the charm quark as massive or massless. 
The explicit expressions for the two-loop hard-scattering kernels are
\bea
  T_1^{(2),re}  &=&\frac{\left(47 u^5-278 u^4+1223 u^3-2316 u^2+2036 u-652\right) \ln ^4(1-u)}{162 (u-1)^2u^3} \nnb\\
&&   -\frac{\left(2 u^3+4 u^2+173 u+16\right) \ln ^3(1-u)}{81 u} -\frac{\left(4 u^3-61 u^2-436 u+16\right) \ln ^2(1-u)}{54 u^2}\nnb\\
&&    +\frac{2 \left(73 u^5+38u^4-1103 u^3+2316 u^2-2036 u+652\right) \ln (u) \ln ^3(1-u)}{81 (u-1)^2u^3} \nnb\\
&&    -\frac{\left(17 u^3+300 u^2-1098 u+978\right) \ln ^2(u) \ln ^2(1-u)}{27u^3}  \nnb\\
&&    -\frac{\pi ^2 \left(9u^5+166 u^4-1167 u^3+2316 u^2-2036 u+652\right) \ln ^2(1-u)}{81 (u-1)^2 u^3} \nnb\\
&&    +\frac{\left(2 u^5-20 u^3+125 u^2-76u-52\right) \ln (u) \ln ^2(1-u)}{27 (u-1)^2 u} +\frac{2}{9} \ln ^3(u) \ln (1-u)\nnb\\
&&    +\frac{7(u-2)^2 \ln (2-u) \ln ^2(1-u)}{9 (u-1)^2}+\frac{16}{9} \text{Li}_2(u) \ln^2(1-u) \nnb\\
&&    +\frac{\left(2 u^6+4 u^5-191 u^4-167 u^3+1022u^2-646 u-6\right) \ln ^2(u) \ln (1-u)}{27 (u-1) u^3} \nnb\\
&&    -\frac{\pi ^2 \left(2 u^5+355 u^3-623 u^2+385 u-140\right) \ln (1-u)}{81(u-1)^2 u} \nnb\\
&&    -\frac{\left(4 u^4-638 u^3+1487 u^2-1597 u+664\right) \ln (u) \ln (1-u)}{27(u-1) u^2} \nnb\\
&&    +\frac{14 (u-2)^2 \text{Li}_2(u-1) \ln (1-u)}{9 (u-1)^2} +\frac{16 \left(6 u^2-16 u-5\right) \text{Li}_3(u) \ln(1-u)}{27 (u-1)^2}\nnb\\
&&    -\frac{2 \left(94u^3-271 u^2+166 u+32\right) \text{Li}_2(u) \ln (1-u)}{27 (u-1)^2 u}+\frac{(1601 u-1172) \ln(1-u)}{54 u}  \nnb\\
&&    +\frac{4 \left(4 u^3-50 u^2+183 u-163\right) \ln (u) \text{Li}_2(u) \ln (1-u)}{27 u^3} \nnb\\
&&    +\frac{\left(2 u^3-436 u^2+657 u-332\right) \ln^2(u)}{27 (u-1) u} -\frac{8 \left(3 u^2-14 u-19\right) \zeta (3) \ln (1-u)}{27(u-1)^2}\nnb\\
&&    +\frac{2 \left(20 u^5-94 u^4+292u^3-579 u^2+509 u-163\right) \text{Li}_2(u)^2}{27 (u-1)^2 u^3} \nnb\\
&&    +\frac{\pi ^2 \left(4u^4-435 u^3+3174 u^2-5346 u+2688\right)}{162 (u-1) u^2} +\frac{64}{9}\text{Li}_3(1-u) \ln (1-u)\nnb\\
&&    -\frac{\pi ^4 \left(225 u^5+378u^4-5914 u^3+11598 u^2-10198 u+3266\right)}{2430 (u-1)^2 u^3} \nnb\\
&&    +\frac{16 \left(u^3+2u-2\right) \text{HPL}(\{-2,2\},1-u)}{27 u^3} -\frac{\left(2 u^3+4 u^2-313 u+273\right) \ln^3(u)}{81 (u-1)} \nnb\\
&&    -\frac{8 (u-2) \left(5 u^2-6 u+2\right)\text{HPL}(\{-1,2\},1-u)}{27 (u-1) u^2} -\frac{47 \ln ^4(u)}{162}\nnb\\
&&    +\frac{\pi ^2 (u-2) \left(53 u^3-104 u^2-16 u+4\right) \ln (2-u)}{81 (u-1)^2u^2} -\frac{65}{81} \pi ^2 \ln ^2(u)\nnb\\
&&    -\frac{\pi ^2 \left(2 u^5+4 u^4-427 u^3+1209 u^2-2140 u+1320\right) \ln (u)}{81(u-1) u^2} \nnb\\
&&    +\frac{(1601 u-942) \ln (u)}{54 (u-1)} -\frac{4\left(u^2\!-3\right) \ln (u) \text{Li}_2(-u)}{27 u^3}-\frac{4 \pi ^2 \left(u^3+2 u-2\right) \text{Li}_2(u-1)}{81 u^3}\nnb\\
&&    +\frac{2 \pi ^2 \left(33 u^3+400 u^2-1472u+1312\right) \ln (1-u) \ln (u)}{81 u^3} -\frac{16}{27}\text{HPL}(\{-2,2\},u)\nnb\\
&&    +\frac{2 \pi ^2 \left(2 u^4+u^3+5 u^2+9 u-9\right) \ln (u+1)}{81 (u-1)u^3} +\frac{16 (u+1)^2 \text{HPL}(\{-1,2\},u)}{27(u-1) u}\nnb\\
&&    -\frac{\left(8 u^4-937 u^3+1523u^2-1509 u+664\right) \text{Li}_2(u)}{27 (u-1) u^2} -\frac{2 \left(u^2-3\right) \ln ^2(u) \ln(u+1)}{27 u^3}\nnb\\
&&    -\frac{2 \pi ^2 \left(10 u^5-208 u^4+1183 u^3-2340 u^2+2060 u-660\right) \text{Li}_2(u)}{81 (u-1)^2 u^3} \nnb\\
&&    -\frac{2 \left(34u^4+530 u^3-1551 u^2+972 u+6\right) \ln (u) \text{Li}_2(u)}{27 (u-1) u^3}+\frac{4}{81} \pi ^2\text{Li}_2(-u) \nnb\\
&&    +\frac{2\left(4 u^7+2 u^6-39 u^5-63 u^4+678 u^3-959 u^2+362 u-6\right) \text{Li}_3(1-u)}{27(u-1)^2 u^3} \nnb\\
&&    -\frac{16 \left(6 u^3+50 u^2-183 u+163\right) \ln (u) \text{Li}_3(1-u)}{27u^3} +\frac{68}{27} \ln ^2(u) \text{Li}_2(u) \nnb\\
&&    -\frac{14 (u-2)^2 \text{Li}_3(u-1)}{9 (u-1)^2} +\frac{4 \left(u^2-3\right)\text{Li}_3(-u)}{27 u^3}-\frac{104}{9} \ln (u) \text{Li}_3(u)  \nnb\\
&&    +\frac{2 \left(4 u^7+2 u^6-39 u^5+743 u^4-2747 u^3+3350 u^2-1292u-6\right) \text{Li}_3(u)}{27 (u-1)^2 u^3} \nnb\\
&&    -\frac{4 \left(86 u^5-406 u^4+1553 u^3-2919 u^2+2569u-823\right) \text{Li}_4(1-u)}{27 (u-1)^2 u^3} \nnb\\
&&    +\frac{4 \left(96 u^5-372 u^4+1283u^3-2316 u^2+2036 u-652\right) \text{Li}_4(u)}{27 (u-1)^2 u^3} \nnb\\
&&    -\frac{16 \left(5 u^5+36u^4-289 u^3+579 u^2-509 u+163\right) \text{Li}_4\left(\frac{u}{u-1}\right)}{27 (u-1)^2u^3} \nnb\\
&&    +\frac{2 \left(62 u^5-397 u^4+1012 u^3-1112 u^2+428 u+2\right) \zeta (3)}{9 (u-1)^2u^3} \nnb\\
&&    +\frac{8 \left(3 u^3+100 u^2-366 u+326\right) \ln (u) \zeta (3)}{27 u^3}-\frac{16\pi ^2 (2 u-1) \ln (2)}{27 (u-1) u}-\frac{3793}{36}\nnb\\
&&    +L\left(-\frac{32}{81} \ln ^3(1-u)+\frac{2 (27 u+4) \ln ^2(1-u)}{27 u}-\frac{32}{27}\ln (u) \ln ^2(1-u)\right. \nnb\\
&&    -\frac{32}{27} \ln ^2(u) \ln (1-u)+\frac{2 (13 u-10) \ln(1-u)}{3 u}+\frac{4 (51 u-4) \ln (u) \ln (1-u)}{27 u} \nnb\\
&&    -\frac{64}{27} \text{Li}_2(u)\ln (1-u)+\frac{32}{81} \pi ^2 \ln (1-u) +\frac{32 \ln ^3(u)}{81}-\frac{2 (51 u-43)\ln ^2(u)}{27 (u-1)} \nnb\\
&&    -\frac{2 \pi ^2 (27 u-23)}{81 (u-1)}+\frac{26 (3 u-2) \ln (u)}{9(u-1)}+\frac{8 \left(39 u^2-39 u+2\right) \text{Li}_2(u)}{27 (u-1) u}\nnb\\
&&    \left.-\frac{64}{27} \ln(u) \text{Li}_2(u) -\frac{64}{27} \text{Li}_3(1-u)+\frac{64\text{Li}_3(u)}{27}-55\right) -\frac{26 L^2}{3} \, ,
\eea
where the functions $\text{HPL}$ denote the harmonic polylogarithms~\cite{Remiddi:1999ew}. We proceed with
\bea
  T_1^{(2),im}  &=&\frac{8}{81} \ln ^3(1-u)+\frac{2 \left(u^3+2 u^2-53 u+10\right) \ln ^2(1-u)}{27 u}+\frac{8}{9} \ln (u) \ln ^2(1-u)  \nnb\\
&&       +\frac{\left(4 u^2+455 u-40\right) \ln (1-u)}{27 u}+\frac{8}{27} \text{Li}_2(u) \ln (1-u)-\frac{64}{81} \pi ^2 \ln (1-u) \nnb\\
&&       -\frac{2 \left(2 u^4+4 u^3-40 u^2+31 u-12\right) \ln (u) \ln (1-u)}{27 (u-1) u}-\frac{56}{27} \ln ^2(u) \ln (1-u)  \nnb\\
&&       +\frac{2 \pi ^2 \left(3 u^3+5 u^2-8 u-5\right)}{81 (u-1)} +\frac{\left(2 u^3+4 u^2+38 u-57\right) \ln^2(u)}{27 (u-1)} +\frac{4 \ln ^3(u)}{27}\nnb\\
&&       -\frac{\left(4u^2+221 u-210\right) \ln (u)}{27 (u-1)}+\frac{40}{81} \pi ^2 \ln (u)-\frac{4}{27} u\text{Li}_2(u)-\frac{8}{3} \ln (u) \text{Li}_2(u)  \nnb\\
&&       +\frac{20 \text{Li}_2(u)}{27(u-1)} -\frac{8 \text{Li}_2(u)}{27 u} +\frac{4 \text{Li}_2(u)}{9}-\frac{80}{27}\text{Li}_3(1-u) +\frac{40 \text{Li}_3(u)}{9}-\frac{1609}{54} \nnb \\
&&       +L\left(-\frac{16}{27} \ln ^2(1-u)-\frac{4 (39 u-35) \ln (u)}{27 (u-1)} -\frac{64\text{Li}_2(u)}{27} +\frac{16 \pi ^2}{81} \right.\nnb\\
&&       \left.+\frac{4 (39 u-4) \ln (1-u)}{27 u}-\frac{32}{27} \ln(u) \ln (1-u)+\frac{16 \ln ^2(u)}{27} -\frac{26}{3}\right) \, ,
\eea
\bea
T_{1,n_l}^{(2),re} &=&\frac{8}{27} \ln ^3(1-u)-\frac{4 (u+12) \ln ^2(1-u)}{81 u}-\frac{16}{27} \ln (u) \ln^2(1-u)\nnb \\
&&           +\frac{16}{27} \ln ^2(u) \ln (1-u)-\frac{4 (51 u-32) \ln (1-u)}{81 u}  \nnb\\
&&           -\frac{8(19 u-16) \ln (u) \ln (1-u)}{81 (u-1)}+\frac{16}{27} \pi ^2 \ln (1-u)+\frac{8 (14 u-11) \ln ^2(u)}{81 (u-1)}  \nnb\\
&&           +\frac{4 \pi ^2 (19 u-16)}{243 (u-1)} -\frac{4 (51 u-32) \ln (u)}{81 (u-1)} -\frac{16}{27} \pi ^2 \ln (u)  -\frac{8 \ln^3(u)}{27}  \nnb\\
&&           -\frac{8 \text{Li}_2(u)}{27(u-1)}-\frac{16 \text{Li}_2(u)}{27 u}-\frac{232 \text{Li}_2(u)}{81}-\frac{16}{27}\text{Li}_3(1-u)+\frac{16 \text{Li}_3(u)}{27}+\frac{250}{27} \nnb \\
&&           +L \left(-\frac{8}{27} \ln ^2(1-u)-\frac{8 (3 u-2) \ln (1-u)}{27 u} -\frac{16}{27} \ln(u) \ln (1-u) \right. \nnb\\
&&           \left. +\frac{8 \ln ^2(u)}{27}-\frac{8 (3 u-2) \ln (u)}{27 (u-1)}-\frac{32\text{Li}_2(u)}{27}+\frac{8 \pi ^2}{81}+\frac{136}{27}\right) +\frac{8 L^2}{9} \, , \\
T_{1,n_l}^{(2),im} &=&\frac{8}{27} \ln ^2(1-u)-\frac{152}{81} \ln (1-u)-\frac{8 \ln ^2(u)}{27}+\frac{80 \ln (u)}{81}+\frac{68}{27} \nnb \\
&&    -\frac{8}{27} \, L \, \bigg(2 \ln (1-u)-2 \ln (u)-3\bigg) \, .
\eea
The terms arising from the bottom quark-loop insertion into the gluon line 
read:
\bea
T_{1,T_f}^{(2),re} &=& L \left(-\frac{8}{27} \ln ^2(1-u)-\frac{8 (3 u-2) \ln (1-u)}{27 u} -\frac{16}{27} \ln(u) \ln (1-u) \right.\nnb\\
&&    \hs{19}\left. +\frac{8 \ln ^2(u)}{27}-\frac{8 (3 u-2) \ln (u)}{27 (u-1)}-\frac{32\text{Li}_2(u)}{27}+\frac{8 \pi^2}{81}+\frac{136}{27}\right) +\frac{8 L^2}{9} \nnb \\
&&    -\frac{4 (51 u-32) \ln (1-u)}{81 u}-\frac{8(19 u-16) \ln (u) \ln (1-u)}{81 (u-1)} -\frac{4 (51 u-32) \ln (u)}{81(u-1)}\nnb\\
&& +\frac{16 (47 u-31) \ln (1-u)}{81 (u-1)^2} +\frac{256}{81 u^2} -\frac{760}{81 u}+\frac{250}{27}\nnb \\
&& +\frac{16 \left(19 u^3-39 u^2+54 u-16\right) \ln (u) \ln (1-u)}{81 u^3}+\frac{56 \text{Li}_2(u)}{27 (u-1)}\nnb \\
&& -\frac{4 \pi ^2 \left(19 u^5-18 u^4-75 u^3+196 u^2-126 u+36\right)}{243(u-1)^3 \, u^2}+\frac{16 (47 u-16) \ln (u)}{81 u^2} \nnb \\
&& -\frac{224 \text{Li}_2(u)}{27 u}+\frac{256 \text{Li}_2(u)}{81 (u-1)^3}-\frac{256 \text{Li}_2(u)}{81 u^3}+\frac{232 \text{Li}_2(u)}{81} +\frac{32 \text{Li}_2(u)}{3 u^2}-\frac{16 \text{Li}_3(u)}{27}\nnb \\
&& +\frac{16\left(u^3-6 u+6\right) \text{Li}_3(1-u)}{27 u^3}+\frac{32 \left(3 u^3-3 u^2+3 u-1\right) \zeta (3)}{9 (u-1)^2 u^3}-\frac{1840}{81 (u-1)}\nnb \\
&& -\frac{256}{81(u-1)^2}-\frac{128 \ln ^3(2)}{27 (u-1)^2}+\frac{64 \pi ^2 \ln (2)}{27 (u-1)^2} + \frac{4 (6\omega_1-1)}{243 \omega_1\sqrt{\omega_1}} \, p_1(\omega_1)\nnb \\
&&  + \frac{32}{81} \, p_2(\omega_1)- \frac{(15\phi_1+23)}{243 \phi_1\sqrt{\phi_1}} \, p_1(\phi_1) - \frac{4 (5\phi_1^2+6\phi_1-3)}{81\phi_1^2} \, p_2(\phi_1) \, , \\
T_{1,T_f}^{(2),im} &=&-\frac{8}{27} L\bigg(2 \ln (1-u)-2 \ln (u)-3\bigg)+\frac{68}{27} -\frac{880}{81 (u-1)}+\frac{128}{81 u}-\frac{64 \ln^2(2)}{9 (u-1)^2}\nnb \\
&&  + \frac{32 (6\omega_1-1)}{81 \omega_1\sqrt{\omega_1}} \, p_3(\omega_1) - \frac{32}{27} \, p_4(\omega_1)- \frac{8(15\phi_1+23)}{81 \phi_1\sqrt{\phi_1}} \, p_3(\phi_1) \nnb \\
&& + \frac{4 (5\phi_1^2+6\phi_1-3)}{27\phi_1^2} \, p_4(\phi_1) \, ,
\eea
with the abbreviations
\bea
\omega_1 \equiv \frac{\frac{u}{4}}{\frac{u}{4}+1} \; , \qquad \phi_1 \equiv \frac{\frac{1-u}{4}}{\frac{1-u}{4}+1} \; ,
\eea
and the auxiliary functions
\bea
p_1(x) &=&12 \ln ^2\left(\frac{\sqrt{x}+1}{2 \sqrt{1-x}}\right)-3 \ln ^2(1-x)+12 \ln\left(\frac{2 \left(\sqrt{x}+1\right)}{\sqrt{1-x}}\right) \ln (1-x) \nnb \\
&& -24 \ln\left(\frac{\sqrt{x}+1}{\sqrt{1-x}}\right) \ln (x)-24 \text{Li}_2\left(\frac{\sqrt{x}+1}{2}\right)-24 \text{Li}_2\left(-\sqrt{x}\right) \nnb \\
&&+24\text{Li}_2\left(\sqrt{x}\right)-24 \ln ^2(2)+2 \pi ^2 \, , 
\nonumber \\
p_2(x) &=& 2 \ln ^3\left(\frac{\sqrt{x}+1}{\sqrt{1-x}}\right)-\pi ^2 \ln \left(\frac{2
   \left(\sqrt{x}+1\right)}{\sqrt{1-x}}\right)-3
   \text{Li}_3\left(\frac{1-\sqrt{x}}{\sqrt{x}+1}\right)+2 \ln ^3(2) \, , 
\nonumber \\
p_3(x) &=& \ln \left(\frac{\sqrt{x}+1}{\sqrt{1-x}}\right) \, , 
\nonumber \\
p_4(x) &=& \ln \left(\frac{\sqrt{x}+1}{2 \sqrt{1-x}}\right) \ln \left(\frac{2\left(\sqrt{x}+1\right)}{\sqrt{1-x}}\right) \, .
\eea
The charm-mass dependent parts mentioned at the beginning of the section read
\bea
T_{1,c}^{(2),re} &=& -\frac{8}{27} \ln ^3(1-u)+\frac{4 (u+12) \ln ^2(1-u)}{81 u}+\frac{16}{27} \ln (u) \ln^2(1-u)-\frac{1840 z}{81 (u-1)} \nnb \\
&& +\frac{8}{27} \ln (z)\ln ^2(1-u)+\frac{8 (u-1)^2 \ln ^2(1-u)}{27 u z}-\frac{16}{27} \ln ^2(u) \ln (1-u) -\frac{256z}{81 (u-1)^2}\nnb \\
&& -\frac{32 (5 u+3) \ln (1-u)}{81 u}+\frac{880 z \ln (1-u)}{81 (u-1)}+\frac{8 (19 u-16) \ln (u) \ln (1-u)}{81 (u-1)}-\frac{760 z}{81 u} \nnb \\
&& +\frac{16 z \ln (u) \ln (1-u)}{9 u}-\frac{8 u^2 \ln (u) \ln (1-u)}{27 (u-1) z}+\frac{8 (3 u-2) \ln(z) \ln (1-u)}{27 u} +\frac{256 z}{81 u^2}\nnb \\
&& +\frac{16}{27} \ln (u) \ln (z) \ln (1-u)-\frac{8 (u-1)^2 \ln(z) \ln (1-u)}{27 u z}-\frac{16 (u-1)^2 \ln (1-u)}{27 u z} \nnb \\
&& -\frac{16}{27} \pi ^2 \ln(1-u)+\frac{8 \ln ^3(u)}{27}-\frac{8 \pi ^2 (u-2) z^2}{27 u^2}-\frac{8 (14 u-11) \ln^2(u)}{81 (u-1)}-\frac{4 \pi ^2 (u+2)}{243 (u-1)} \nnb \\
&& -\frac{16 z \ln ^2(u)}{9 u}+\frac{8 u^2 \ln ^2(u)}{27 (u-1)z}-\frac{4 \left(19 u^3-42 u^2+39 u-8\right) z \ln ^2(z)}{81 (u-1)^2 u^2} -\frac{4 \ln^2(z)}{9}\nnb \\
&& -\frac{4 \pi ^2 \left(55 u^3-132 u^2+111u-26\right) z}{243 (u-1)^2 u^2}-\frac{8 \left(149 u^3-192 u^2+75 u-16\right) z \ln (z)}{81 (u-1)^2 u^2}\nnb \\
&& +\frac{16 (19 u-16) \ln (u)}{81 (u-1)} +\frac{8 (u-1) f_{44}(z,u)}{27z}+\frac{8 \left(u^3-6 z^2 u+6 z^2\right)f_{44}(z,1-u)}{27 u^2 z}\nnb \\
&& +\frac{304 z \ln (u)}{81 u}+\frac{8 \left(3 u^2-2 z u-6 u+6 z+3\right)\left(u^2+4 z u-2 u+1\right) f_{45}(z,u)}{81 (u-1)^2 z} \nnb \\
&& -\frac{8 \left(3 u^2-7 z u+4 z\right) \left(u^2-4 z u+4 z\right) f_{45}(z,1-u)}{81 u^2 z}-\frac{16 u^2 \ln (u)}{27 (u-1) z}+\frac{16}{27} \pi ^2 \ln (u) \nnb \\
&& -\frac{8}{27} \ln ^2(u) \ln (z)+\frac{8 (3 u-2) \ln (u) \ln(z)}{27 (u-1)}+\frac{16 z \ln (u) \ln (z)}{9 u}-\frac{8 u^2 \ln (u) \ln (z)}{27(u-1) z} \nnb \\
&& -\frac{8}{9} \ln (1-z) \ln (z)-\frac{8\pi ^2}{81}  \ln (z)+\frac{4 \left(3 u^2-3 u+1\right) \left(\pi ^2 u-6\text{Li}_2(u)\right)}{81 (u-1) u z} +\frac{16 z \text{Li}_2(u)}{9 u}\nnb \\
&& +\frac{8 \left(29 u^2-20 u-6\right) \text{Li}_2(u)}{81 (u-1) u}-\frac{8 \left(u^3+12 u^2-15 u+10\right) z \text{Li}_2(1-z)}{81 (u-1)^2 u^2}-\frac{8 \text{Li}_2(z)}{9} \nnb \\
&& +\frac{32}{27} \ln (z)\text{Li}_2(u)-\frac{8 (2 u-1) \text{Li}_2(1-z)}{27 z} +\frac{4 (u-2) z^2 q_7(z)}{9u^2}+\frac{8 (2u-1) q_8(z)}{27 \sqrt{z}}\nnb \\
&& -\frac{16 \left(2 u^3-21 u^2+29 u-14\right) z^{3/2} q_8(z)}{81 (u-1)^2u^2}-\frac{8 \left(9 u^2-5 u-2\right) \!\sqrt{z} \, q_8(z)}{81 (u-1) u}+\frac{64 z^2 \zeta (3)}{9 (u-1)^2}\nnb \\
&& -\frac{128 z^2 \ln^3(2)}{27 (u-1)^2}+\frac{64 \pi ^2 z^2 \ln (2)}{27 (u-1)^2} +\frac{16}{27}\text{Li}_3(1-u)-\frac{16 \text{Li}_3(u)}{27}+ \frac{4 (6\omega_z-1)}{243 \omega_z\sqrt{\omega_z}} \, p_1(\omega_z)\nnb \\
&&  + \frac{32}{81} \, p_2(\omega_z)- \frac{(15\phi_z+23)}{243 \phi_z\sqrt{\phi_z}} \, p_1(\phi_z) - \frac{4 (5\phi_z^2+6\phi_z-3)}{81\phi_z^2} \, p_2(\phi_z) \, , \label{eq:Tcre}
\eea
\bea
T_{1,c}^{(2),im} &=&-\frac{64 \ln ^2(2) z^2}{9 (u-1)^2}-\frac{16 (47 u+8) z}{81 (u-1)u}-\frac{8}{27} \ln ^2(1-u)+\frac{8 \ln ^2(u)}{27}+\frac{16}{27} \ln (1-u) \ln (z) \nnb \\
&& +\frac{152}{81} \ln (1-u)-\frac{80\ln (u)}{81}-\frac{16}{27} \ln (u) \ln (z)-\frac{8\ln (z)}{9}+ \frac{32 (6\omega_z-1)}{81 \omega_z\sqrt{\omega_z}} \, p_3(\omega_z) \nnb \\
&&  - \frac{32}{27} \, p_4(\omega_z)- \frac{8(15\phi_z+23)}{81 \phi_z\sqrt{\phi_z}} \, p_3(\phi_z) + \frac{4 (5\phi_z^2+6\phi_z-3)}{27\phi_z^2} \, p_4(\phi_z) \, ,
\eea
where now
\bea
\omega_z \equiv \frac{\frac{u}{4z}}{\frac{u}{4z}+1} \; , \qquad \phi_z \equiv \frac{\frac{1-u}{4z}}{\frac{1-u}{4z}+1} \; ,
\eea
as well as $z = m_c^2/m_b^2$. The other functions, notably $f_{44}$, $f_{45}$, $q_7$, and $q_8$, can be found in (\ref{eq:f44}),
(\ref{eq:f45}), and (\ref{eq:q8}), respectively.
The two-loop expressions for the kernel $T_2^{(2)}$ 
related to the insertion of the 
operator $Q_2$ are simpler. They are
\bea
  T_2^{(2),re}  &=& \frac{4 (u-4) \left(u^4-2 u^3+7 u^2-8 u+3\right) \ln ^4(1-u)}{27 (u-1)^2 u^3} +\frac{32 (2 u+1) \text{Li}_3(u) \ln (1-u)}{9 (u-1)^2} \nnb \\
&&    +\frac{2 \left(u^3+2 u^2+7 u-4\right) \ln ^3(1-u)}{27 u} +\frac{\left(2 u^3-27 u^2-12 u+8\right) \ln ^2(1-u)}{9 u^2}   \nnb \\
&&    -\frac{16 \left(2 u^5-8 u^4+16 u^3-36 u^2+35 u-12\right) \ln (u) \ln ^3(1-u)}{27 (u-1)^2 u^3} +\frac{4 \ln ^4(u)}{27} \nnb \\
&&    +\frac{4 \left(u^3-6 u^2+33 u-36\right) \ln ^2(u) \ln ^2(1-u)}{9 u^3}  +\frac{2 (u-2)^2 \ln (2-u) \ln ^2(1-u)}{3 (u-1)^2} \nnb \\
&&    -\frac{8 \pi ^2 \left(u^5+2 u^4-13 u^3+36 u^2-35 u+12\right) \ln ^2(1-u)}{27 (u-1)^2 u^3} -\frac{8}{9} \text{Li}_2(u) \ln ^2(1-u)\nnb \\
&&    -\frac{2 \left(u^5-23 u^3+87 u^2-92 u+30\right) \ln (u) \ln ^2(1-u)}{9 (u-1)^2 u} -\frac{8}{27} \ln ^3(u) \ln (1-u) \nnb \\
&&    -\frac{2 \left(u^6+2 u^5-5 u^4+12 u^3-63 u^2+51 u-3\right) \ln ^2(u) \ln (1-u)}{9 (u-1) u^3} -\frac{4}{9} \pi ^2 \ln ^2(u)\nnb \\
&&    -\frac{(151 u-86) \ln (1-u)}{9 u}+\frac{2 \pi ^2 \left(u^5-8 u^3+75 u^2-95 u+30\right) \ln (1-u)}{27(u-1)^2 u} \nnb \\
&&    +\frac{32 \pi ^2 \left(u^3+2 u^2-11 u+12\right) \ln (u) \ln (1-u)}{27 u^3} +\frac{4 (u-2)^2 \text{Li}_2(u-1) \ln (1-u)}{3 (u-1)^2} \nnb \\
&&    +\frac{2 \left(2 u^4-50 u^3+14 u^2+81 u-42\right) \ln (u) \ln (1-u)}{9 (u-1) u^2} -\frac{16}{3} \text{Li}_3(1-u) \ln (1-u)\nnb \\
&&    +\frac{4 \left(12 u^3-53 u^2+58 u-20\right) \text{Li}_2(u) \ln (1-u)}{9 (u-1)^2 u} -\frac{32 (2 u+1) \zeta(3) \ln (1-u)}{9 (u-1)^2} \nnb \\
&&    +\frac{8 \left(4 u^3-2 u^2+11 u-12\right) \ln (u) \text{Li}_2(u) \ln (1-u)}{9 u^3} +\frac{2 \left(u^2-3\right) \ln ^2(u) \ln (u+1)}{9 u^3}\nnb \\
&&    +\frac{4 \left(8 u^5-16 u^4+24 u^3-36 u^2+35 u-12\right)\text{Li}_2(u)^2}{9 (u-1)^2 u^3} +\frac{2 \pi ^2 \left(u^2-3\right) \ln (u+1)}{9 u^3} \nnb \\
&&    -\frac{\pi ^2 \left(2 u^4+3 u^3-100 u^2+298 u-168\right)}{27 (u-1) u^2}-\frac{2 \left(u^3-34 u^2+7 u+21\right) \ln ^2(u)}{9 (u-1) u} \nnb \\
&&    -\frac{\pi ^4 \left(62 u^5+2 u^4-195 u^3+720 u^2-700 u+240\right)}{405 (u-1)^2 u^3} +\frac{4 \left(u^2-3\right) \ln (u)\text{Li}_2(-u)}{9 u^3}\nnb \\
&&    +\frac{2 \pi ^2 (u-2)^2 \ln (2-u)}{3 (u-1)^2}-\frac{(151 u-102) \ln (u)}{9 (u-1)}+\frac{2 (u-2) \left(u^2+4 u-3\right)\ln ^3(u)}{27 (u-1)} \nnb \\
&&    +\frac{2 \pi ^2 \left(u^5+2 u^4-23 u^3-15 u^2+136 u-96\right)\ln (u)}{27 (u-1) u^2}-\frac{4 (u-2)^2 \text{Li}_3(u-1)}{3 (u-1)^2} \nnb \\
&&    -\frac{4 \pi ^2 \left(4 u^5-22 u^4+61 u^3-144 u^2+140 u-48\right) \text{Li}_2(u)}{27(u-1)^2 u^3} \nnb \\
&&    +\frac{4 \left(6 u^4-26 u^3+97 u^2-75 u+3\right) \ln (u) \text{Li}_2(u)}{9(u-1) u^3} +\frac{8}{9} \ln ^2(u)\text{Li}_2(u)\nnb \\
&&    -\frac{4 \left(2 u^7+u^6-18 u^5+79 u^4-111 u^3+63 u^2-10 u-3\right)\text{Li}_3(1-u)}{9 (u-1)^2 u^3} \nnb \\
&&    -\frac{32 (u-4) (2 u-3) \ln (u) \text{Li}_3(1-u)}{9 u^3}  +\frac{2 \left(4 u^4-97 u^3+51 u^2+83 u-42\right) \text{Li}_2(u)}{9 (u-1)u^2}            \nnb \\
&&    -\frac{4 \left(2 u^7+u^6-12 u^5-12 u^4+139 u^3-220 u^2+102 u-3\right) \text{Li}_3(u)}{9 (u-1)^2 u^3} \nnb \\
&&    -\frac{16}{3} \ln (u) \text{Li}_3(u)+\frac{8\left(2 u^5+14 u^4-69 u^3+180 u^2-175 u+60\right) \text{Li}_4(1-u)}{9 (u-1)^2u^3} \nnb \\
&&    +\frac{8 \left(2 u^5-22 u^4+57 u^3-144 u^2+140 u-48\right) \text{Li}_4(u)}{9 (u-1)^2 u^3} +\frac{1507}{18}\nnb \\
&&    -\frac{32 \left(4 u^4-14 u^3+36 u^2-35 u+12\right)\text{Li}_4\left(\frac{u}{u-1}\right)}{9 (u-1)^2 u^3}+\frac{32 (u-4) (2 u-3) \ln (u)\zeta (3)}{9 u^3} \nnb \\
&&    -\frac{4 \left(36 u^5-86 u^4-67 u^3+227 u^2-110 u+3\right) \zeta (3)}{9 (u-1)^2 u^3}-\frac{4 \left(u^2-3\right) \text{Li}_3(-u)}{9 u^3} \nnb \\
&&    +L \left(-\frac{4}{3} \ln ^2(1-u)-\frac{4 (3 u-2) \ln (1-u)}{3 u}-\frac{8}{3} \ln (u) \ln (1-u) \right. \nnb \\
&&    \left. +\frac{4 \ln ^2(u)}{3}-\frac{4 (3 u-2) \ln (u)}{3(u-1)}-\frac{16 \text{Li}_2(u)}{3}+\frac{4 \pi ^2}{9}+34\right)  +4 L^2 \, ,
\eea
\bea
  T_2^{(2),im}  &=& \frac{8}{27} \ln ^3(1-u)-\frac{2 \left(u^3+2 u^2-5 u+6\right) \ln ^2(1-u)}{9u} +\frac{16}{9} \ln (u) \ln ^2(1-u)\nnb \\
&&  -\frac{4\left(u^2+29 u-5\right) \ln (1-u)}{9 u}+\frac{4 \left(u^4+2 u^3-5 u^2+6 u-2\right) \ln(u) \ln (1-u)}{9 (u-1) u} \nnb \\
&&  +\frac{32}{9} \text{Li}_2(u) \ln (1-u)-\frac{8}{9} \pi ^2\ln (1-u) -\frac{16}{9} \ln ^2(u) \ln (1-u) \nnb \\
&&  +\frac{8 \ln ^3(u)}{27}-\frac{2 \left(u^3+2 u^2+u-1\right) \ln ^2(u)}{9(u-1)} -\frac{2 \pi ^2 \left(3 u^3+5 u^2-11 u+4\right)}{27 (u-1)}\nnb \\
&&  +\frac{4}{9} (u+21) \ln(u) -\frac{8}{27} \pi ^2 \ln (u)  +\frac{4}{9} u \text{Li}_2(u)-\frac{32}{9} \ln (u) \text{Li}_2(u)+\frac{4 \text{Li}_2(u)}{9 (u-1)} \nnb \\
&&  -\frac{8 \text{Li}_2(u)}{9 u}-\frac{4\text{Li}_2(u)}{3}+\frac{16}{9} \text{Li}_3(1-u)+\frac{16\text{Li}_3(u)}{9}+\frac{155}{9}  \nnb \\
&&  +L \left(-\frac{8}{3} \ln (1-u)+\frac{8 \ln(u)}{3}+4\right) \, .
\eea
\subsection{Charming functions}
\label{ap:charmfunc}

In this Appendix we list the expressions for the functions that 
depend on $m_c$. We define $z=m_c^2/m_b^2$ and 
\bea
\eta &\equiv& \frac{4 z}{\left(\sqrt{4 z+1}+1\right)^2} \; ,
\nonumber \\
\kappa &\equiv& 1+\sqrt{z} \; .
\label{eq:defeta}
\eea
In terms of these variables our building blocks read
\bea
q_1(z) &=& \ln^3(\eta) + 6 \ln(\eta) \PL{2}{\eta}-6\PL{3}{\eta}+6\zeta(3) \; ,
\nonumber \\
q_2(z) &=& \ln(\eta) \sqrt{4 z+1} \; ,\nonumber \\
q_3(z) &=& 9 \ln^4(\eta)-12\pi^2\ln^2(\eta)+
144 \, {\rm Li}_{2}^2\!\left({\eta}\right) - 
48 \pi^2 \PL{2}{\eta}+4 \pi^4 \; ,\nonumber \\
q_4(z) &=& \ln^3(\eta) - 2 \pi^2 \ln(\eta) +12 \PL{3}{\eta}-12\zeta(3) \; ,
\nonumber \\
q_5(z) &=& -\ln^3(\eta) -4 \ln^2(\eta) \PL{2}{\eta} + 2 \pi^2 \ln(\eta) -12 \PL{3}{\eta}+12\zeta(3) \; ,\nonumber \\
q_6(z) &=& \left[ -3 \ln^2(\eta) -12 \PL{2}{\eta} +2 \pi^2\right] 
\sqrt{4 z+1} \; ,\nonumber \\
q_7(z) &=& \ln^2(z) + 2 \PL{2}{1-z} +\pi^2 \; ,\nonumber \\
q_8(z) &=& 2 \ln(\kappa) \ln(z)-\ln(1-z)\ln(z)-4 \PL{2}{\sqrt{z}}+\PL{2}{z}+\pi^2 \; ,\nonumber\\
q_9(z) &=&  \PL{3}{1-z} -2 \PL{3}{1-\sqrt{z}}  - 2 \PL{3}{\sqrt{z}}+ \PL{3}{z}
         + 2 \PL{3}{\frac{\sqrt{z}}{\sqrt{z}+1}} -2 \zeta(3) \; , \nnb\\ 
q_{10}(z) &=& -\ln^3(z) + 3 \ln(1-z)\ln^2(z)+ 6 \ln^2(\kappa) \ln(z)-4 \pi^2 \ln(z)-4\ln^3(\kappa) \nnb \\
       &&  +8 \pi^2 \ln(\kappa) +12 \, q_9(z) \; , \nonumber\\
q_{11}(z) &=& \ln^4(z) + 24 \ln^2(\kappa)\ln^2(z) -48 \ln(1-z)\ln(\kappa)\ln^2(z) +8 \pi^2 \ln^2(z) \nnb \\
       && +12 \PL{2}{z}\ln^2(z) - 8 \pi^2 \ln(1-z)\ln(z)+16 \pi^2 \ln(\kappa)\ln(z) +16 \ln^3(\kappa)\ln(z)\nnb\\
       && -96 \ln(1-\sqrt{z}) \ln(z) \PL{2}{-\sqrt{z}}-192 \PL{2}{-\sqrt{z}}\PL{2}{\sqrt{z}}+40 \pi^2 \PL{2}{z}\nnb\\
       && - 96 \ln(\kappa) \ln(z) \PL{2}{\sqrt{z}}-96\pi^2 \PL{2}{\sqrt{z}} +8 \pi^4 - 48 \ln(z) \, q_9(z) \; , \nonumber\\
q_{12}(z) &=& -4 \ln^3(1-z) +6 \ln(z) \ln^2(1-z) +12 \ln(2\kappa) \ln^2(1-z)+3 \ln^2(z)\ln(1-z) \nnb \\
       && -12 \ln^2(2\kappa)\ln(1-z)-24 \ln(z)\ln(\kappa)\ln(1-z)-4 \pi^2\ln(1-z)+4 \ln^3(\kappa)\nnb \\
       && + 18 \ln^2(\kappa)\ln(z) - 12 \ln^2(2)\ln(z) + 8 \pi^2 \ln(8 z/\kappa)-6 \ln(\kappa)\ln^2(z)\nnb \\
       && + 24 \ln(2)\ln(2 z)\ln(\kappa)-12 \ln(z) \PL{2}{z}-24 \ln(z) \PL{2}{\frac{\kappa}{2}}-24 \PL{3}{\frac{1-\sqrt{z}}{2}}\nnb \\
       && +24 \PL{3}{\frac{2\sqrt{z}}{\sqrt{z}-1}}-24 \PL{3}{\frac{2\sqrt{z}}{\sqrt{z}+1}}+24 \PL{3}{\frac{\sqrt{z}}{\sqrt{z}+1}}
           +24 \PL{3}{\frac{\kappa}{2}}\nnb\\
       && +24 \PL{3}{1-\sqrt{z}}-72 \PL{3}{\sqrt{z}}+48 \ln(z) \PL{2}{\sqrt{z}}+12 \PL{3}{z}-24\zeta(3) \; .
\label{eq:q8}
\eea

\end{document}